\documentclass[%
 reprint,
nofootinbib,
 amsmath,amssymb,
 aps,
]{revtex4-2}

\usepackage{array}[=2016-10-06]
\usepackage{amsmath}
\usepackage{multirow}
\usepackage{graphicx}
\usepackage{dcolumn}
\usepackage{tabularx}
\usepackage{bm}
\usepackage{newtxtext,newtxmath}
\usepackage{subcaption}
\usepackage[T1]{fontenc}
\usepackage{makecell}
\usepackage{animate}
\usepackage{mathtools}
\usepackage{url}
\usepackage[hidelinks,colorlinks]{hyperref}
\usepackage{xcolor}
\hypersetup{
    colorlinks=true,
    linkcolor={blue},
    citecolor={blue},
    urlcolor={blue}
}

\def\hlinewd#1{%
\noalign{\ifnum0=`}\fi\hrule \@height #1 %
\futurelet\reserved@a\@xhline} 
\usepackage{graphicx}	
\usepackage{amsmath}	

\def\apjl{Astrophys. J. Lett.}
\def\apjs{Astrophys. J. Suppl. Ser.}
\def\aap{Astron. Astrophys.}

\begin{document}

\preprint{APS/123-QED}


\title{Bypassed Core Formation in Milky Way-Mass SIDM Halos: Implications for the Local Group Past-Pericenter Scenario}


\author{Zhichao Carton Zeng$^{1}$}\email{E-mail: zczeng@tamu.edu}
\author{Odelia V. Hartl$^{1}$}
\author{Louis E. Strigari$^{1}$}
\author{Annika H. G. Peter$^{2,3,4}$}
\author{Xiaolong Du$^{5,6}$}
\author{Charlie Mace$^{2,3}$}
\author{Andrew Benson$^{5}$}

\affiliation{$^{1}$Department of Physics and Astronomy, Mitchell Institute for Fundamental Physics and Astronomy, Texas A\&M University, College Station, Texas 77843, USA \\
$^{2}$Department of Physics, The Ohio State University, 191 W. Woodruff Ave., Columbus OH 43210, USA \\
$^{3}$Center for Cosmology and Astroparticle Physics, The Ohio State University, 191 W. Woodruff Ave., Columbus OH 43210, USA\\
$^{4}$Department of Astronomy, The Ohio State University, 140 W. 18th Ave., Columbus OH 43210, USA\\
$^{5}$ Carnegie Observatories, 813 Santa Barbara Street, Pasadena CA 91101, USA\\
$^{6}$ Department of Physics and Astronomy, University of California, Los Angeles, CA 90095, USA\\
}

\date{\today}

\begin{abstract}
We consider a scenario in which the Milky Way (MW) and M31 have had a previous pericentric passage, and investigate its compatibility with self-interacting dark matter (SIDM). Using initial conditions sampled from Local Group (LG) analogues in the IllustrisTNG simulation, we perform controlled re-simulations of the MW-M31 orbit, evolving the system under both standard cold dark matter (CDM) and various SIDM cross-sections. 
We find that the deep baryonic potential of the MW preconditions the halo's thermal structure, establishing an initial negative temperature gradient. This drives SIDM halos to bypass the standard core-formation phase and enter immediate core-collapse, resulting in monotonically increasing central densities. In full orbital simulations, the compact stellar component (disk/bulge) of the MW analog remains robust against tidal disruption for pericenter distances as close as $r_{\rm peri}\lesssim20$ kpc during an encounter at cosmic time $\sim8$ Gyr. The diffuse stellar halo is comparatively more susceptible, facing disruption for $r_{\rm peri}\lesssim100$ kpc. 
Our results demonstrate a dichotomy in structural evolution: the compact disk/bulge is sensitive to intrinsic SIDM thermodynamics but dynamically robust against the pericenter encounter, whereas the diffuse stellar halo is largely independent of the specific SIDM model but more vulnerable to orbital tidal disruptions.
\end{abstract}

\maketitle


\section{Introduction}

The Local Group (LG) is defined as the approximately spherical volume with a $\sim 1$ Mpc radius centered on the Milky Way (MW)--M31 barycenter. Estimates of the LG mass lie in the range $\sim 2$--$4 \times 10^{12}$ M$_\odot$, which is dominated by the mass contained within the MW and M31~\cite{1959ApJ...130..705K,2008MNRAS.384.1459L,hartl21,2023MNRAS.521.4863S}. In addition to the MW and M31, the LG contains dozens of less luminous dwarf galaxies, bound to either the MW, to M31, or to the LG as a system~\cite{1998ARA&A..36..435M}. Because of its proximity, individual stars within galaxies in the LG are resolved, allowing a unique window into the kinematics, formation, and dark matter properties~\cite{simon2019}. 

The classical model for the formation of the LG is that the MW and M31 first formed with initial conditions given by the Hubble flow, then detached and have been on a first-infalling, radial approaching orbit. Recent Gaia and HST measurements have provided an important insight into this simple scenario and to the 3D nature of the MW--M31 orbit~\cite{vdm2019ApJ...872...24V,Salomon2021MNRAS.507.2592S}. In particular, the preference for a non-negligible tangential velocity has challenged the long-standing model of a pure MW--M31 radial orbit~\cite{strigari25}. The existence of a tangential velocity not only provides clues as to the past history of the orbit, it also has interesting implications for the future evolution of the LG~\cite{sawala25}. 

Given the measured kinematics, there is a wide range of possible orbits, and in fact, it has been recognized that a past pericentric passage is consistent with the data~\cite{hartl24}. Such a model has been long ago speculated, and has recently been re-considered in the context of MW and M31 mass measurements \cite{benisty21}. Though not often discussed in the literature, a previous MW--M31 pericentric passage may leave observational imprints on the MW and M31. However, because there are no significant tidal shocks observed in the LG, a very close pericentric passage between the galaxies in the recent past is unlikely. 

In addition, such an event may be impacted by the nature of dark matter, in particular whether it is a standard cold dark matter (CDM) model, or if dark matter is self-interacting. Self-interacting dark matter (SIDM) can lower the central density of halos through particle scattering and subsequent thermalization, leading to the formation of constant-density cores \cite{spergel00}. In the absence of strong baryonic contraction, this would render the MW (and M31) central potentials shallower, potentially making the galaxies more susceptible to tidal disruption during the pericenter passage. Beyond gravitational tides, the direct scattering between MW and M31 dark matter particles, a process known as `evaporation' \cite{kahlhoefer14, kummer18, mv19, nadler20, slone21, zzc22, klemmer26}, can unbind dark matter from the halos, exacerbating the disruption. These factors motivate a detailed exploration of the compatibility between the LG past-pericenter scenario and SIDM physics.

SIDM has undergone a recent renaissance, driven by the prediction of a two-phase evolutionary track: halos may evolve from a lower-density, cored configuration (core formation phase) to a cuspy, ultra-high-density state (core collapse phase) \cite{balberg02, balberg02b, essig19}. This mechanism naturally generates significant diversity in central halo densities, a variance further amplified by the hierarchical assembly of subhalos within host halos \cite{sameie20, nishikawa20, slone21, zzc22, zzc23, zzc24, dnyang22, nadler23, dnyang23, sashimi-sidm, shirasaki22}. Consequently, SIDM offers the potential to explain a series of emerging observational anomalies on small scales, including the diversity problem of dwarf galaxies \cite{oman15, hayashi20, santos20}, the presence of individual ultra-dense perturbers in strongly lensed systems \cite{minor21b, nadler23, despali24, enzi25, tajalli25, sbli25, vegetti26}, statistical excess in galaxy-galaxy lensing signals \cite{meneghetti20, dnyang21, meneghetti23}, dense low-mass perturbers affecting stellar streams \cite{bonaca18, xyzhang24, nibauer25}, and the formation of high-redshift supermassive black holes beyond the Eddington limit \cite{balberg02b, pollack15, wxfeng21, fz25, roberts25, wxfeng25}. 

Conversely, existing constraints on the self-interaction cross-section $\lesssim\mathcal{O}(1)\rm\ cm^2/g$ are derived primarily from galaxy clusters, whose characteristic velocities $\gtrsim1000$ km/s (cluster central density \cite{rocha13, elbert18, andrade22, odonnell26}; cluster ellipticity \cite{Peter13, robertson19, McDaniel21}; DM-galaxy offset and halo survival after cluster mergers \cite{merten11, kahlhoefer14, kim17, robertson17, valdarnini23}; see Table I in \cite{tulin17} for an overview). Fewer studies exist at the Milky Way mass scale compared to dwarfs or clusters \cite{robles19, sameie21, nadler25, despali25}. This is partly because the stellar-to-halo mass ratio peaks at this scale \cite{kravtsov14, behroozi19}, introducing a strong degeneracy between baryonic physics and SIDM physics. Recent hydrodynamical simulations suggest that at this mass scale, baryons can deepen the potential, causing SIDM halos to contract rather than form cores \cite{sameie21, rose22, despali25}, complicating the interpretation of isolated MW-like systems. Studies placing SIDM within the full dynamical environment of the Local Group remain even scarcer. This work therefore serves as a diagnostic study of MW-system evolution within the unique context of the Local Group environment, utilizing the past-pericenter scenario to leverage potential constraining power on SIDM models.

In this paper, we perform a detailed analysis of the MW-M31 pericentric passage scenario. We start from cosmological initial conditions for a LG analogue system with kinematics consistent with the observed LG. We extract the initial conditions of these analogs to perform controlled, idealized re-simulations, isolating the binary system from large-scale cosmological perturbations. By populating these halos with both dark matter and stellar particles, we study the co-evolution of the dark matter halo and the stellar components. This approach allows us to systematically evolve the MW and M31 analogs under various self-interacting dark matter (SIDM) models, isolating the gravitational interplay between SIDM dynamics and stellar potentials from complex hydrodynamical feedback processes. We thereby identify specific observables sensitive to the pericentric passage and quantify their dependence on the underlying dark matter physics.

This paper is organized as follows. In Sec. \ref{sec:method}, we outline the selection criteria for LG analogs in TNG300-1 that exhibit a past pericenter and select the optimal candidate for high-resolution follow-up. In Sec. \ref{sec:iso}, we simulate the MW analog in isolation to characterize the interplay between SIDM dynamics and baryonic potential depth. Specifically, we examine the evolution of compact (disk/bulge) versus diffuse (stellar halo) stellar distributions under different cross-sections, independent of environmental tidal effects. In Sec. \ref{sec:lgsim}, we model the full LG interaction, evolving both the MW and M31 analogs simultaneously across a range of SIDM models. We utilize these controlled simulations to quantify the tidal disruption experienced by the MW analog as a function of the minimum past pericenter distance. Finally, in Sec. \ref{sec:summary}, we summarize our conclusions and discuss pathways for future investigation.

\section{Methods}\label{sec:method}

\subsection{Sample selection of LG analogs}


The sampled LG analogs selected for re-simulation in this work are drawn from the Illustris TNG300-1
simulation \cite{pillepich17, springel17, nelson18}, following the criteria as outlined in \cite{hartl21} and \cite{hartl24}: a) each member of the LG analog (the MW and M31 analogs) has a B-band magnitude $-22.3<M_B<-19.3$; b) the MW--M31 analog pairs are approaching each other at $z=0$ with $v_r<0$ km/s; c) the MW-M31 analog pairs have a present-day separation $500\ \rm kpc < r < 1000\ \rm kpc$. Of the 597 pairs meeting these criteria, 77 (13\%) are found to have experienced a pericenter passage in the past prior to their current approach \cite{hartl24}. This relatively wide range of parameters was chosen in order to increase the sample size of LG analogues. By further inspecting the smoothness of the orbital history, 15 out of these 77 pairs appear to have evolved in a relatively isolated environment (see the green dashed lines in Fig. \ref{fig:orbit-pairs-peris}), with pericenter distances ranging from 60 to 700 kpc. We adopt the properties of these 15 LG analogs from TNG300-1 to initialize the idealized simulations described in the next section.

\subsection{Idealized simulations of LG analogs}\label{sec:orbitsim}

For each of these 15 LG analogs in TNG, we perform idealized simulations of their two constituents (the MW and M31 analogs) in isolation using \texttt{Arepo} \cite{Springel10} with an established SIDM module \cite{mv12, mv14, mv19}. In this section, the LG halos are composed of dark matter (DM) only at relatively low resolution with $5\times10^5$ DM particles in total, with the softening length $\epsilon$ assigned according to the recommended criterion in \cite{vdb18} to counter artificial disruption. This is because the primary objective of this section is to study the orbital evolution of the MW--M31 analog pairs, examining to what extent their orbits in the idealized simulation (free from the large-scale structure present in TNG) can reproduce the TNG orbits with the past-pericenter feature. This allows us to identify the `best' LG analog for further investigations in the next section. The detailed properties of each LG analog simulated in this section are summarized in Table \ref{table:allpairs}.

\begin{table*} 
	\centering
	\begin{tabular*}{\textwidth}{@{\extracolsep{\fill}}l|ccc|ccc|cc|ccc} 
		\hline
        \multicolumn{1}{c|}{} & 
        \multicolumn{3}{c|}{M31 analog} & 
        \multicolumn{3}{c|}{MW analog} &
        \multicolumn{2}{c|}{} &
        \multicolumn{3}{c}{orbital info in TNG at $z=0$}\\ \hline
		  alias & TNG ID & $M_{\rm bound}$ & $c_{200c}$ & TNG ID & $M_{\rm bound}$ & $c_{200c}$ & $\epsilon_{\rm soft}$[kpc] & $M_p$ &  separation [kpc] & $v_r$ [km/s] & $v_{\rm tan}$ [km/s] \\
		\hline
        \multicolumn{12}{c}{Past pericenter LG analogs}  \\
        \hline
        pair0 & 1297153 & $5.6\times10^{12}$ & 6.8 & 1297154 & $5.4\times10^{11}$ & 9.0 & 0.81 & $1.2\times10^{7}$ & 650.9 & $-83.8$ & 143.7 \\
        pair1 & 1305068 & $2.5\times10^{12}$ & 7.5 & 1305069 & $1.2\times10^{12}$ & 8.2 & 0.69 & $7.5\times10^{6}$ & 891.2 & $-99.7$ & 185.8 \\
        pair2 & 1432332 & $4.1\times10^{12}$ & 7.1 & 2070493 & $6.3\times10^{11}$ & 8.8 & 0.73 & $9.5\times10^{6}$ & 825.2 & $-132.9$ & 79.1 \\
        pair3 & 1501189 & $2.3\times10^{12}$ & 7.6 & 1501190 & $6.2\times10^{11}$ & 8.9 & 0.60 & $5.8\times10^{6}$ & 587.1 & $-144.0$ & 129.0 \\
        pair4 & 1541350 & $2.5\times10^{12}$ & 7.5 & 1541351 & $4.8\times10^{11}$ & 9.1 & 0.60 & $5.9\times10^{6}$ & 643.3 & $-142.3$ & 99.5 \\
        pair5 & 1546393 & $1.6\times10^{12}$ & 7.9 & 1546394 & $5.0\times10^{11}$ & 9.1 & 0.53 & $4.3\times10^{6}$ & 715.1 & $-98.7$ & 101.0 \\
        pair6 & 1629357 & $1.6\times10^{12}$ & 7.9 & 1629358 & $4.5\times10^{11}$ & 9.2 & 0.52 & $4.2\times10^{6}$ & 552.9 & $-65.4$ & 97.8 \\
        pair7 & 1654070 & $2.7\times10^{12}$ & 7.5 & 1978541 & $7.6\times10^{11}$ & 8.7 & 0.66 & $7.0\times10^{6}$ & 747.4 & $-82.3$ & 144.6 \\
        pair8 & 1668556 & $1.0\times10^{12}$ & 8.4 & 1668557 & $6.7\times10^{11}$ & 8.8 & 0.49 & $3.4\times10^{6}$ & 572.8 & $-94.7$ & 80.3 \\
        pair9 & 1717129 & $1.1\times10^{12}$ & 8.3 & 1717130 & $3.0\times10^{11}$ & 9.6 & 0.43 & $2.8\times10^{6}$ & 691.2 & $-64.8$ & 17.1 \\
        pair10 & 1831231 & $1.3\times10^{12}$ & 8.2 & 2002567 & $9.9\times10^{11}$ & 8.4 & 0.56 & $4.5\times10^{6}$ & 781.2 & $-69.3$ & 65.8 \\
        pair11 & 1903149 & $1.0\times10^{12}$ & 8.4 & 1961840 & $8.4\times10^{11}$ & 8.6 & 0.51 & $3.8\times10^{6}$ & 804.7 & $-73.0$ & 90.2 \\
        pair12 & 1936476 & $8.5\times10^{11}$ & 8.6 & 2273413 & $3.5\times10^{11}$ & 9.4 & 0.41 & $2.4\times10^{6}$ & 648.6 & $-94.7$ & 26.7 \\
        pair13 & 1964657 & $8.6\times10^{11}$ & 8.5 & 1937742 & $8.2\times10^{11}$ & 8.6 & 0.49 & $3.4\times10^{6}$ & 649.0 & $-65.3$ & 112.5 \\
        pair14 & 2116605 & $5.0\times10^{11}$ & 9.1 & 2050295 & $4.0\times10^{11}$ & 9.3 & 0.37 & $1.8\times10^{6}$ & 509.2 & $-47.1$ & 25.0 \\
        \hline
        \multicolumn{12}{c}{First infall LG analogs} \\ \hline
        pair0 & 1452703 & $4.5\times10^{12}$ & 7.0 & 1932977 & $9.2\times10^{11}$ & 8.5 & 0.79 & $1.1\times10^{7}$ & 913.2 & $-161.8$ & 43.1 \\
        pair1 & 1706871 & $1.1\times10^{12}$ & 8.3 & 1706872 & $7.6\times10^{11}$ & 8.7 & 0.51 & $3.7\times10^{6}$ & 527.0 & $-116.6$ & 59.3 \\
        pair2 & 1767994 & $1.1\times10^{12}$ & 8.3 & 1767995 & $4.8\times10^{11}$ & 9.1 & 0.47 & $3.2\times10^{6}$ & 549.5 & $-90.8$ & 72.2 \\
	\end{tabular*}
	\caption{Properties of each LG analog we simulate in Sec.~\ref{sec:orbitsim}. We always choose the more massive halo of the pair to be the M31 analog, and the less massive one to be the MW analog. Concentrations $c_{200c}$ follow \cite{diemer19} at $M_{\rm bound}$. The force softening lengths $\epsilon_{\rm soft}$ are assigned according to the criterion in \cite{vdb18}. All masses are in $M_\odot$.}
    \label{table:allpairs}
\end{table*}

The MW and M31 analogs' bound masses $M_{\rm bound}$ listed in Table \ref{table:allpairs} are extracted from $z=0$ snapshot of TNG. The corresponding halo concentrations for all 15 analogs in this section are drawn from the empirical mass-concentration relation of \cite{diemer19} at each subhalo's $M_{\rm bound}$. We adopt this uniform procedure because the LG analog members in TNG are identified as subhalos, for which individual concentration measurements are not provided; concentration is instead reported for Friends-of-Friends (FoF) groups. Whether the two galaxies share the same final FoF group at $z=0$ can be diagnosed from their TNG subhalo IDs in Table \ref{table:allpairs} (consecutive IDs indicate a shared group). Eight of the 15 past-pericenter analogs (pairs 0, 1, 3, 4, 5, 6, 8, and 9) share one group and thus only one group concentration; the remaining seven (pairs 2, 7, 10, 11, 12, 13, and 14) belong to two distinct groups. A consistent, TNG-based concentration assignment for every pair is therefore not available for this low-resolution orbital-reconstruction survey. The macroscopic orbital histories are not highly sensitive to the precise central concentration.
For the high-resolution follow-up in Sec. \ref{sec:iso} and \ref{sec:lgsim}, we focus exclusively on pair14. Because its two galaxies reside in distinct FoF groups at $z=0$ and are the central galaxies of their respective groups, we assign each halo the concentration of its host group ($c_{\rm M31}\approx 7.4$, $c_{\rm MW}\approx 12.1$; see Table \ref{table:pair14}), rather than the mass-concentration relation \cite{diemer19} used here.

\begin{figure*}
    \centering
    \begin{subfigure}[t]{0.29\textwidth}
        \centering
        \includegraphics[width=\textwidth, clip,trim=0.2cm 0cm 0.2cm 0cm]{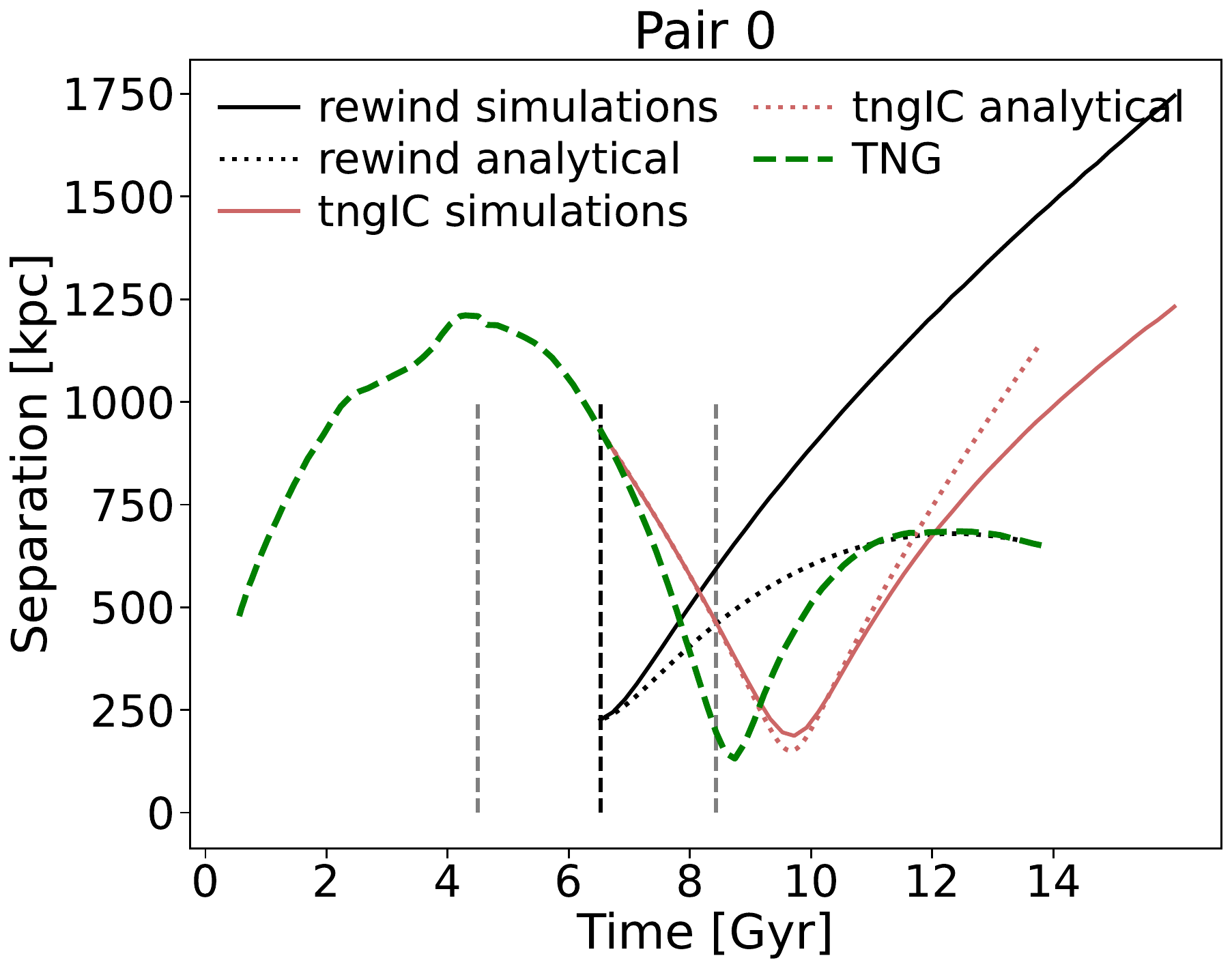}
        \label{fig:peris-0}
    \end{subfigure}
    ~
    \begin{subfigure}[t]{0.29\textwidth}
        \centering
        \includegraphics[width=\textwidth, clip,trim=0.2cm 0cm 0.2cm 0cm]{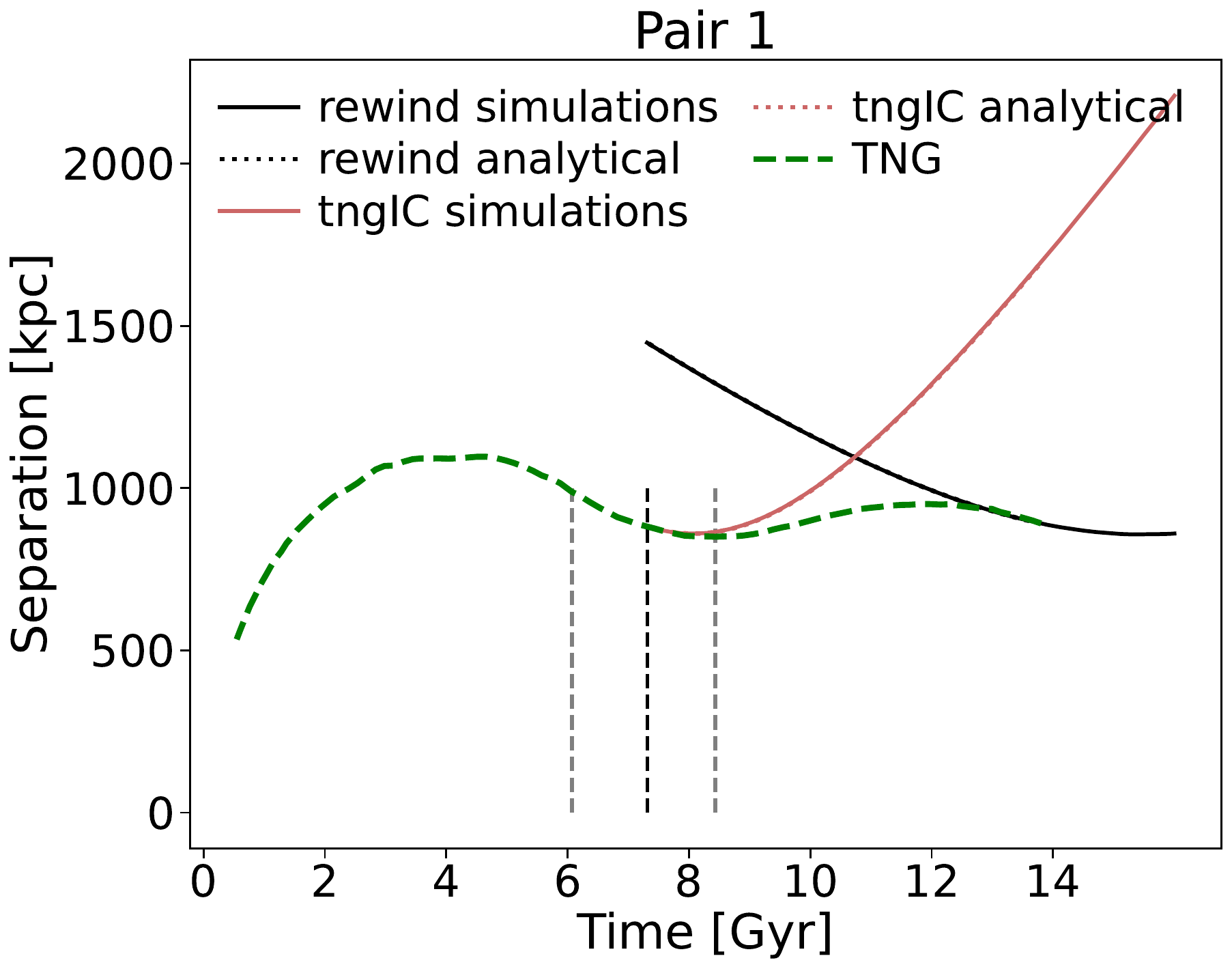}
        \label{fig:peris-1}
    \end{subfigure}
    ~
    \begin{subfigure}[t]{0.29\textwidth}
        \centering
        \includegraphics[width=\textwidth, clip,trim=0.2cm 0cm 0.2cm 0cm]{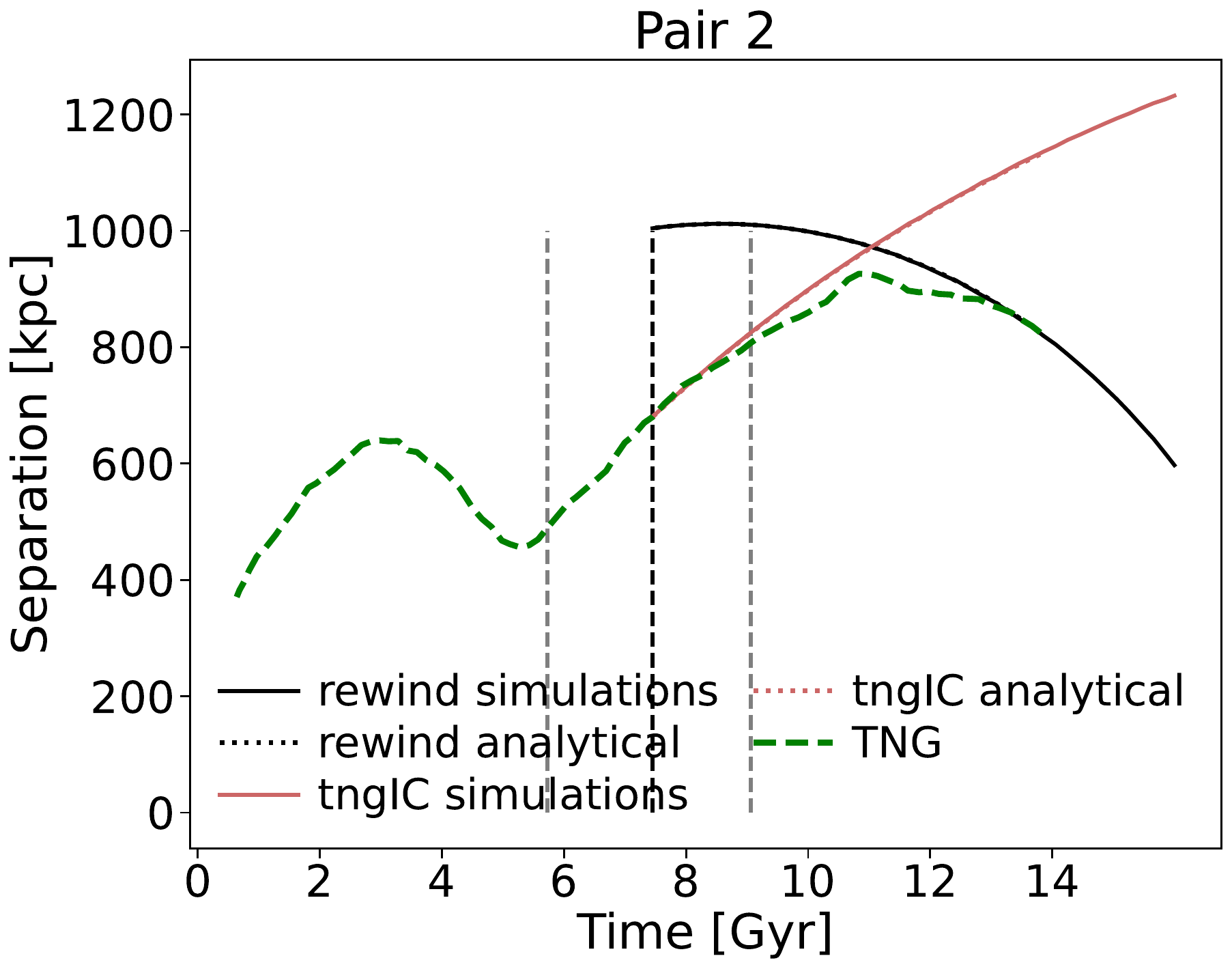}
        \label{fig:peris-2}
    \end{subfigure}
    ~
    \begin{subfigure}[t]{0.29\textwidth}
        \centering
        \includegraphics[width=\textwidth, clip,trim=0.2cm 0cm 0.2cm 0cm]{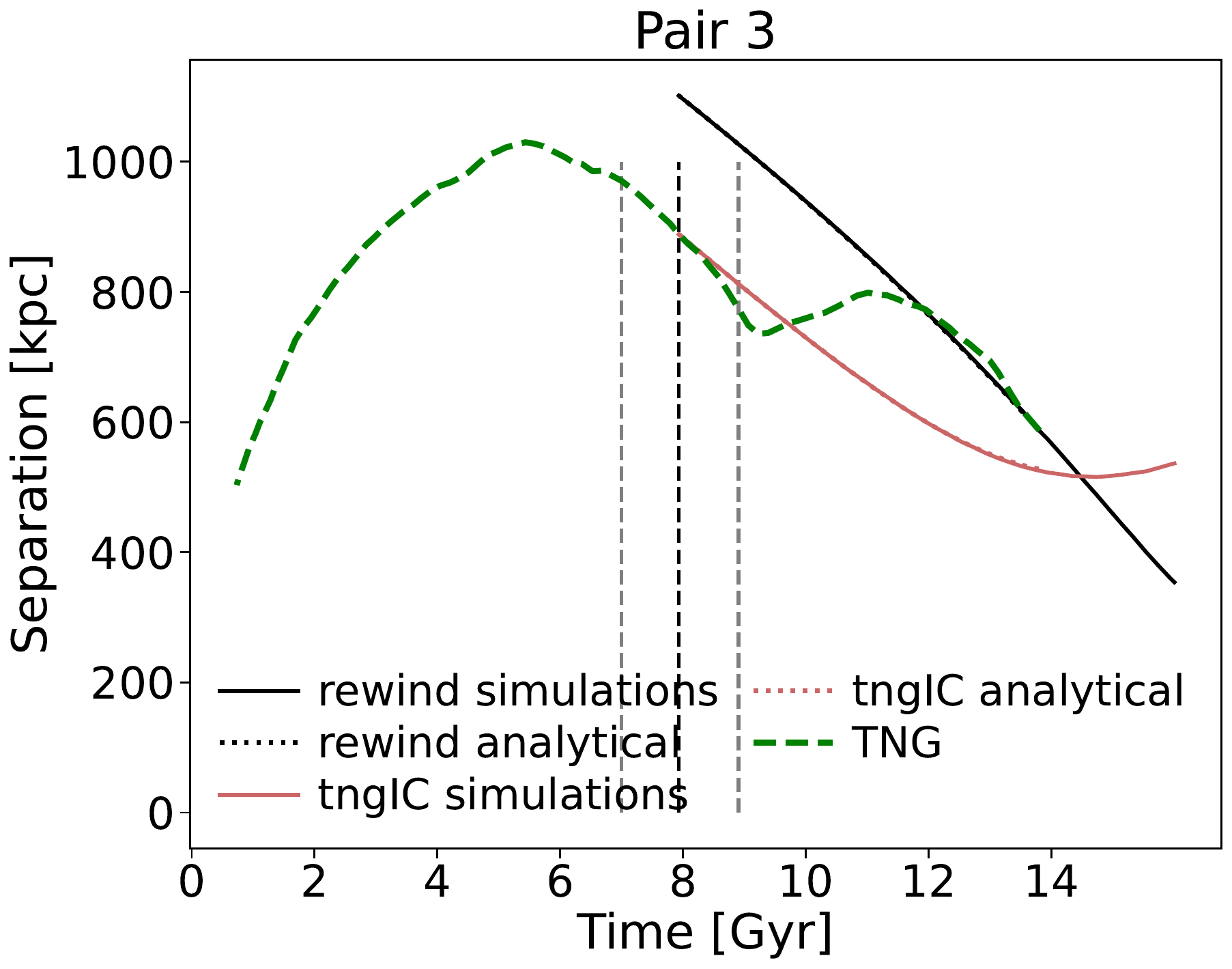}
        \label{fig:peris-3}    
    \end{subfigure}
    ~
    \begin{subfigure}[t]{0.29\textwidth}
        \centering
        \includegraphics[width=\textwidth, clip,trim=0.2cm 0cm 0.2cm 0cm]{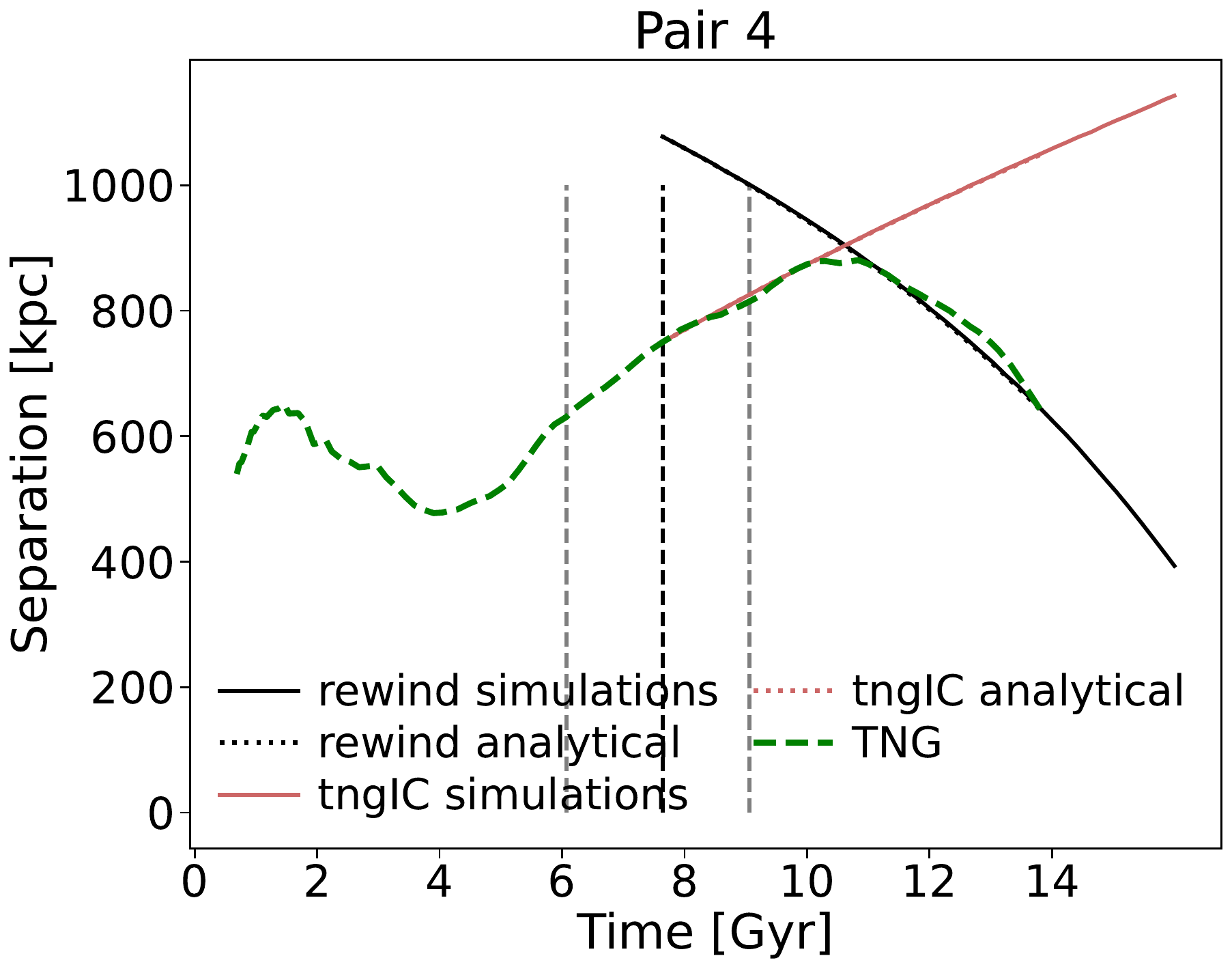}
        \label{fig:peris-4}    
    \end{subfigure}
    ~
    \begin{subfigure}[t]{0.29\textwidth}
        \centering
        \includegraphics[width=\textwidth, clip,trim=0.2cm 0cm 0.2cm 0cm]{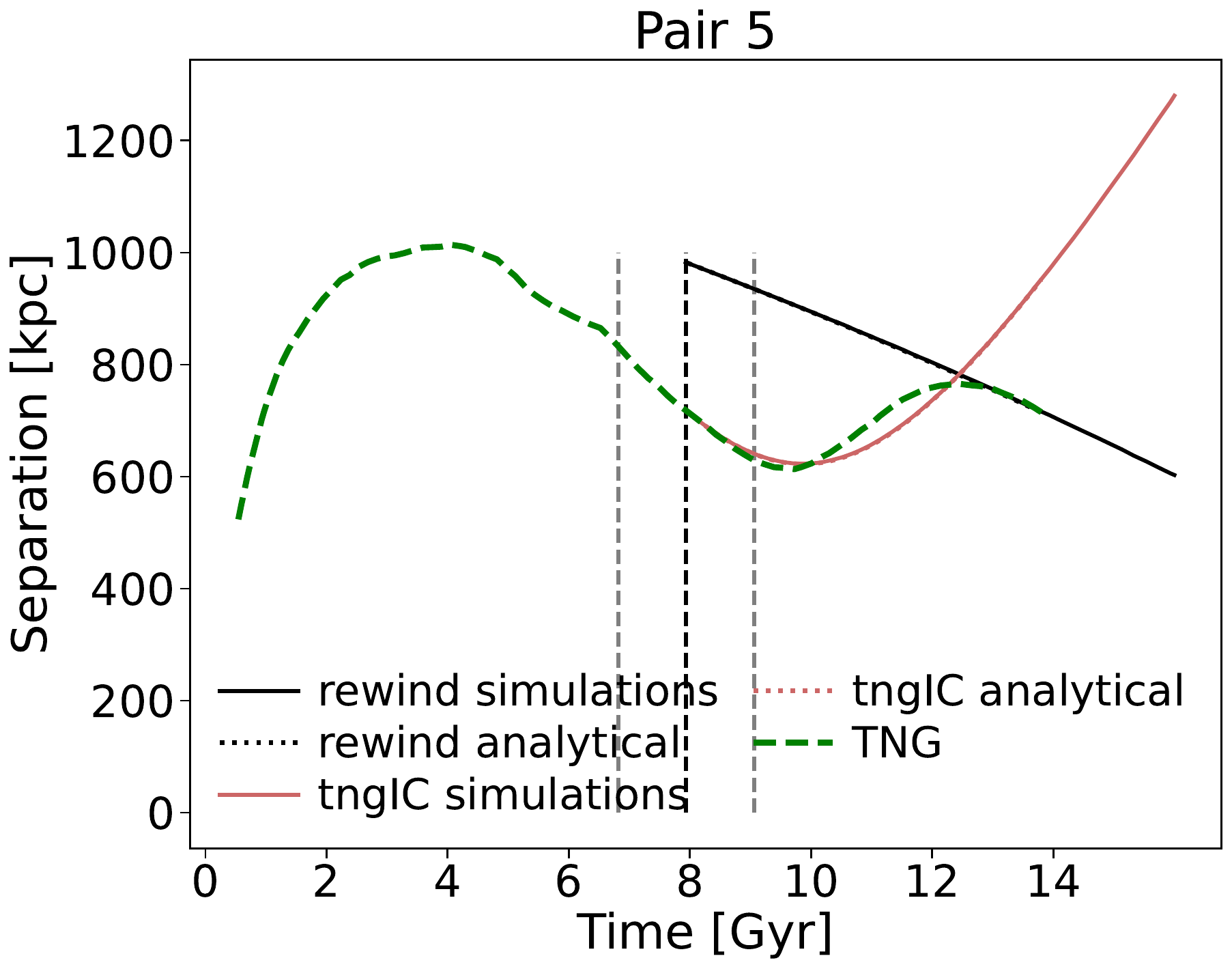}
        \label{fig:peris-5}    
    \end{subfigure}
    ~
    \begin{subfigure}[t]{0.29\textwidth}
        \centering
        \includegraphics[width=\textwidth, clip,trim=0.2cm 0cm 0.2cm 0cm]{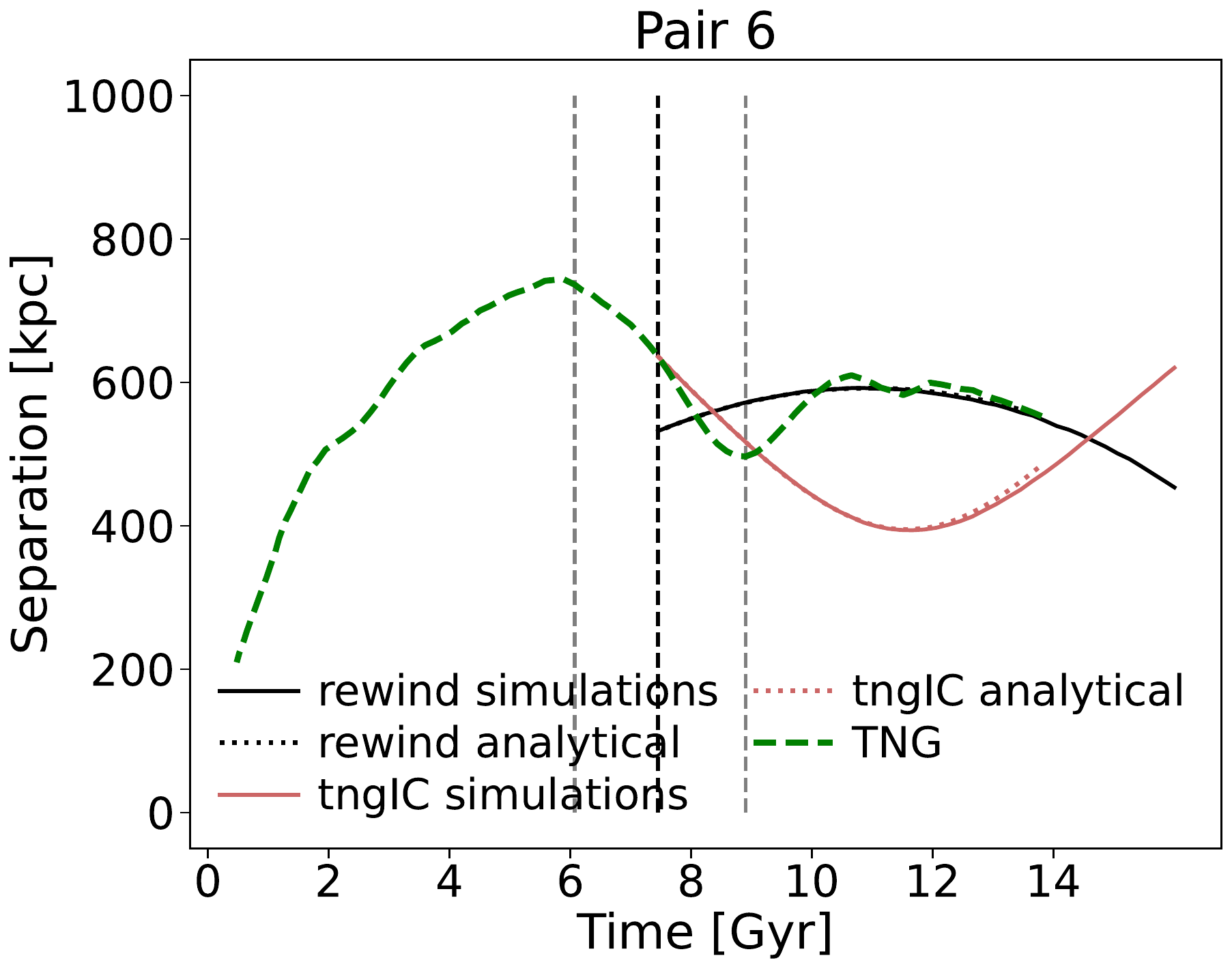}
        \label{fig:peris-6}    
    \end{subfigure}
    ~
    \begin{subfigure}[t]{0.29\textwidth}
        \centering
        \includegraphics[width=\textwidth, clip,trim=0.2cm 0cm 0.2cm 0cm]{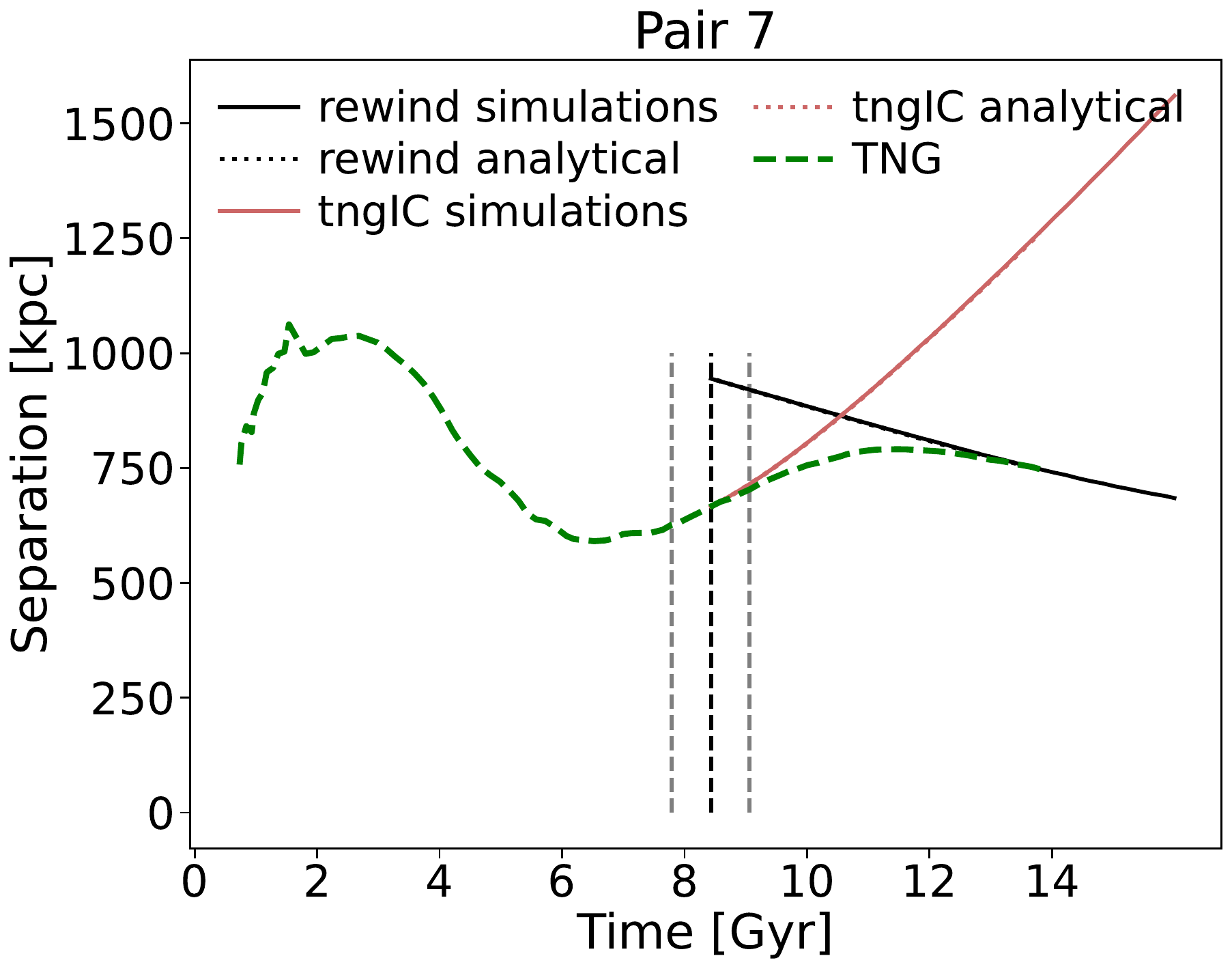}
        \label{fig:peris-7}    
    \end{subfigure}
    ~
    \begin{subfigure}[t]{0.29\textwidth}
        \centering
        \includegraphics[width=\textwidth, clip,trim=0.2cm 0cm 0.2cm 0cm]{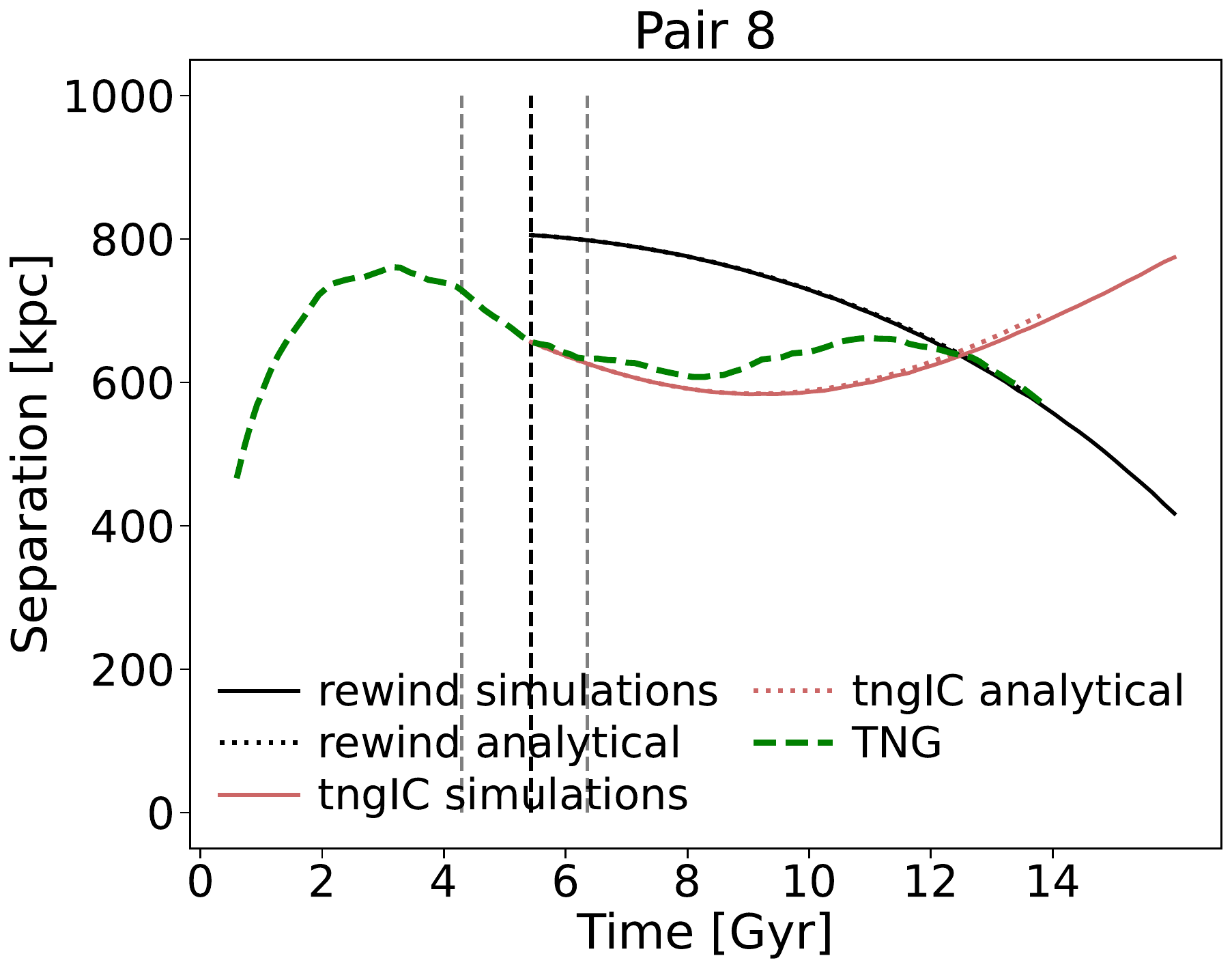}
        \label{fig:peris-8}    
    \end{subfigure}
    ~
    \begin{subfigure}[t]{0.29\textwidth}
        \centering
        \includegraphics[width=\textwidth, clip,trim=0.2cm 0cm 0.2cm 0cm]{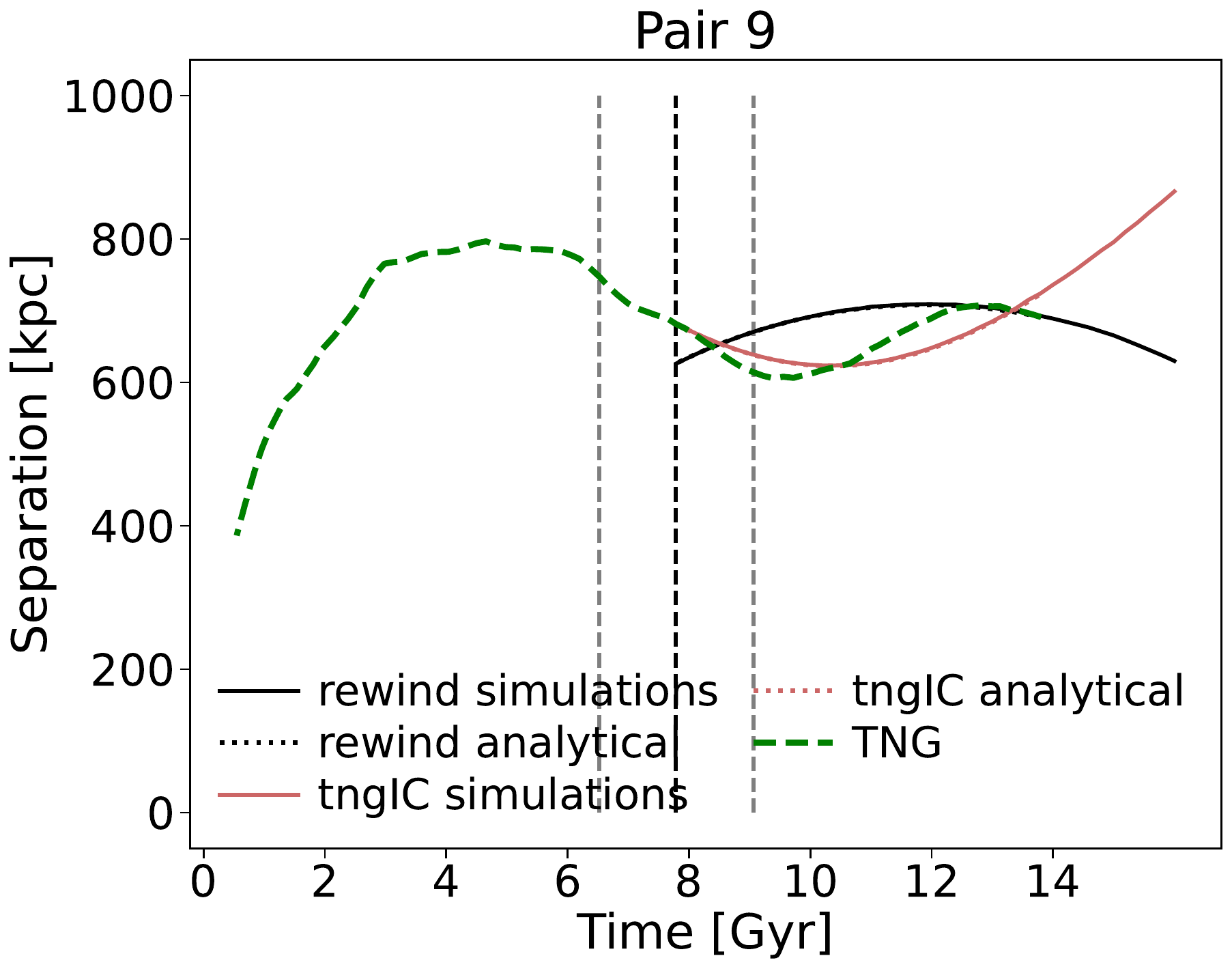}
        \label{fig:peris-9}    
    \end{subfigure}
    ~
    \begin{subfigure}[t]{0.29\textwidth}
        \centering
        \includegraphics[width=\textwidth, clip,trim=0.2cm 0cm 0.2cm 0cm]{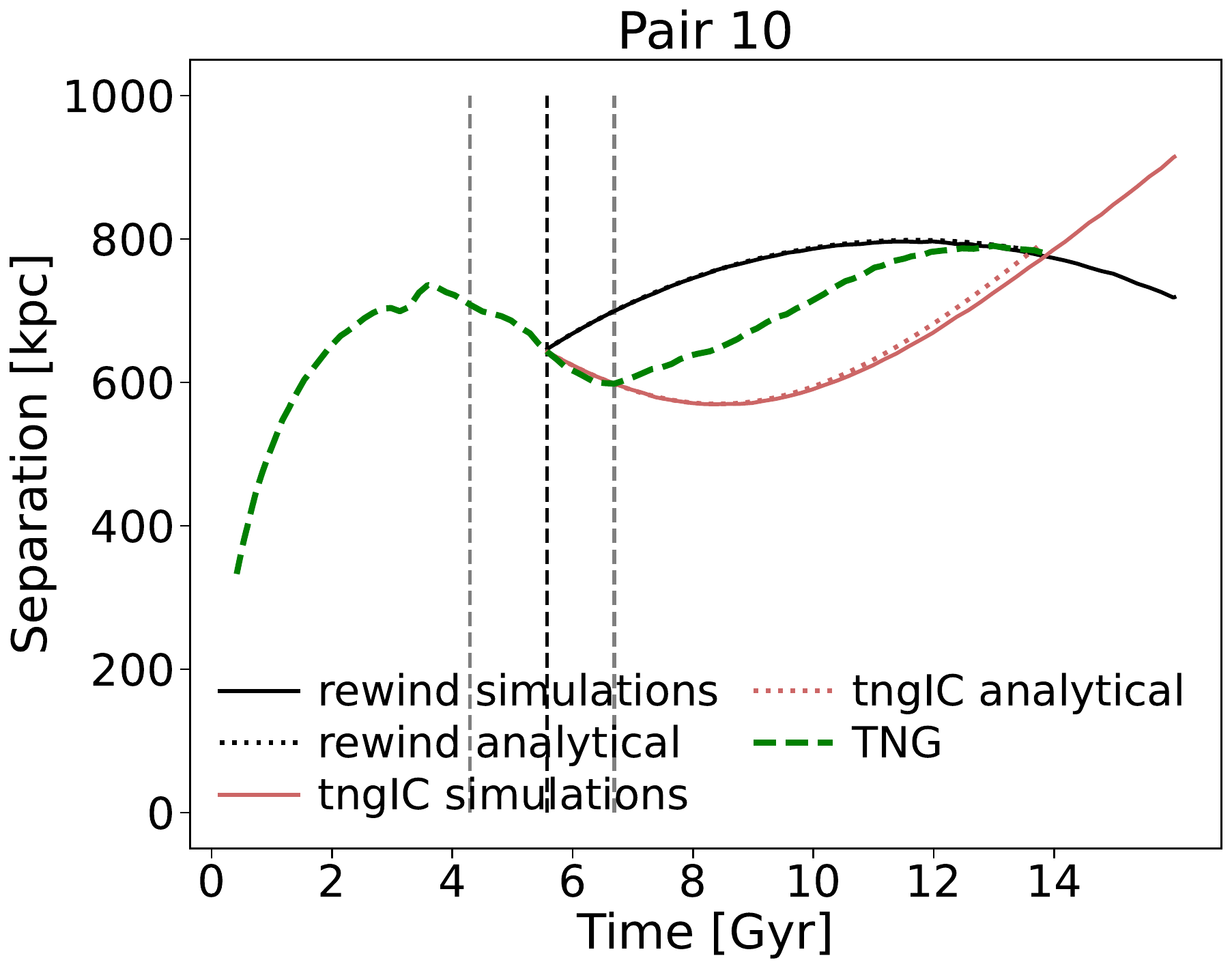}
        \label{fig:peris-10}    
    \end{subfigure}
    ~
    \begin{subfigure}[t]{0.29\textwidth}
        \centering
        \includegraphics[width=\textwidth, clip,trim=0.2cm 0cm 0.2cm 0cm]{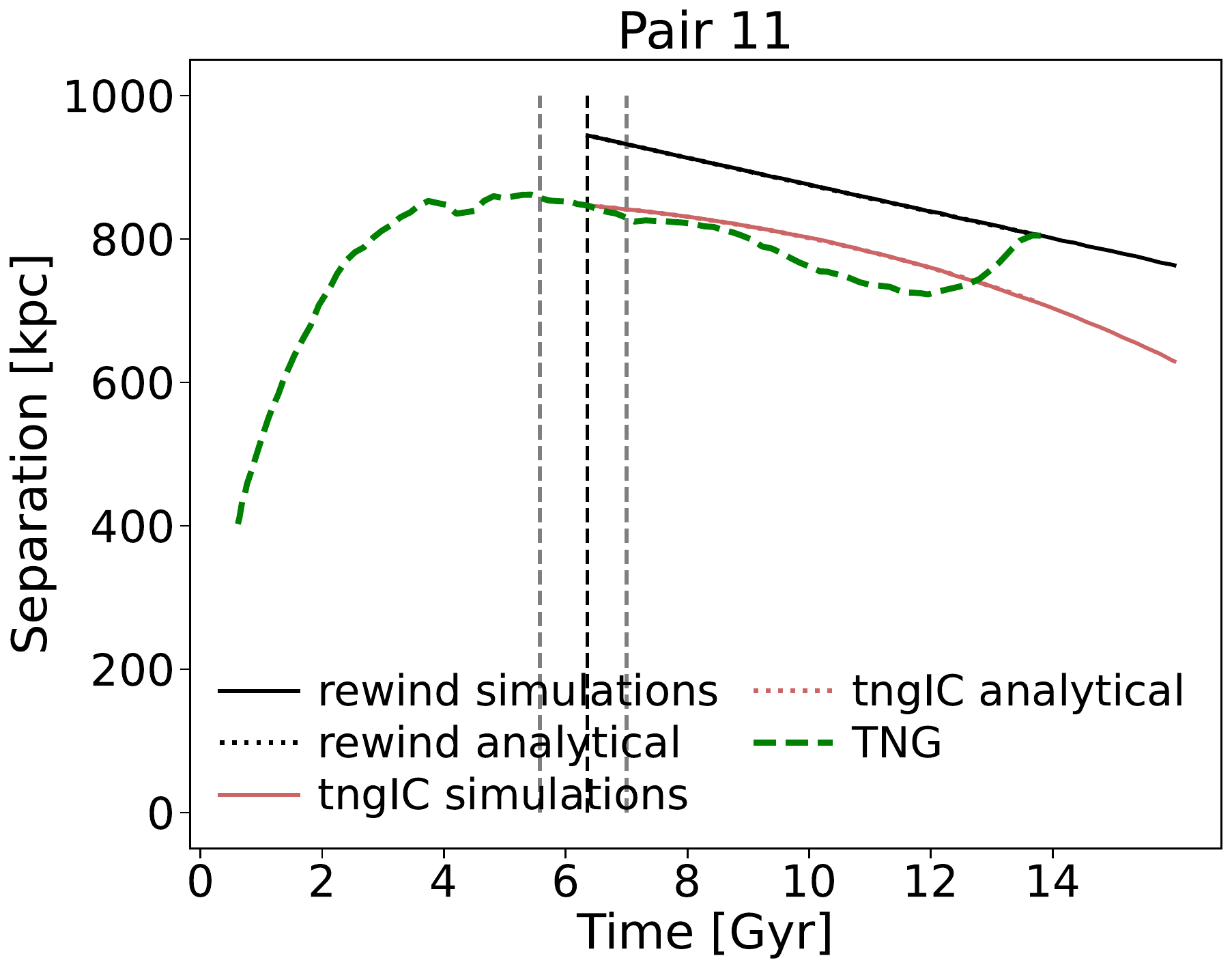}
        \label{fig:peris-11}    
    \end{subfigure}
    ~
    \begin{subfigure}[t]{0.29\textwidth}
        \centering
        \includegraphics[width=\textwidth, clip,trim=0.2cm 0cm 0.2cm 0cm]{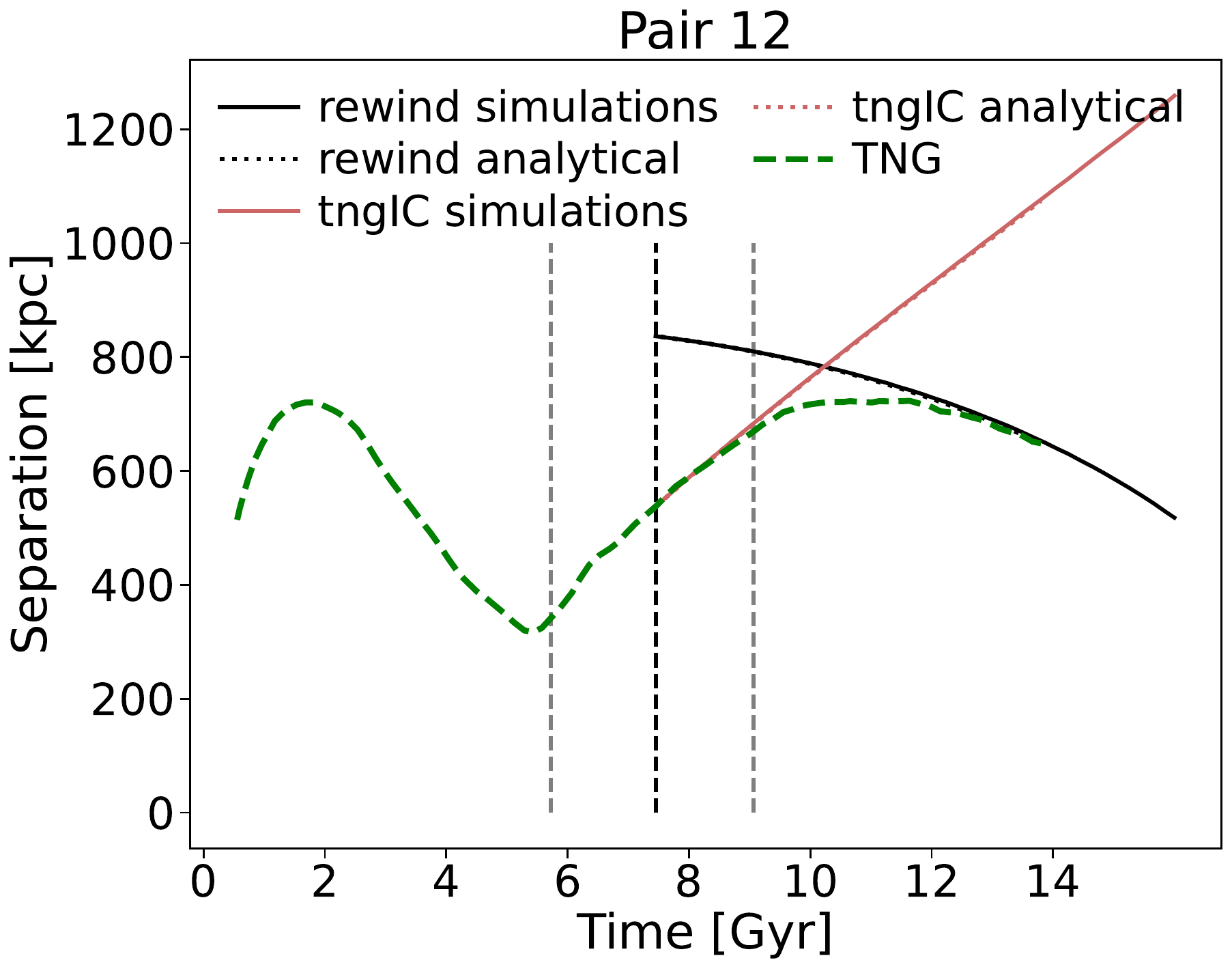}
        \label{fig:peris-12}    
    \end{subfigure}
    ~
    \begin{subfigure}[t]{0.29\textwidth}
        \centering
        \includegraphics[width=\textwidth, clip,trim=0.2cm 0cm 0.2cm 0cm]{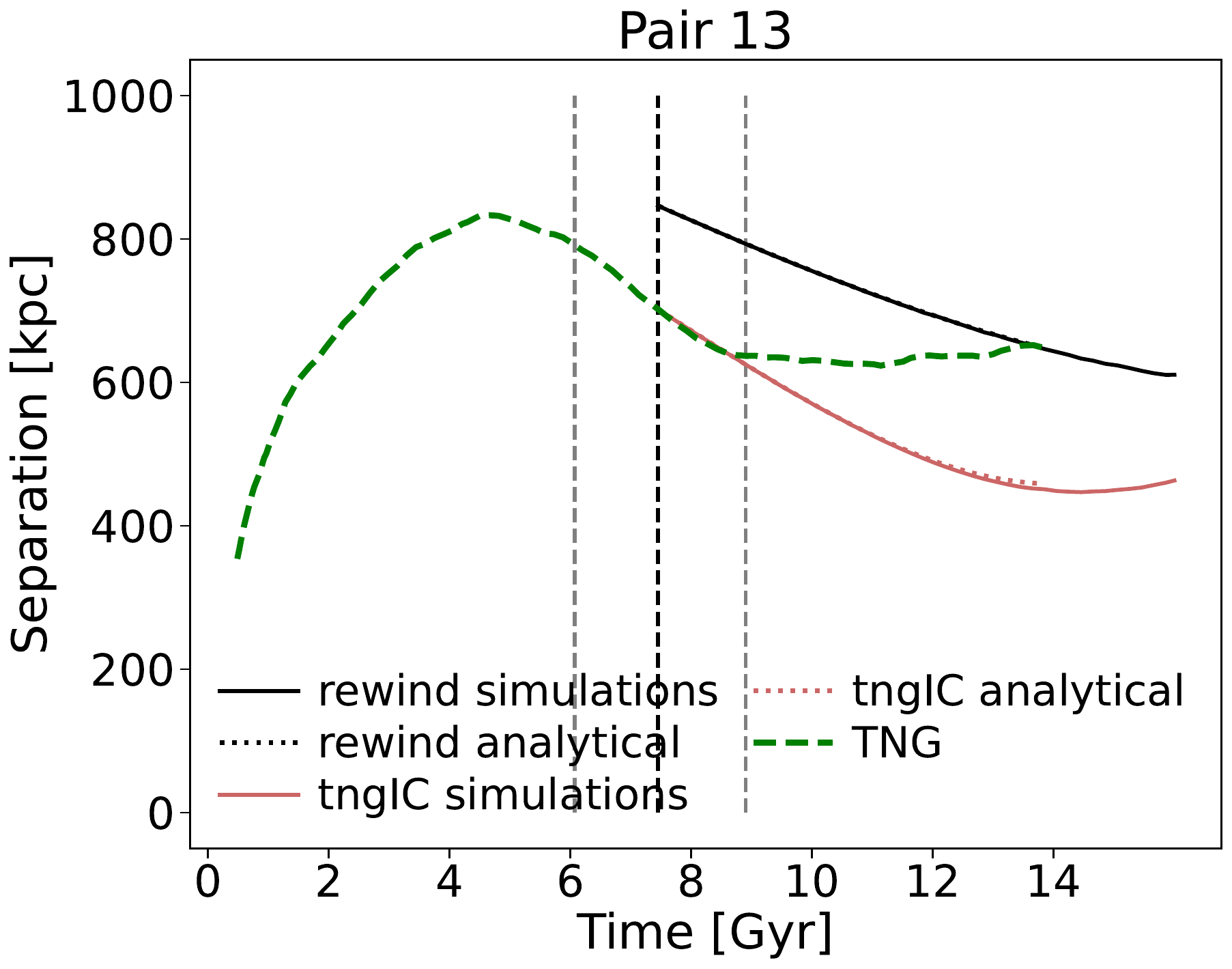}
        \label{fig:peris-13}    
    \end{subfigure}
    ~
    \begin{subfigure}[t]{0.29\textwidth}
        \centering
        \includegraphics[width=\textwidth, clip,trim=0.2cm 0cm 0.2cm 0cm]{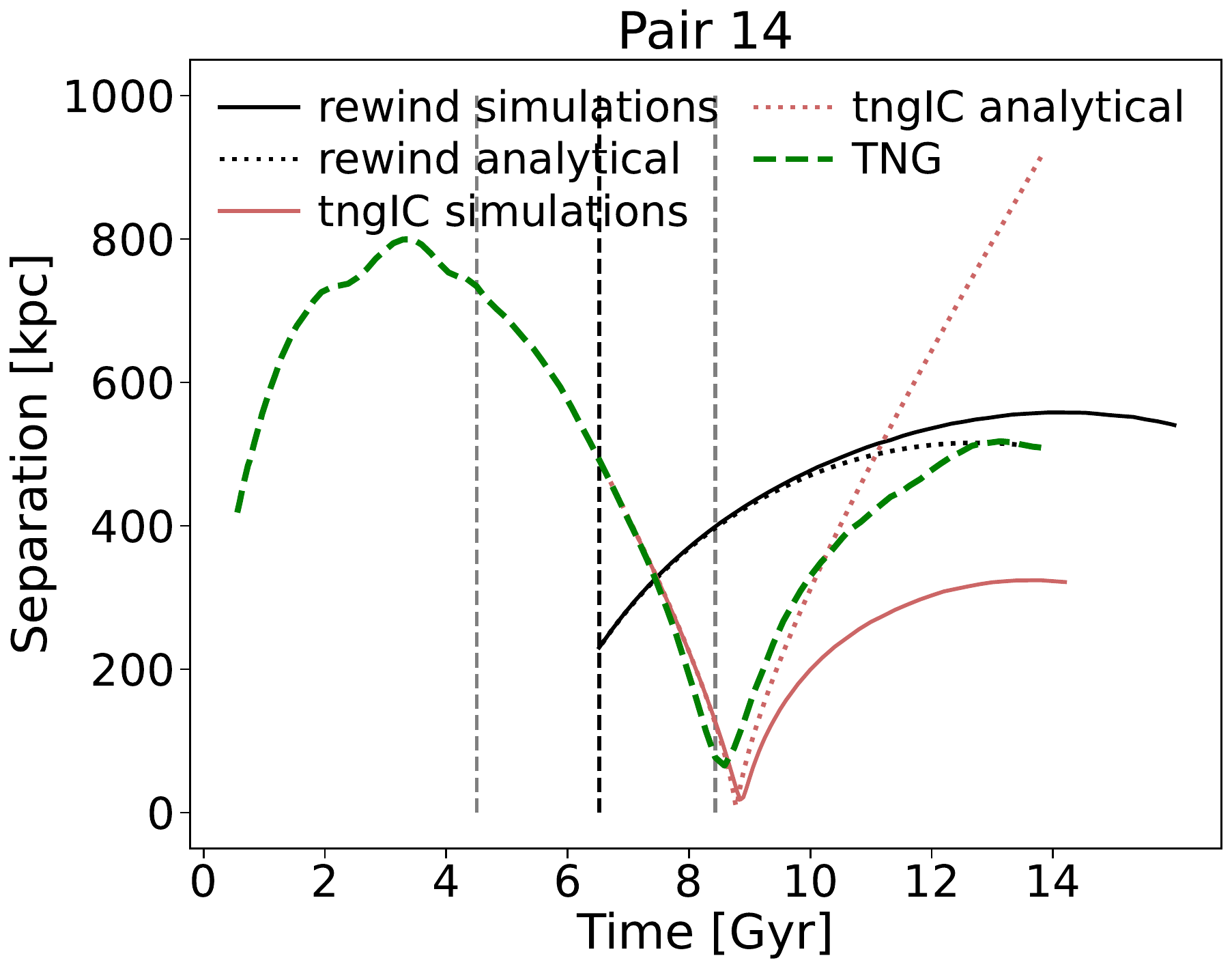}
        \label{fig:peris-14}    
    \end{subfigure}
    \caption{The orbital histories of all 15 LG analogs in TNG that have a past pericenter \cite{hartl24}. The green dashed lines show the original TNG orbit of each LG analog. The black lines show the orbits from the rewound IC. The red lines show the orbits from the IC displacements and velocities directly taken from the TNG data at $t_{\rm ini}$. For each IC choice, the solid lines show the orbits recorded in our idealized simulations, while the dotted lines show the orbits calculated analytically with Eqn. \ref{eqn:TA}. The vertical black dotted lines mark $t_{\rm ini}$, which is the middle point of the longest smooth segment (enclosed by the two vertical grey dotted lines) of the TNG orbits. The time axis represents the time since the Big Bang. }
    \label{fig:orbit-pairs-peris}
\end{figure*}

\begin{figure*}
    \centering
    \begin{subfigure}[t]{0.29\textwidth}
        \centering
        \includegraphics[width=\textwidth, clip,trim=0.2cm 0cm 0.2cm 0cm]{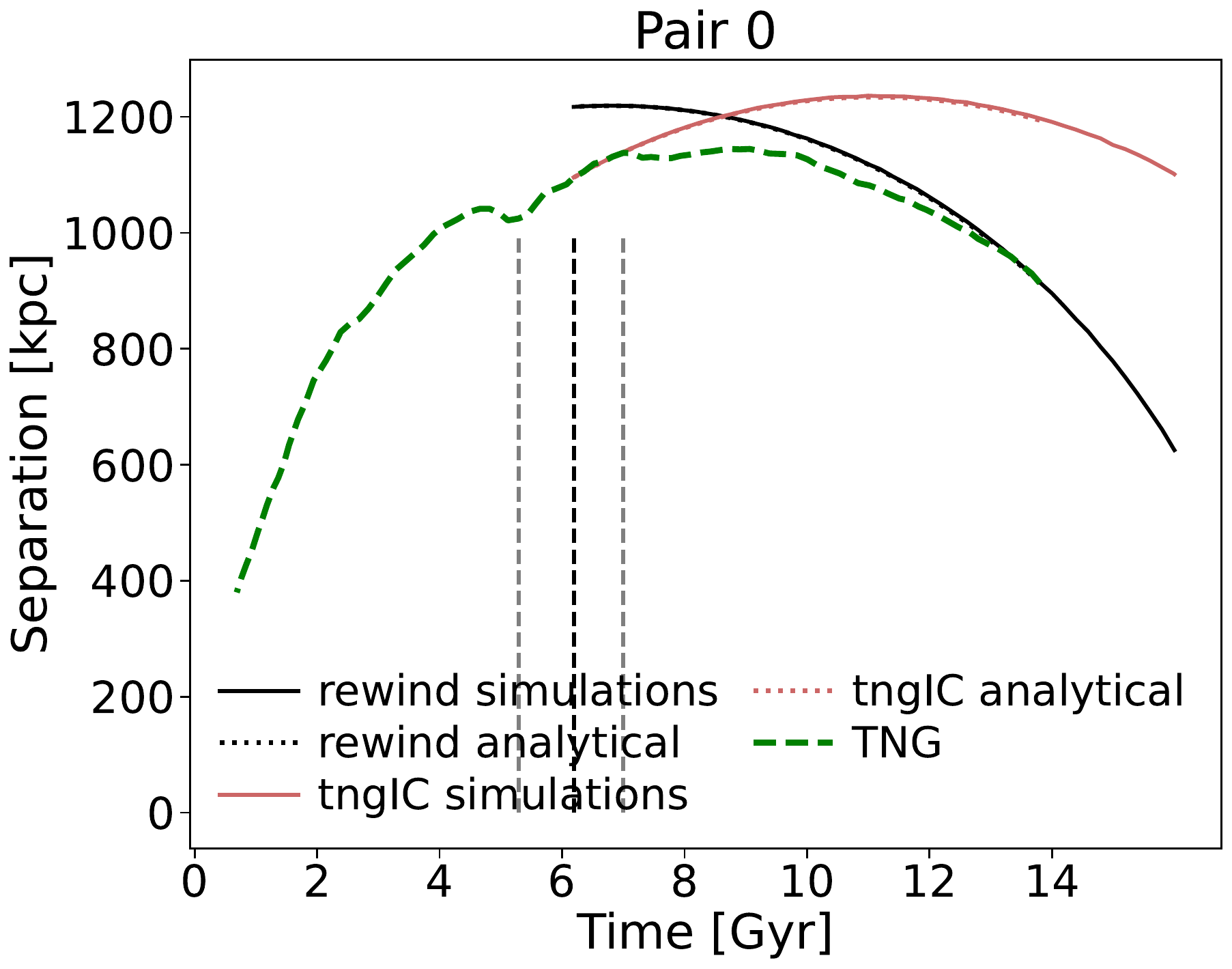}
        \label{fig:multiround-all}
    \end{subfigure}
    ~
    \begin{subfigure}[t]{0.29\textwidth}
        \centering
        \includegraphics[width=\textwidth, clip,trim=0.2cm 0cm 0.2cm 0cm]{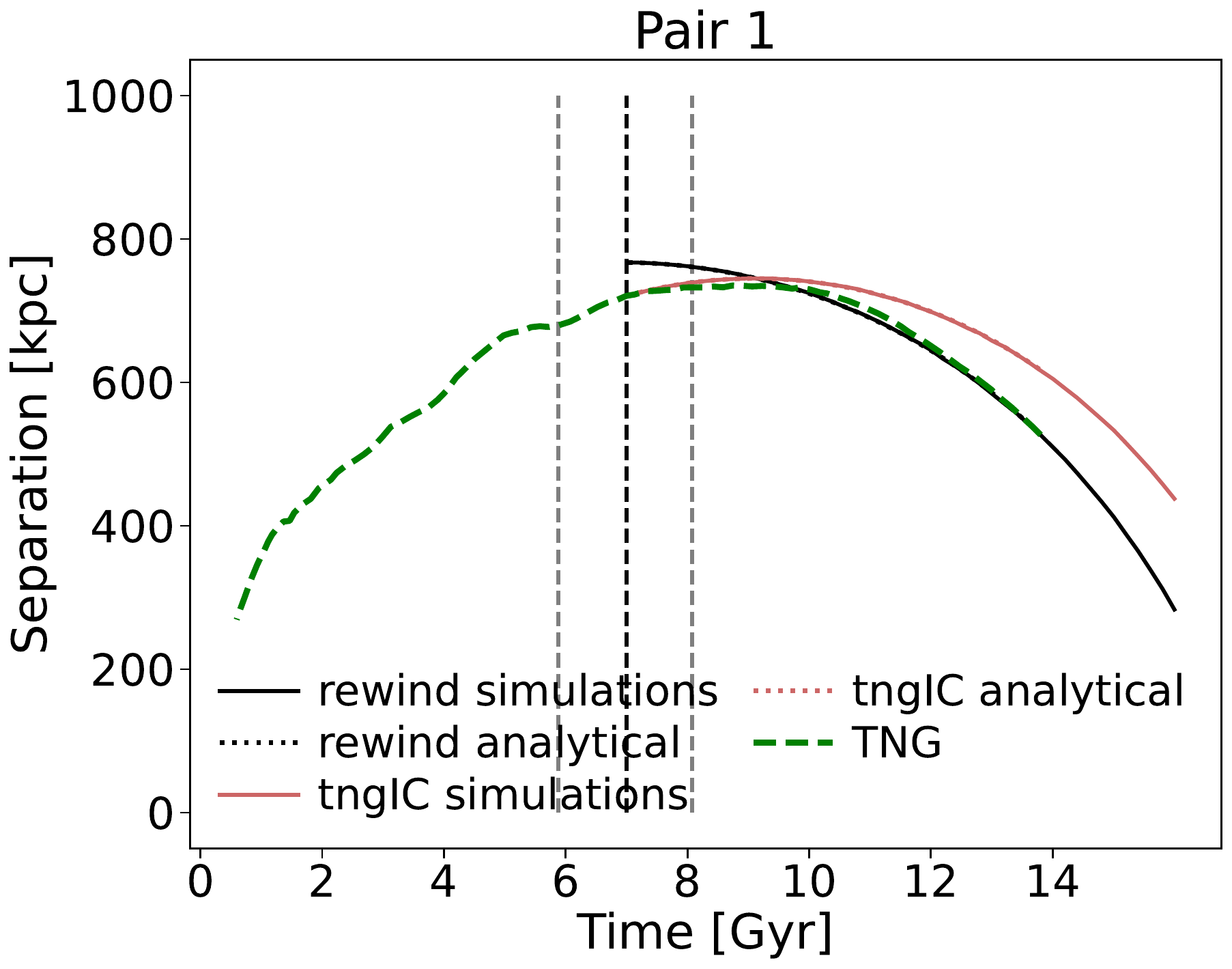}
        \label{fig:multiround-kn0}
    \end{subfigure}
    ~
    \begin{subfigure}[t]{0.29\textwidth}
        \centering
        \includegraphics[width=\textwidth, clip,trim=0.2cm 0cm 0.2cm 0cm]{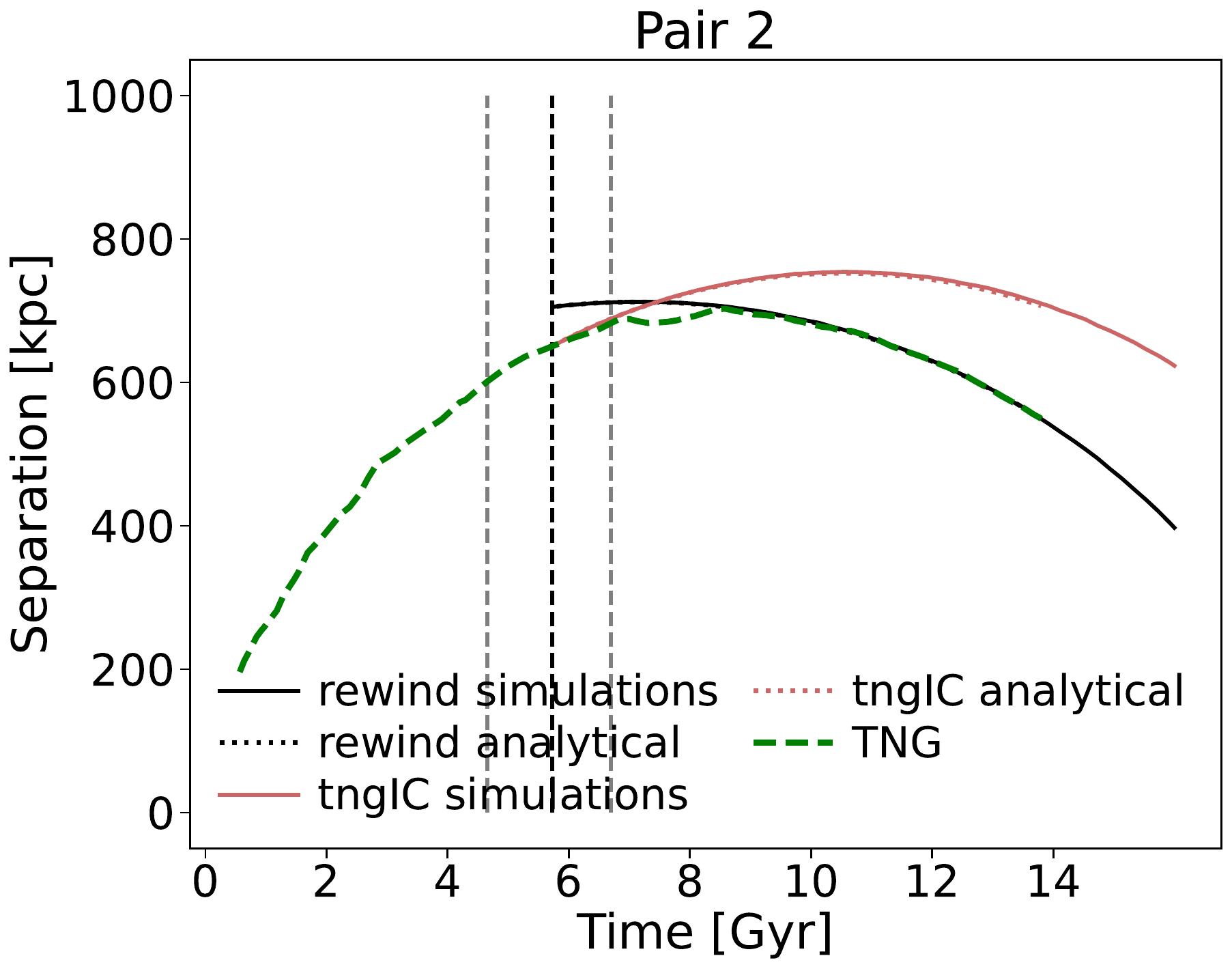}
        \label{fig:multiround-kn1}
    \end{subfigure}
    \caption{Same as Fig. \ref{fig:orbit-pairs-peris}, but for three examples of LG analogs that are on the first infall in TNG.} 
    \label{fig:orbit-pairs-1stinfalls}
\end{figure*}

These properties serve as inputs for the N-body initial condition (IC) generator \texttt{SpherIC} \cite{gk2013}, which creates two spherically symmetric NFW halos corresponding to the MW and M31 analogs. We note a structural difference in mass definition: TNG subhalo catalogs provide the total bound masses $M_{\rm bound}$ whereas the isolated halos generated by \texttt{SpherIC} feature an exponential truncation starting at $r_{200c}$. To conserve the total mass budget, we rescale the input $M_{\rm bound}$ in Table \ref{table:allpairs} to obtain the corresponding $M_{200c}$ such that the sum of the truncated virial mass and the extended mass tail equals the original TNG value: $M_{200c}^{\rm SpherIC}+M^{\rm SpherIC}(>r_{200c})=M^{\rm TNG}_{\rm bound}$. The two independently generated halos are then combined into a single IC file for the isolated LG simulation. The initial displacement and relative velocity of these two halos, as well as the starting time of this idealized simulation $t_{\rm ini}$, are determined via two methods:  a) directly obtaining the displacement and relative velocity from an early TNG snapshot at the desired $t_{\rm ini}$; b) rewinding the LG analog's orbit by integrating the two-body Newtonian equations of motion (Eqn. \ref{eqn:TA}) backward in time from the present-day displacement and relative velocity to $t_{\rm ini}$. The starting time $t_{\rm ini}$ of each LG analog is defined as the midpoint of the longest smooth, monotonic segment of its TNG orbital history. This selection strategy aims to minimize random noise in the IC setup arising from local perturbations along the TNG orbit. The rewinding approach, stemming from the classic Timing Argument (TA) model \cite{kahn59}, approximates the LG's orbital motion as that of two point masses under Newtonian gravity and cosmic expansion \cite{partridge13, pena14, hartl21, hartl24}:

\begin{equation}
    \begin{split}
    \ddot{\textbf{x}}_{\rm MW} = -\frac{GM_{\rm MW}M_{\rm M31} \cdot\textbf{r}} {r^3} + H^2_{0}\Omega_{\Lambda} \textbf{x}_{\rm MW} \\   
    \ddot{\textbf{x}}_{\rm M31} = \frac{GM_{\rm MW}M_{\rm M31} \cdot\textbf{r}} {r^3} + H^2_{0}\Omega_{\Lambda} \textbf{x}_{\rm M31},
    \end{split}
    \label{eqn:TA}
\end{equation}

where the relative distance vector $\textbf{r}=\textbf{x}_{\rm M31} - \textbf{x}_{\rm MW}$ denotes the displacement vector between the two galaxies, with $\textbf{x}_{\rm M31}$ and $\textbf{x}_{\rm MW}$ being the absolute coordinate vectors of the two halos within the TNG simulation box, while $H_0$ and $\Omega_\Lambda$ represent the Hubble parameter and dark energy density parameter at $z=0$ respectively. As we shall see later in Fig. \ref{fig:orbit-pairs-peris} and Fig. \ref{fig:orbit-pairs-1stinfalls}, this simplified model is generally sufficient when the MW and M31 analogs are outside of each other's virial radius. However, when the possibility of pericenter encounter is taken into consideration, effects arising from the extended halo mass distributions and dynamical friction can become significant. Previous studies such as \cite{petts15, sawala25} have employed a modified Chandrasekhar formula to account for this gravitational influence on orbital decay. Nevertheless, the (modified) Chandrasekhar formalism may still exhibit limited accuracy compared to N-body simulations \cite{zzc22}, particularly in major merger scenarios such as this. Therefore, we should expect the orbit analytically calculated via Eqn. \ref{eqn:TA} to deviate from the simulated orbit when the LG undergoes a close pericenter passage.

Figure \ref{fig:orbit-pairs-peris} presents the orbital histories of all 15 LG analogs identified with a past pericenter, while Fig. \ref{fig:orbit-pairs-1stinfalls} displays three representative examples of first-infall systems. In each panel, we plot five distinct orbital trajectories. The original TNG orbit is shown as a green dashed line. Orbits derived from the `rewind IC' are depicted in black, while those initialized from the TNG snapshot at $t_{\rm ini}$ (`TNG IC') are shown in red. For both initialization methods, the solid lines represent the results of our N-body simulations, and the dotted lines represent the analytical predictions using Eqn. \ref{eqn:TA}. The deviation between the solid (simulated) and dotted (analytical) lines quantifies the impact of dynamical friction. This effect is particularly significant for cases with a close pericenter in the past, such as `pair0' and `pair14'. 
Comparing the idealized simulations (black and red lines) with the original TNG trajectory (green dashed lines) highlights discrepancies between a full cosmological context and an isolated environment. These discrepancies likely stem from environmental factors absent in the idealized runs, such as large-scale structure, other (dwarf) members in the LG analogs, the smooth mass accretion of each halo (since our setups have to assume a fixed $z=0$ mass in IC), or a combination of some of these missing ingredients. As such, this comparison can be viewed as a diagnostic for the degree of isolation of each LG analog within the TNG simulation \cite{hartl24}.  As shown in Fig. \ref{fig:orbit-pairs-1stinfalls}, first-infall analogs generally show good agreement between the original TNG orbits and those reconstructed via both the `rewind' and `TNG IC' methods. However, among the 15 LG analogs that have a past pericenter in TNG, we only find `pair0' and `pair14' using the `TNG IC' method successfully capture the qualitative pericenter feature of the TNG orbit (i.e., a deep pericentric bounce), although with slight offsets in the exact time and distance. This suggests that the majority of the past-pericenter LG analogs (13 out of 15) are strongly influenced by their cosmological environment, preventing accurate reproduction of their orbits in an idealized, isolated setting.  We subsequently focus on `pair14' for the remainder of this work, as it offers the highest fidelity reproduction of the TNG orbital history in our idealized simulations. Additionally, for `pair0' and `pair14', the `TNG IC' runs provide a better match to the TNG orbit than the `rewind IC' runs. The failure of the latter to capture the past pericenter is attributed to the absence of dynamical friction in the backward integration to generate the initial conditions.

\section{SIDM evolution of the Milky Way analog in isolation}\label{sec:iso}

Before investigating the full Local Group orbital encounter, we first study the co-evolution of the MW analog and its stellar components in strict isolation.
The aim is to understand the mutual effects of the stellar and SIDM components, before introducing the complexities of the MW--M31 interaction. We note that this approach mirrors the study of isolated SIDM dwarfs in \cite{zzc24}, where the evolution of dwarf galaxies' stellar properties was explored under different SIDM models, both in isolation and in orbits of larger systems.
However, a key difference here is that the stellar mass fraction of the MW/M31 analogs is about $\sim 1-2\%$, much higher than that of the dwarf system in \cite{zzc24}, which accounts for $\sim 0.06\%$. While these stellar mass fractions are typical for their respective mass scales (e.g. see Fig. 9 of \cite{behroozi19}), this large difference hints that in MW-sized systems the stellar component plays a more important role in dictating the co-evolution of SIDM and stars. As we demonstrate below, the core formation and core collapse processes in SIDM halos are highly sensitive to the detailed distribution of the stellar component in these MW-sized systems.

Following the discussion in Sec. \ref{sec:orbitsim}, we focus on simulating one specific LG analog, labeled `pair14', selected from the past-pericenter sample for the remainder of this work. This system was chosen because its orbit in our idealized simulation provides the best reproduction of the original TNG orbital history. The detailed properties of the initial conditions for the MW and M31 analogs of `pair14' are listed in Table \ref{table:pair14}. We assume an NFW distribution for the DM components of both MW and M31 analogs, and Hernquist distribution for their stellar components. Similar to \cite{zzc24}, we use \texttt{SpherIC} to generate the initial condition of this MW analog with two species of simulation particles, DM and stars, both of which follow spherically symmetric distribution. To minimize the spurious heating of the less massive simulation particles caused by mass segregation \cite{bt08, ludlow19, ludlow23}, we adopt the configuration from \cite{zzc24} (see their Fig. 2) to use a 1:1 mass ratio between DM and star particles, instead of the baryon-to-DM ratio of 1:5 often used in cosmological simulations. 
Consequently, both the DM and stellar particles share an identical mass of $M_p = 4.5 \times 10^5 M_\odot$. The isolated MW analog simulated in this section thus contains a total of $\sim 8.9 \times 10^5$ particles. Following the numerical criteria recommended by \cite{mace24}, this resolution safely places our setups within the well-converged regime for SIDM core-collapse physics, ensuring that the structural evolution remains free from severe numerical artifacts.

\begin{figure*}
    \centering
    \begin{subfigure}[t]{0.3\textwidth}
        \centering
        \includegraphics[width=\textwidth, clip,trim=0.2cm 0cm 0.2cm 0cm]{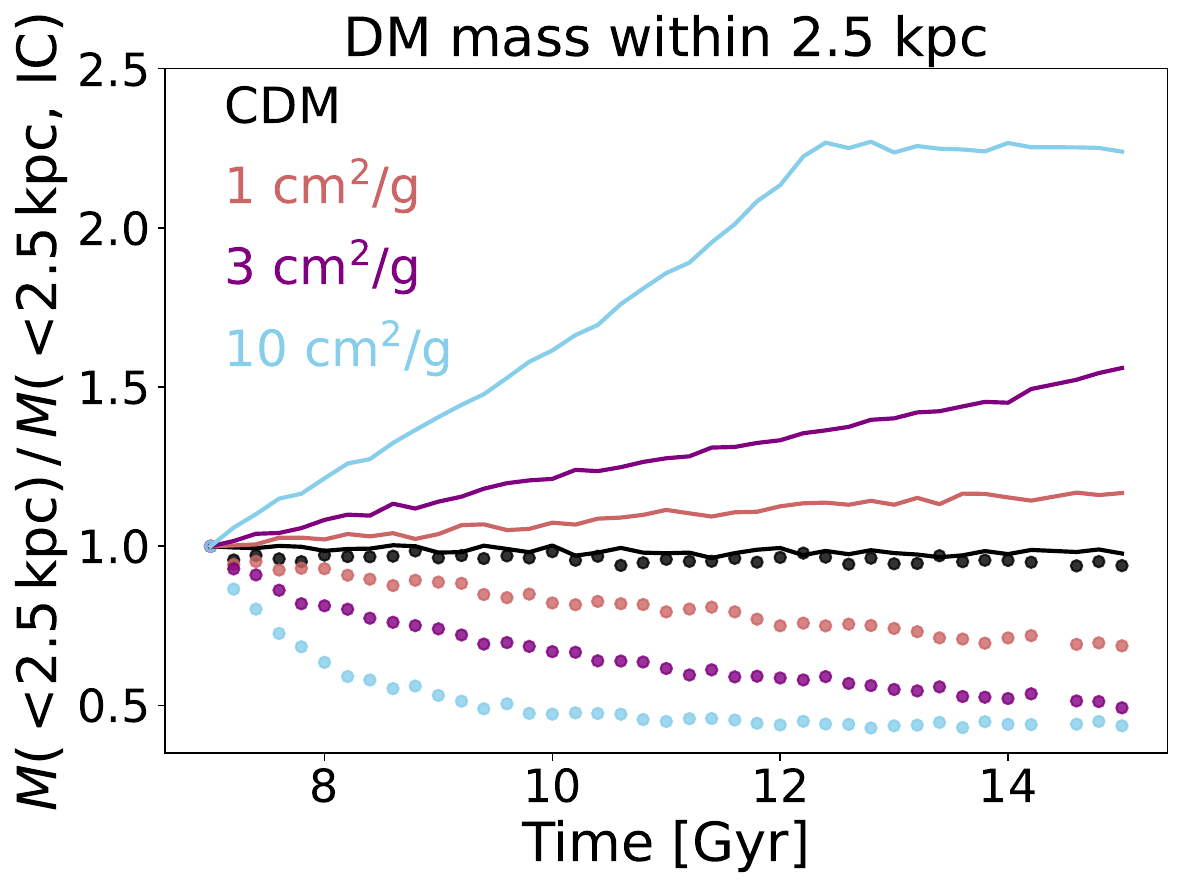}
        \caption{}
        \label{fig:iso-tevo-a}
    \end{subfigure}
    ~
    \begin{subfigure}[t]{0.3\textwidth}
        \centering
        \includegraphics[width=\textwidth, clip,trim=0.2cm 0cm 0.2cm 0cm]{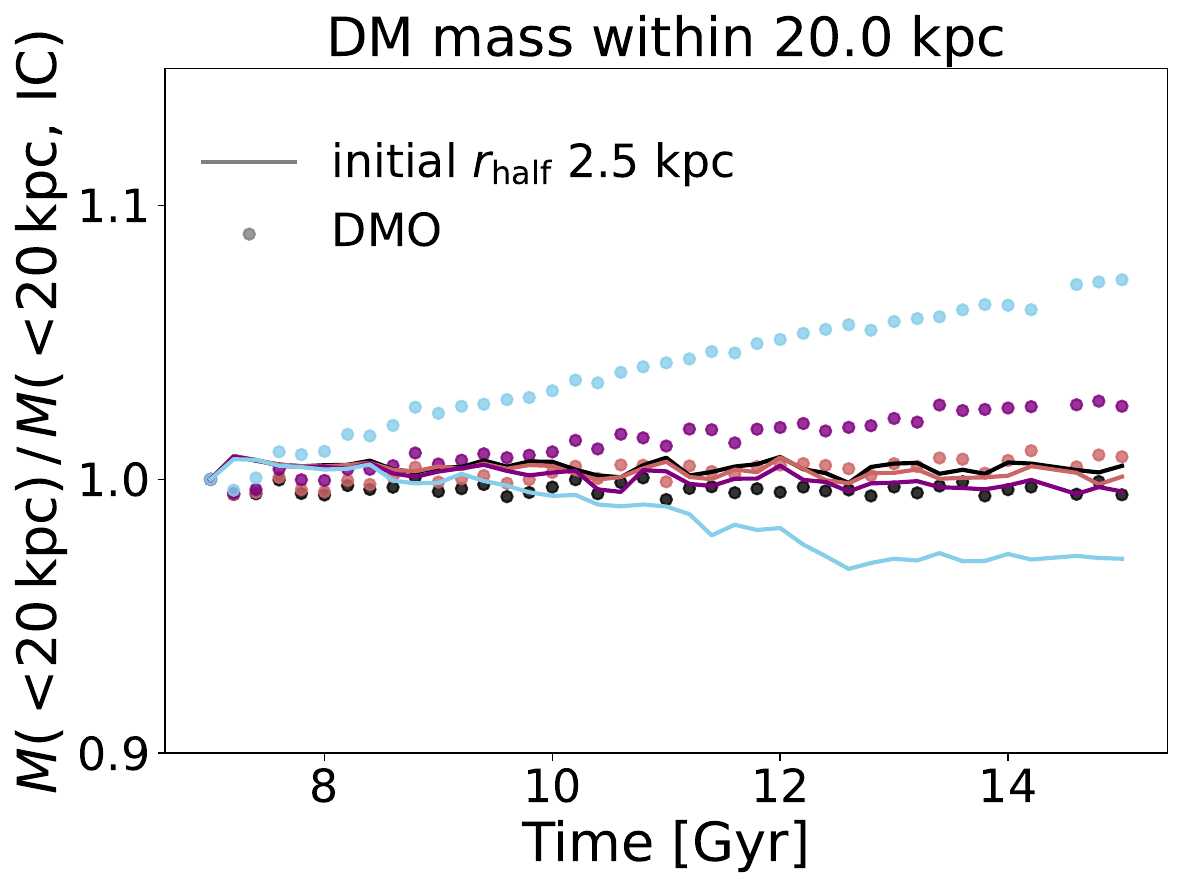}
        \caption{}
        \label{fig:iso-tevo-b}
    \end{subfigure}
    ~
    \begin{subfigure}[t]{0.3\textwidth}
        \centering
        \includegraphics[width=\textwidth, clip,trim=0.2cm 0cm 0.2cm 0cm]{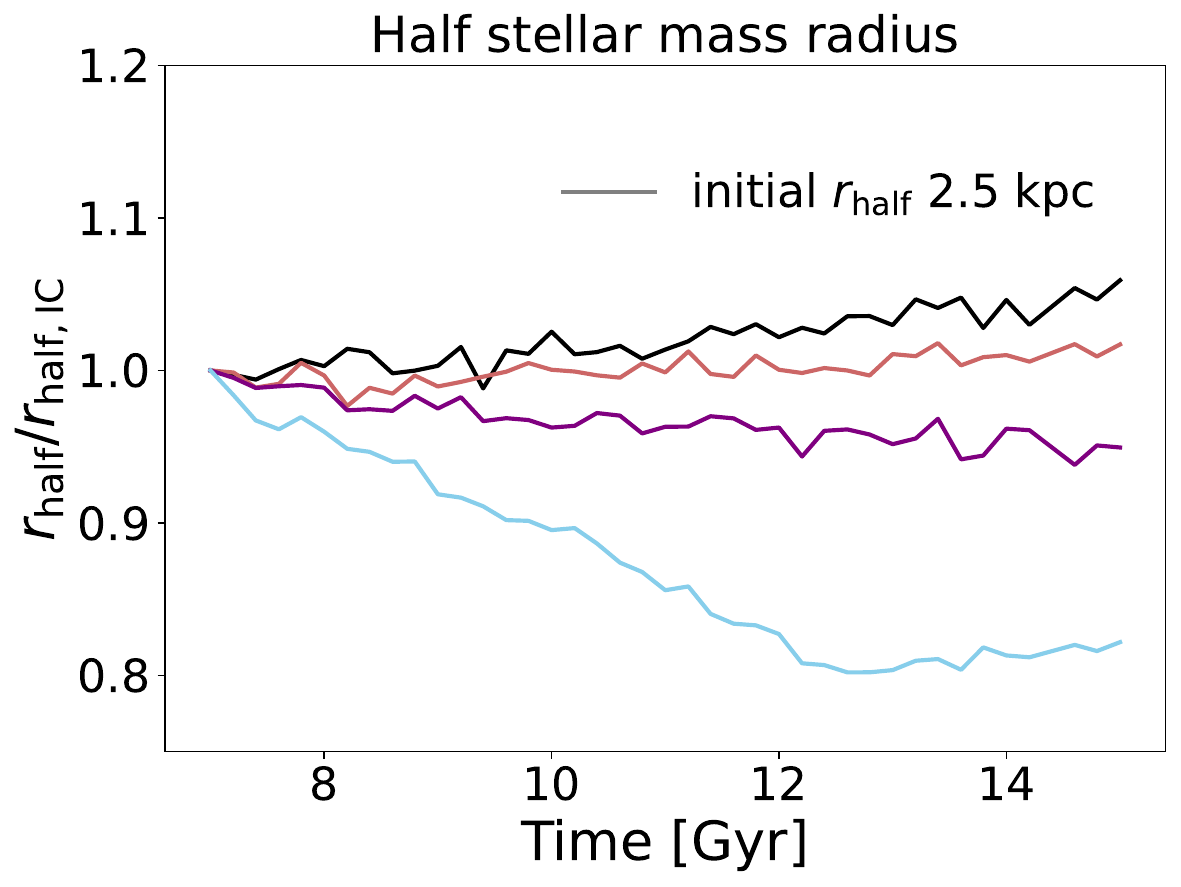}
        \caption{}
        \label{fig:iso-tevo-c}
    \end{subfigure}
    ~
    \begin{subfigure}[t]{0.3\textwidth}
        \centering
        \includegraphics[width=\textwidth, clip,trim=0.2cm 0cm 0.2cm 0cm]{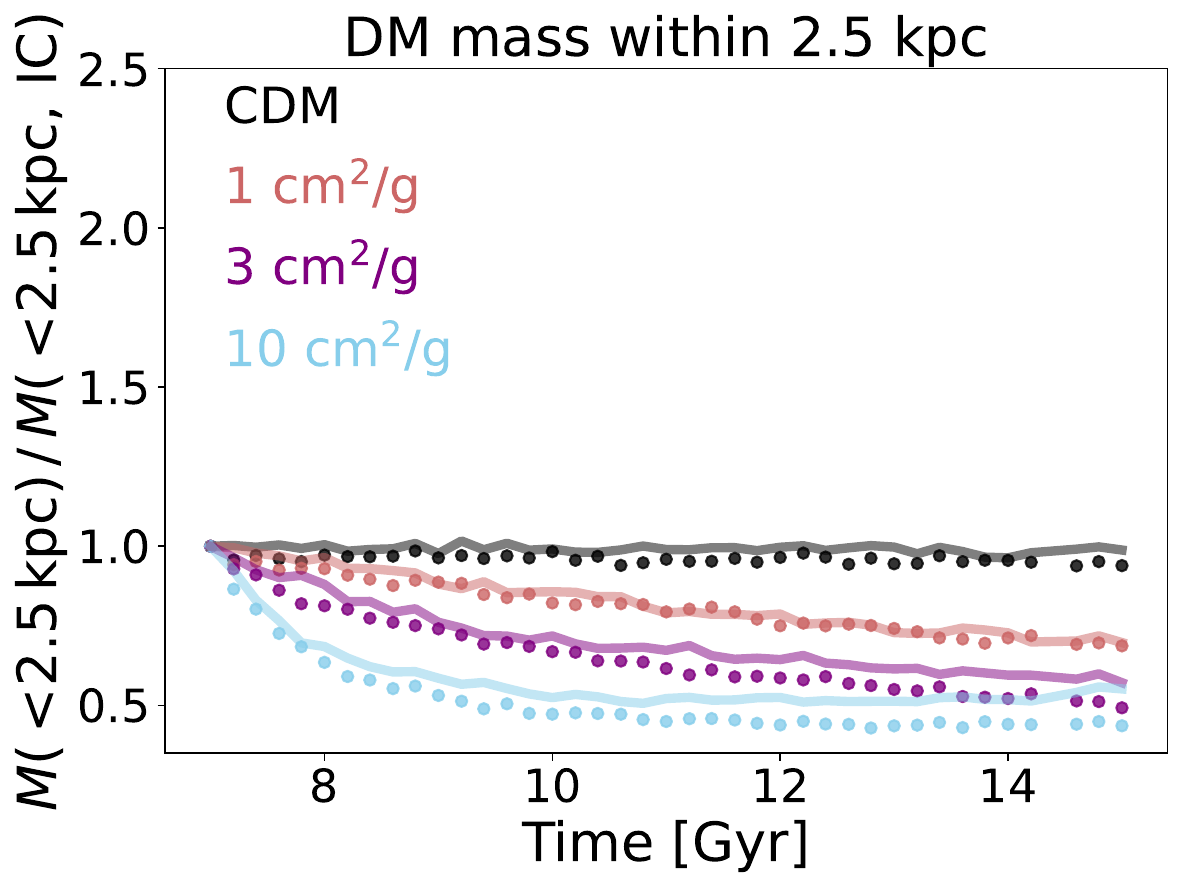}
        \caption{}
        \label{fig:iso-tevo-d}
    \end{subfigure}
    ~
    \begin{subfigure}[t]{0.3\textwidth}
        \centering
        \includegraphics[width=\textwidth, clip,trim=0.2cm 0cm 0.2cm 0cm]{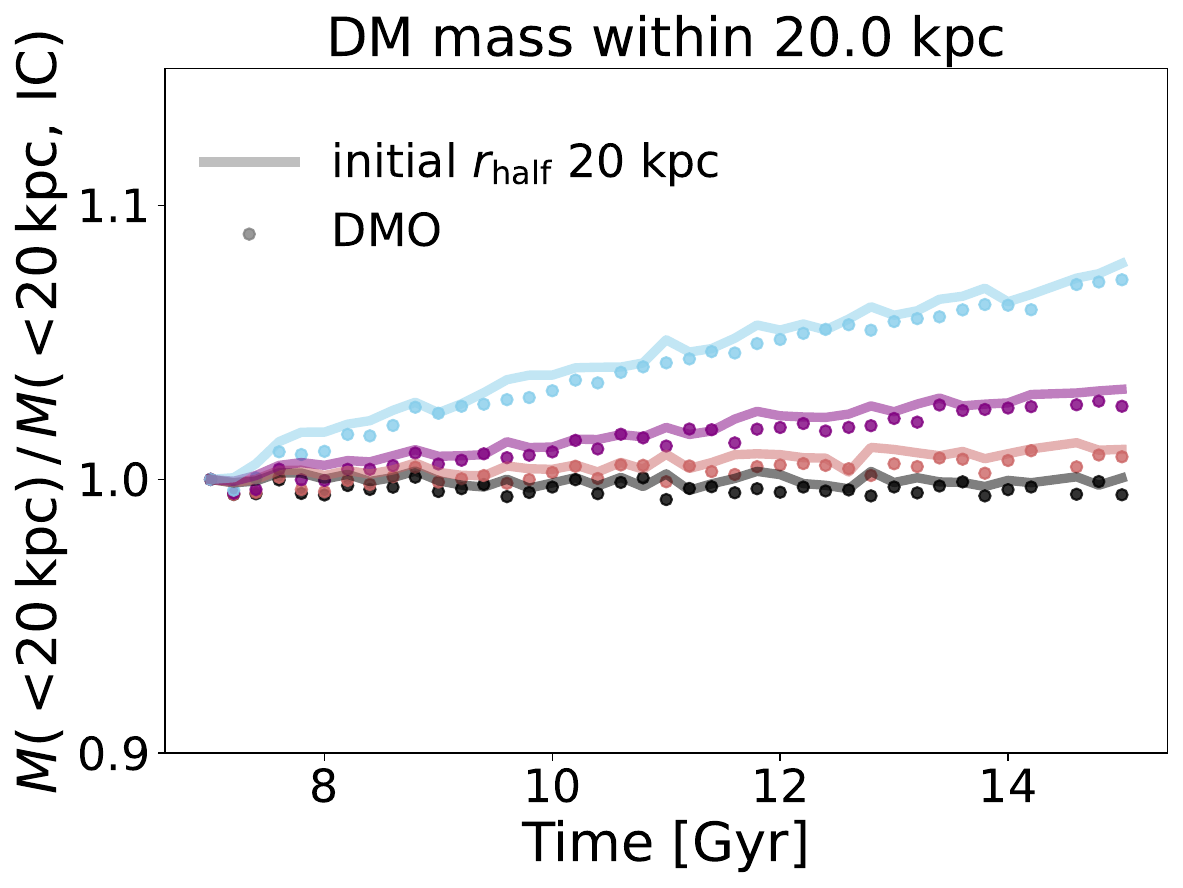}
        \caption{}
        \label{fig:iso-tevo-e}
    \end{subfigure}
    ~
    \begin{subfigure}[t]{0.3\textwidth}
        \centering
        \includegraphics[width=\textwidth, clip,trim=0.2cm 0cm 0.2cm 0cm]{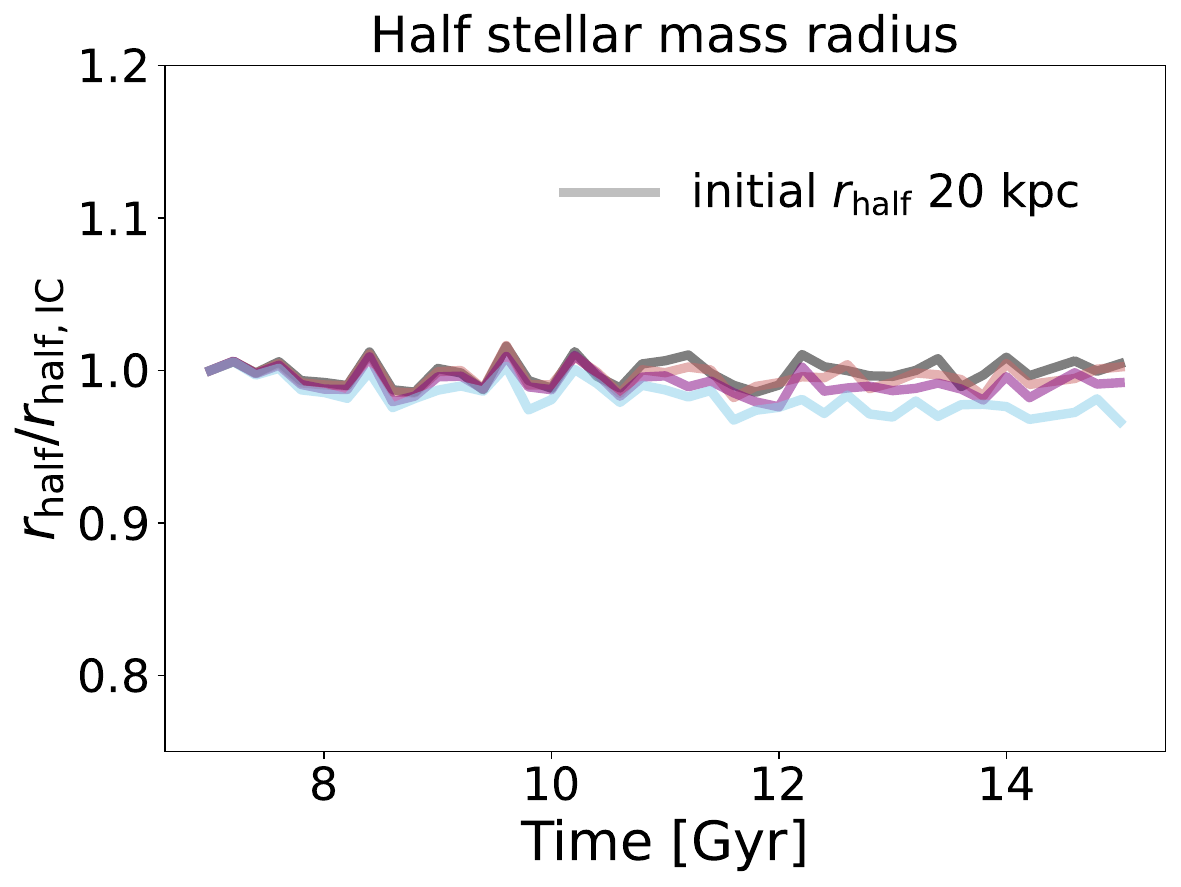}
        \caption{}
        \label{fig:iso-tevo-f}
    \end{subfigure}
    \caption{Evolution of the isolated MW analog with different DM models, including CDM (black), 1 $\rm cm^2/g$ (red), 3 $\rm cm^2/g$ (purple) and 10 $\rm cm^2/g$ (blue). The top row shows the evolution of the MW analog with an initial stellar $r_{\rm half}=2.5$ kpc (compact-stellar; thin lines) and the DMO case (scattered dots). The bottom row shows the MW analog with $r_{\rm half}=20$ kpc (diffuse-stellar; thick lines) and the DMO case. For each row from left to right: the evolution of the enclosed DM mass within 2.5 kpc $M_{\rm DM}(<2.5\rm kpc)$; the evolution of $M_{\rm DM}(<20\rm kpc)$; the evolution of the MW analog's stellar $r_{\rm half}$. }
    \label{fig:iso-timeevo}
\end{figure*}

We emphasize that given the user-specified density profiles (NFW for DM and Hernquist for stars in this work), \texttt{SpherIC} self-consistently calculates the corresponding velocity distribution function to ensure the multi-component system is initialized in strict dynamical equilibrium. Consequently, the inclusion of the central stellar mass does not modify the initial DM density profile, but it elevates the central DM velocity dispersion to support the deepened combined potential. As demonstrated by the pure CDM control runs in Fig.~\ref{fig:iso-timeevo} (black datasets), both the enclosed DM masses and the stellar half-mass radii remain stable over $\sim8$ Gyr of collisionless evolution, confirming that our initial conditions are free from early-stage numerical relaxation artifacts.
Following the setup in \cite{zzc23}, we monitor the progress of halo core collapse by tracking $\rho_{\rm cen50}$, the mean density of the 50 densest particles, at every timestep. We define a halo as core-collapsed when $\rho_{\rm cen50}$ increases by a factor of five relative to its initial value. At this threshold, we disable dark matter self-interactions, effectively treating the halo as CDM for the remainder of the simulation. This intervention is necessary because, as core collapse proceeds, the computational cost becomes prohibitive, and the validity of the N-body approximation in this regime becomes questionable.

\begin{table*} 
	\centering
	\begin{tabular*}{\textwidth}{@{\extracolsep{\fill}}lcccccccccc} 
		\hline
		   & $M_{200c}[M_\odot]$(DM+star) & $r_{200c}$ [kpc] & $r_s$ [kpc] & $c_{200c}$ & $M_\star[M_\odot]$ & $N_{\rm DM}$ & $N_\star$ & $M_p[M_\odot]$ & $r_{\rm half}$ [kpc] & $\epsilon_{\rm soft}$ [kpc]\\
		\hline
        \multirow{2}{*}{MW analog} & \multirow{2}{*}{$3.04\times10^{11}$} & \multirow{2}{*}{140.32} & \multirow{2}{*}{11.57} & \multirow{2}{*}{12.1} & \multirow{2}{*}{$8.93\times10^9$} & \multirow{2}{*}{870573} & \multirow{2}{*}{19844} & \multirow{2}{*}{$4.5\times10^5$} & 2.5 & 0.16 \multirow{3}{*}{} \\ \cline{10-10}
        & & & & & & & & & 20 & \\ \hline
        M31 analog & $3.57\times10^{11}$ & 148.17 & 20.08 & 7.4 & $9.99\times10^9$ & 1087371 & 22209 & $4.5\times10^5$ & 2.7 & 0.16 \\ \hline
	\end{tabular*}
	\caption{Initial condition parameters for the `pair14' LG analog, simulated in Secs.~\ref{sec:iso} and~\ref{sec:lgsim} with stellar particles under various SIDM models. We note that the total particle mass $M_p\times N_{\rm DM}$ does not strictly equal the DM portion of $M_{200c}$ due to the exponential truncation of the halos at $r_{200c}$, which results in an extended mass distribution beyond this radius. Furthermore, as detailed in Sec.~\ref{sec:orbitsim}, the $M_{200c}$ values listed here have been rescaled so that the truncated NFW total mass matches the TNG bound mass after subtracting the stellar component; they therefore differ from the entries in Table~\ref{table:allpairs}. Halo concentrations $c_{200c}=r_{200c}/r_s$ are taken from TNG FoF group catalogs at $z=0$ ($c_{\rm M31}\approx 7.4$, $c_{\rm MW}\approx 12.1$), rather than from the mass-concentration relation used in Sec.~\ref{sec:orbitsim}. The force softening $\epsilon_{\rm soft}$ follows \cite{vdb18}. This choice of mass resolution satisfies the convergence criteria for SIDM core-collapse fidelity established in \cite{mace24, palubski24}.}
    \label{table:pair14}
\end{table*}

As detailed in Table \ref{table:pair14}, we prepare two distinct configurations for the MW analog, fixed at the same total stellar mass but initialized with different stellar half-mass radii $r_{\rm half}=$2.5 kpc and 20 kpc. This choice is motivated by the structural composition of MW-mass galaxies, which typically features a compact central component (disk and bulge) and a diffuse extended component (stellar halo). 
While our idealized models do not capture the exact morphology of the true Milky Way, they serve as representatives of MW-mass systems to explore the unique SIDM physics at this mass scale.
Thus these two types of star distribution serve as extremes to set bounds on the dynamical co-evolution of the SIDM and stellar components, both in the isolated evolution studied here and the LG pericenter scenario discussed in the next section. We emphasize that a realistic MW-mass system, in which the majority of the stellar mass resides in the disk and bulge \cite{Bland-Hawthorn16}, corresponds more closely to the compact test case ($r_{\rm half}=$2.5 kpc).

\begin{figure*}
    \centering
    \begin{subfigure}[t]{0.3\textwidth}
        \centering
        \includegraphics[width=\textwidth, clip,trim=0.2cm 0cm 0.2cm 0cm]{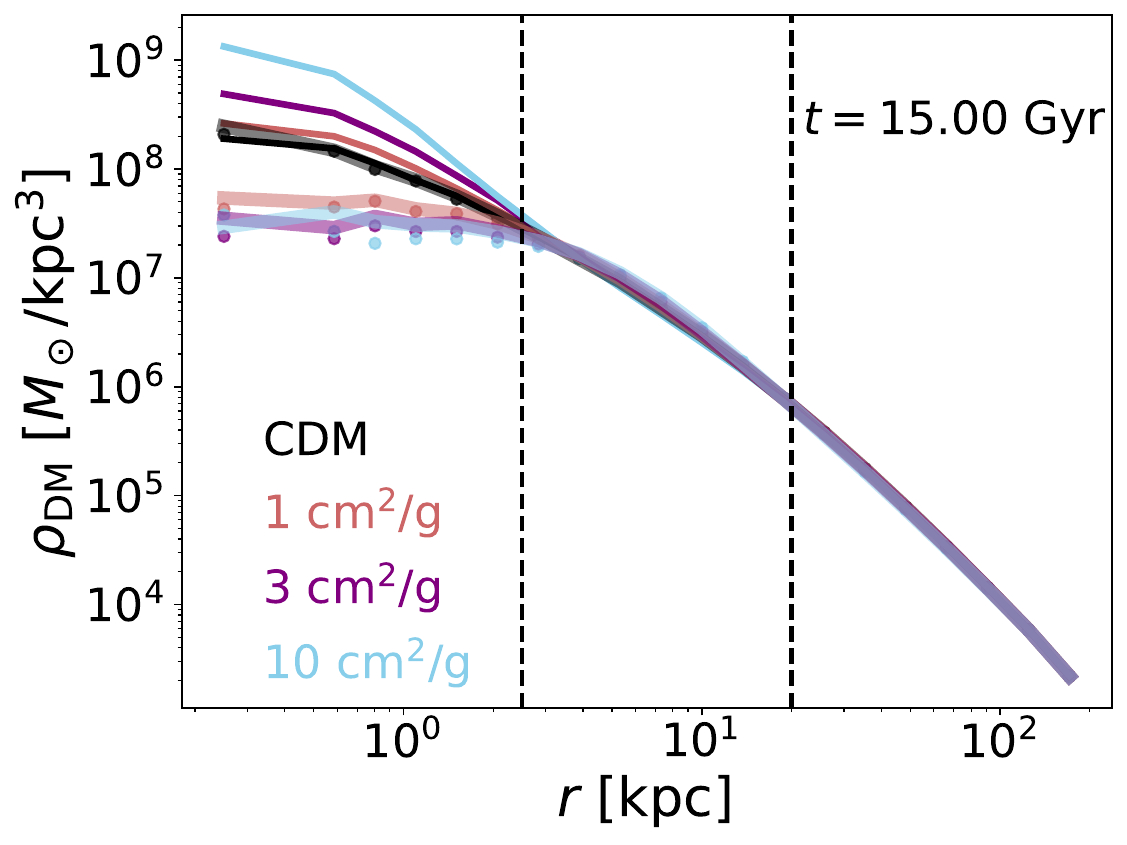}
        \caption{}
        \label{fig:final-pro-a}
    \end{subfigure}
    ~
    \begin{subfigure}[t]{0.3\textwidth}
        \centering
        \includegraphics[width=\textwidth, clip,trim=0.2cm 0cm 0.2cm 0cm]{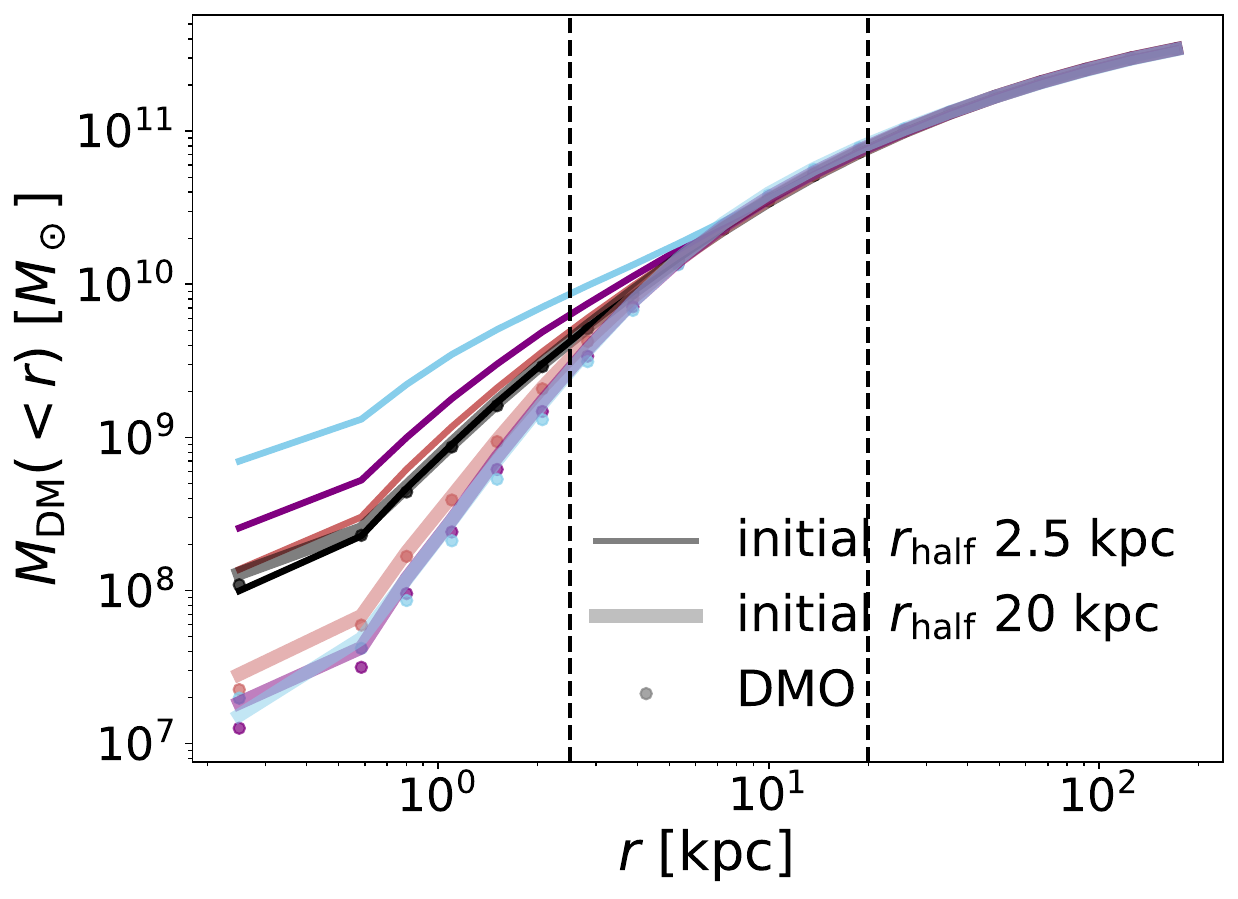}
        \caption{}
        \label{fig:final-pro-b}
    \end{subfigure}
    ~
    \begin{subfigure}[t]{0.3\textwidth}
        \centering
        \includegraphics[width=\textwidth, clip,trim=0.2cm 0cm 0.2cm 0cm]{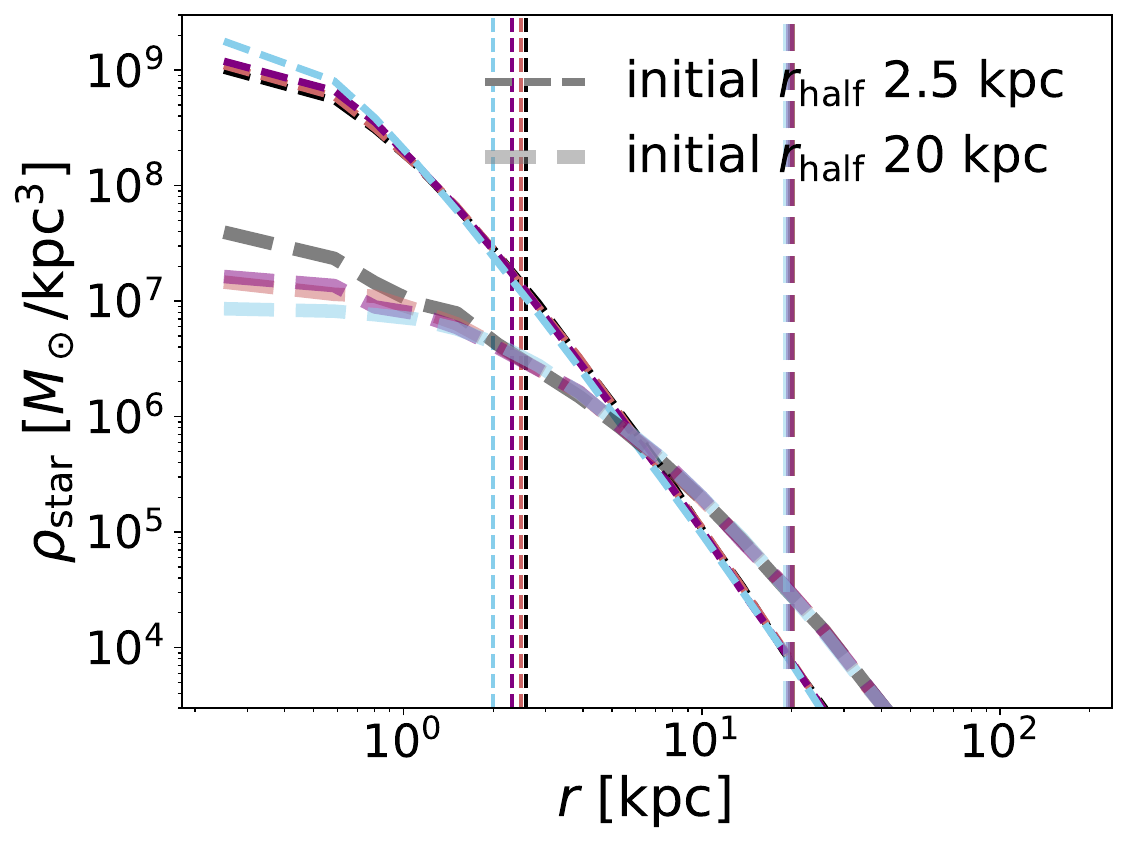}
        \caption{}
        \label{fig:final-pro-c}
    \end{subfigure}
    ~
    \begin{subfigure}[t]{0.3\textwidth}
        \centering
        \includegraphics[width=\textwidth, clip,trim=0.2cm 0cm 0.2cm 0cm]{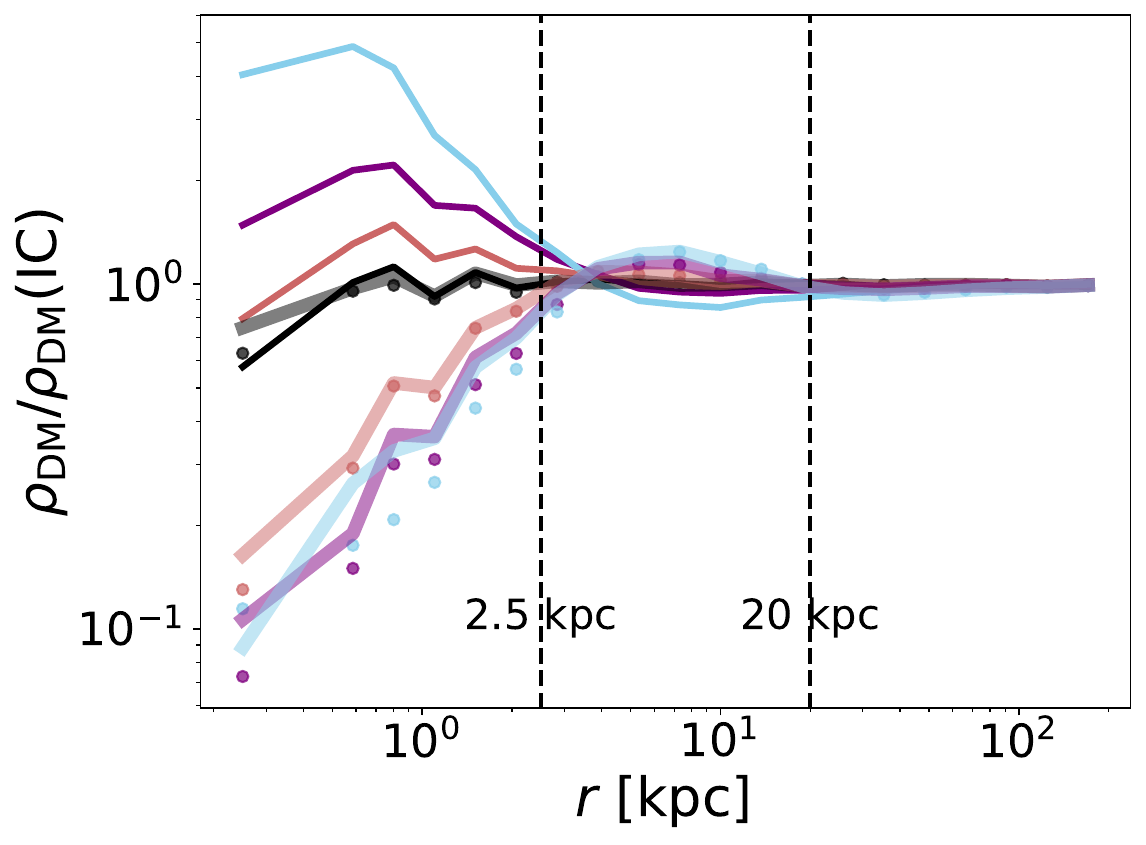}
        \caption{}
        \label{fig:final-pro-d}
    \end{subfigure}
    ~
    \begin{subfigure}[t]{0.3\textwidth}
        \centering
        \includegraphics[width=\textwidth, clip,trim=0.2cm 0cm 0.2cm 0cm]{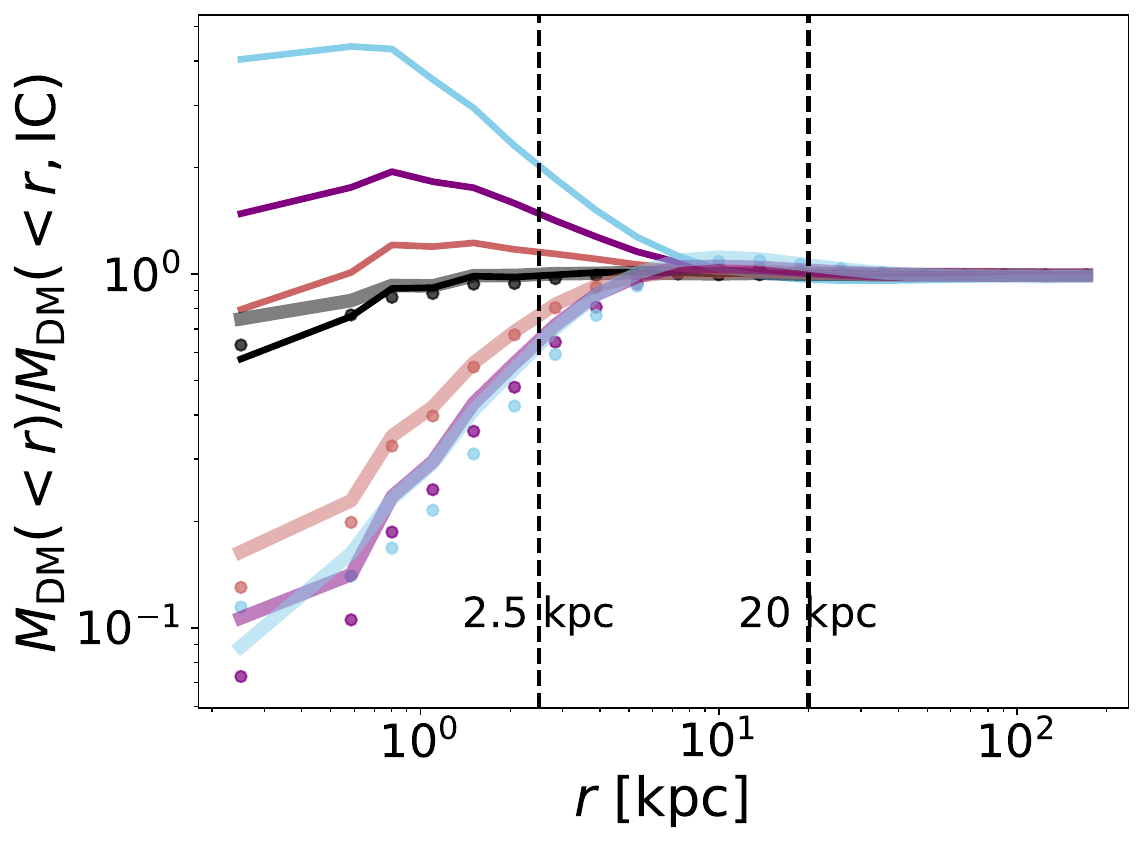}
        \caption{}
        \label{fig:final-pro-e}
    \end{subfigure}
    ~
    \begin{subfigure}[t]{0.3\textwidth}
        \centering
        \includegraphics[width=\textwidth, clip,trim=0.2cm 0cm 0.2cm 0cm]{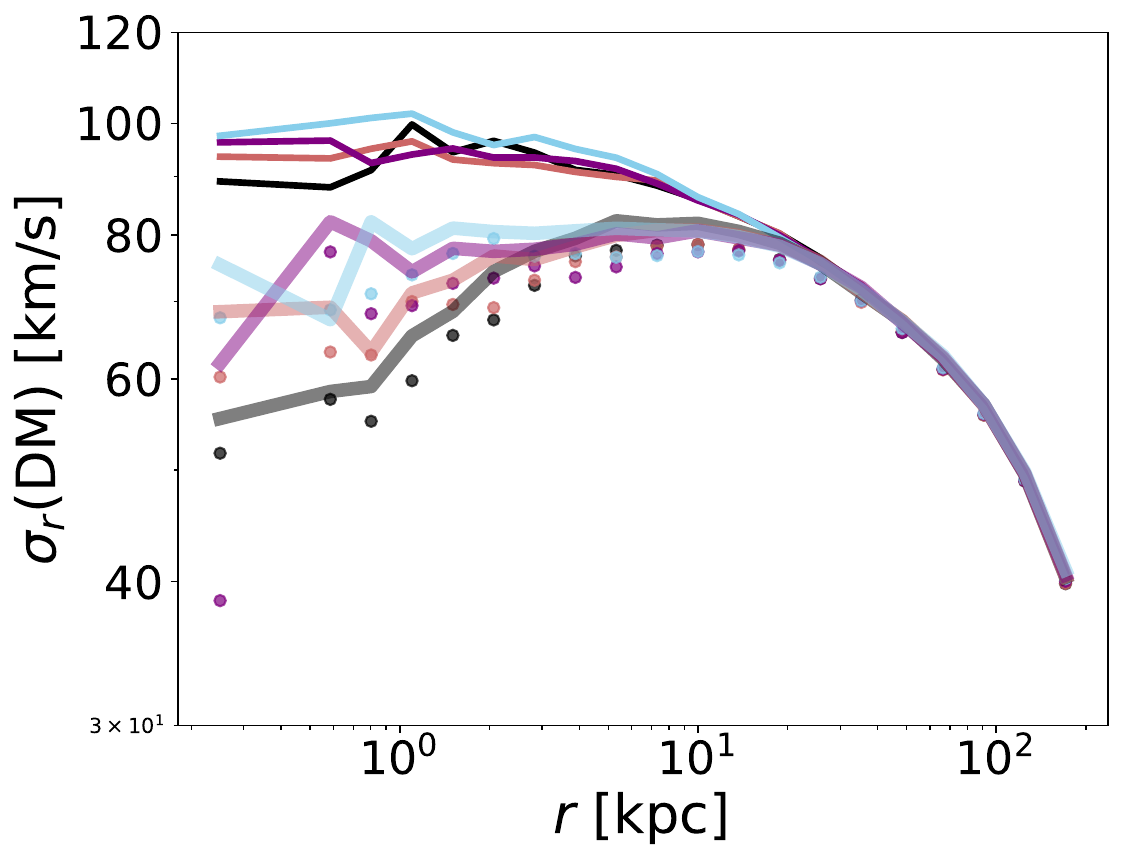}
        \caption{}
        \label{fig:final-pro-f}
    \end{subfigure}
    \caption{Radial profiles of the (isolated) MW analog at the end of the simulation, with different combinations of DM models and stellar IC. a) DM density profiles; b) enclosed DM mass profiles; c) stellar density profiles; d) DM density profiles, normalized by IC; e) enclosed mass profiles, normalized by IC; f) DM velocity dispersion profiles. The line colors and line thickness are the same as in Fig. \ref{fig:iso-timeevo}. The vertical lines in a), b), d) and e) show the initial $r_{\rm half}$ of 2.5 and 20 kpc, while the vertical lines in c) show the final $r_{\rm half}$ of the corresponding DM models and compact/disperse stellar configurations. In f), the CDM lines (black), since they remain nearly unchanged during the collisionless simulation, can be regarded as a proxy for the initial condition velocity dispersion. An explicit demonstration of the $\sigma_v$ profiles for ICs is included in Fig. \ref{fig:ic-sigmav}.}
    \label{fig:final-pro}
\end{figure*}

We present the time evolution of these two test cases of the selected MW analog in Fig. \ref{fig:iso-timeevo}, with either a compact stellar component (initial $r_{\rm half}=2.5$ kpc) or a diffuse stellar component (initial $r_{\rm half}=20$ kpc), evolved under different SIDM models. For comparison, a dark-matter-only (DMO) simulation of the same system is included as a control baseline. In Fig. \ref{fig:final-pro}, we provide a detailed analysis of the density, mass, and velocity dispersion profiles for these MW analogs at the final simulation snapshot. This comparison aims to elucidate two key aspects of the co-evolution between the SIDM halo and the stellar component: a) how different stellar potentials impact the SIDM halo's transition from core-formation to core-collapse, and b) how the evolving SIDM halo reciprocally governs the structural evolution of the stellar component.

\subsection{Evolution of DM within 2.5 kpc}\label{sec:iso-a}

Fig. \ref{fig:iso-tevo-a} and \ref{fig:iso-tevo-d} show the evolution of the enclosed mass of DM within an aperture of 2.5 kpc,  corresponding to the initial $r_{\rm half}$ of the compact stellar component. This radius also approximately coincides with the typical core size formed in the DMO simulations (see Fig. \ref{fig:final-pro-a}). In the DMO simulations (scattered points in Fig. \ref{fig:iso-timeevo}), the enclosed mass within 2.5 kpc remains constant for CDM, as expected. In contrast, it decreases over time for the SIDM models, reflecting the core formation process that pushes the DM outward from the halo central region. The magnitude of this reduction in DM central mass within 2.5 kpc, as well as the DM core density in Fig. \ref{fig:final-pro-a} and \ref{fig:final-pro-d}, scales with the self-interaction cross section. For the MW analog with a diffuse stellar component (thick lines, Fig. \ref{fig:iso-tevo-d}), the central DM mass evolution generally follows the trend of the DMO counterparts, with slightly moderated reduction over time, since the additional stellar potential counteracts the expansion of the SIDM central core. However, for the largest cross section of 10 $\rm cm^2/g$ (cyan thick line), the central DM mass begins to reverse the trend and increase at late times, signaling the onset of core-collapse. This accelerated onset of core collapse relative to the DMO counterpart (cyan thick line vs. cyan dots) demonstrates that the presence of a central baryon potential accelerates the collapse process, consistent with previous studies \cite{wxfeng21, fz22, ymzhong23, sq23, zzc24, dnyang24}. 

As for the MW analogs with a compact stellar component (thin lines in Fig. \ref{fig:iso-tevo-a}), however, all SIDM cases deviate from their DMO counterparts immediately upon the start of the simulation. Across all cross sections (1 $\rm cm^2/g$ to 10 $\rm cm^2/g$), the enclosed DM mass within 2.5 kpc increases monotonically, suggesting that \textit{the core formation phase is effectively bypassed}, with the halo directly entering the core-collapse phase. Notably, the 10 $\rm cm^2/g$ case reaches our core-collapse threshold ($\rho_{\rm cen50}$ growing by a factor of 5), within only $\sim6$ Gyrs. Apart from the established effect of the baryon potential directly accelerating core collapse, which is more prominent in these test cases with a compact stellar component, we note that a distinct factor contributes to this `no-core' behavior. This deep stellar potential boosts the initial velocity dispersion at small radii by nearly a factor of two, altering the profile shape to establish a \textit{negative temperature gradient at the beginning of the simulation} (see Fig. \ref{fig:final-pro-f}, and the explicitly plotted initial velocity dispersion profiles in Fig. \ref{fig:ic-sigmav}).  A negative temperature gradient is a prerequisite and primary driver for runaway gravothermal collapse \cite{balberg02, essig19, nishikawa20, zzc22}, since it boosts the heat outflow in a self-strengthening way. Consequently, assuming dynamical equilibrium in the ICs, these systems with a compact stellar component are essentially initialized in a thermodynamic state that is already unstable to gravothermal collapse. 

This phenomenon of apparently bypassing the SIDM core-formation stage, as we have shown, appears to be highly sensitive to the compactness of the stellar component. While this result physically represents MW-mass systems dominated by a compact disk and bulge with effective $r_{\rm half}\sim$2.5 kpc, it contrasts with our previous findings for SIDM dwarf galaxies \cite{zzc24}, which demonstrate a universal core-formation phase along the evolutionary tracks. This discrepancy is likely attributable to the significant difference in stellar mass fraction between these regimes (dwarf $\lesssim0.1\%$; MW-like $\gtrsim1\%$). To investigate this potential mass dependence, we perform additional simulations varying the stellar mass of this MW analog.

\subsubsection{Further testing the bypassed core-formation with varying stellar mass fraction}

\begin{figure}
    \centering
    \includegraphics[width=\columnwidth, clip,trim=0.2cm 0cm 0.2cm 0cm]{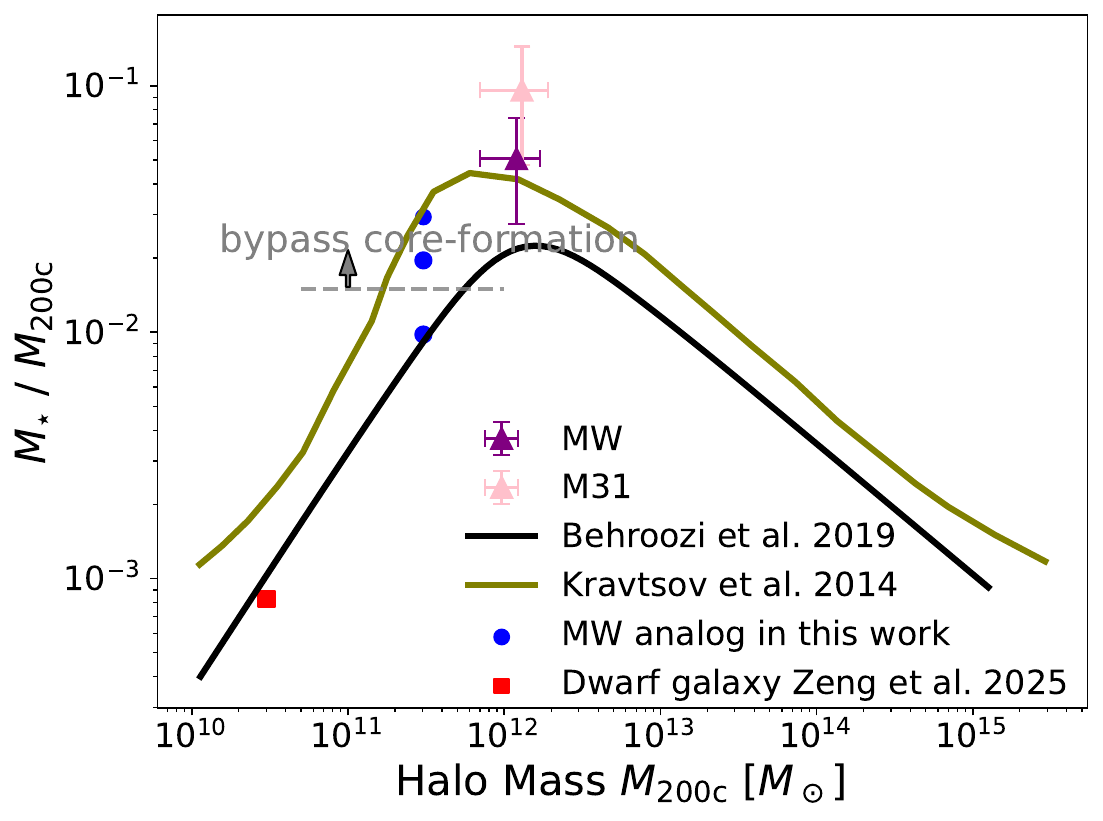}
    \caption{The stellar-mass-halo-mass (SHM) parameter space. The default MW analog we simulate in this work is marked by the top blue dot, and two variants with 2/3 $M_\star$ and 1/3 $M_\star$ are shown by the two lower blue dots. Two popular SHM relations from \cite{behroozi19} and \cite{kravtsov14} are shown in the black and olive lines respectively. The observationally constrained properties of the real MW and M31 are shown in purple and pink triangles, with MW data taken from \cite{strigari25} (total mass) and \cite{licquia15} (stellar mass) and M31 data taken from \cite{strigari25} (total mass) and \cite{tamm12} (stellar mass). A dwarf system simulated with a very similar approach in \cite{zzc24} but displays a vastly different behavior is shown as the red square. }
    \label{fig:SHMR}
\end{figure}

We propose that this phenomenon of bypassing the core-formation process and having the central DM density nearly monotonically increasing over time, even at cross sections as low as 1 $\rm cm^2/g$, is likely unique to MW-sized systems. MW-sized systems have two characteristics relevant for this matter: a) the (central) stellar mass fraction peaks at a halo mass of $\sim 10^{12}M_\odot$, as illustrated in Fig. \ref{fig:SHMR} (see also \cite{kravtsov14, behroozi19, girelli20}); b) such systems do have compact stellar components (disks and bulges), resulting in effective stellar half-mass radii $r_{\rm half}$ comparable to the `compact stellar component' configuration modeled here. 

The stellar mass fraction of our `pair14' MW analog is shown in Fig. \ref{fig:SHMR}, compared against two standard stellar-to-halo mass relations (SHMR) from \cite{kravtsov14} and \cite{behroozi19}. The uppermost blue point in Fig. \ref{fig:SHMR} indicates the fiducial stellar mass fraction of the `pair14' MW analog, derived directly from the TNG data. We find that this fiducial stellar mass fraction, with a value of $3\%$ at a halo mass of $3 \times 10^{11} M_\odot$, is slightly elevated compared to the \cite{behroozi19} (also known as the UniverseMachine) prediction. However, it remains consistent with the SHMR from \cite{kravtsov14}. Furthermore, observational constraints for the actual MW and M31 (also plotted) suggest even higher stellar mass fractions. This suggests that the bypassed core formation and monotonically increasing central density, observed here at cross sections as low as $\sim 1\rm cm^2/g$, may be even more pronounced in the real LG.  

To determine the lower limit of the stellar mass fraction required to sustain this bypassed core formation and thus verify the uniqueness of this phenomenon to MW-mass systems, we perform a controlled experiment. We run additional isolated simulations of this MW analog, fixing its DM mass, halo concentration and stellar half-mass radius ($r_{\rm half}$=2.5 kpc), while scaling the total stellar mass to 2/3 and 1/3 of the fiducial value (indicated by the lower blue points in Fig. \ref{fig:SHMR}). The initial conditions for these variants are regenerated using \texttt{SpherIC} to ensure dynamical equilibrium. 

\begin{figure}
    \centering
    \includegraphics[width=\columnwidth, clip,trim=0.2cm 0cm 0.2cm 0cm]{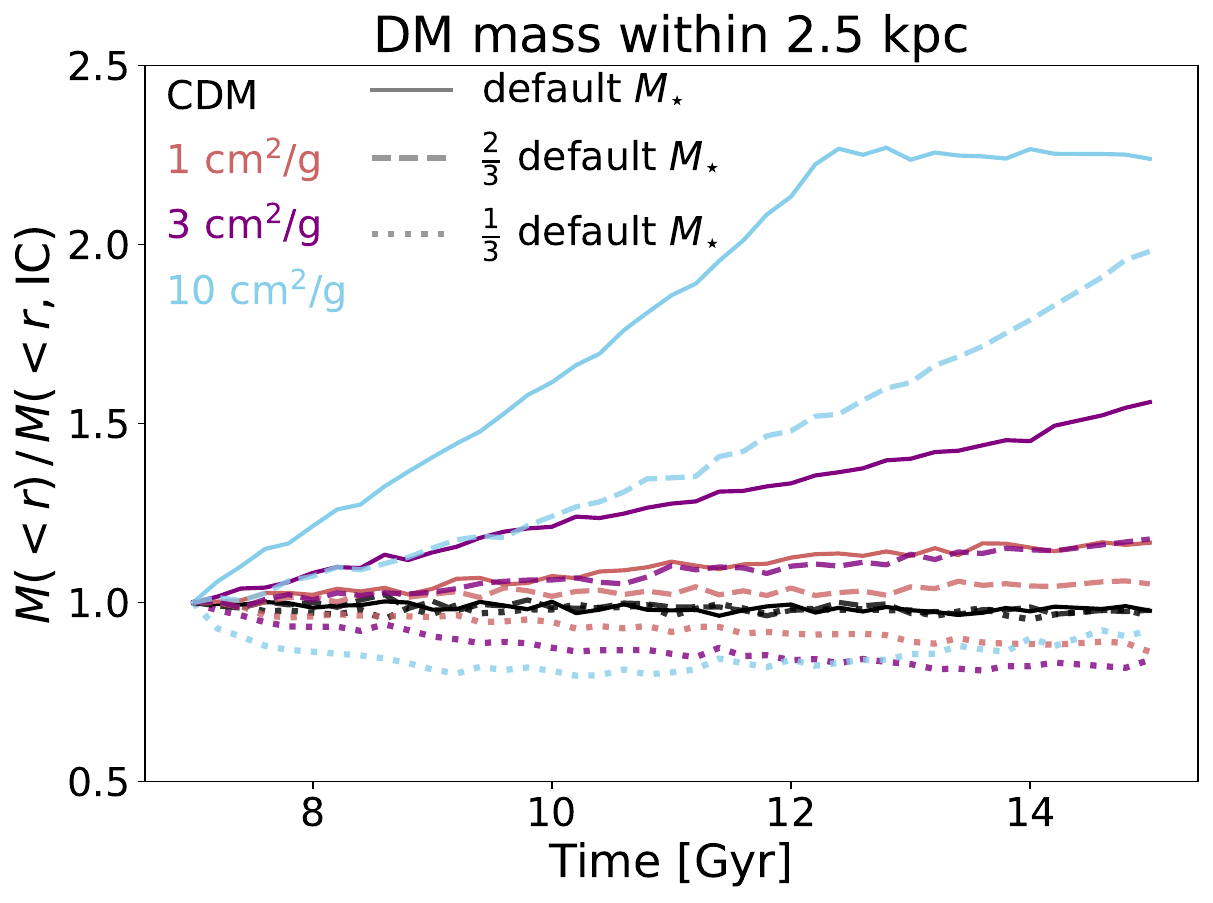}
    \caption{The enclosed DM mass within 2.5 kpc for the compact-stellar case of the MW analog, similar to Fig. \ref{fig:iso-tevo-a}, but with varying stellar masses.}
    \label{fig:vary-starmass}
\end{figure}

Fig. \ref{fig:vary-starmass} presents the evolution of the enclosed DM mass within 2.5 kpc for these three cases. Compared to the fiducial MW analog (solid lines), the $2/3M_\star$ variant (dashed lines) retains the signature of bypassed core formation, with $M(<2.5\rm kpc)$ increasing monotonically over time, albeit at a reduced rate. In contrast, the $1/3M_\star$ case exhibits a standard core-formation phase, similar to the DMO or diffuse-stellar scenarios, and consistent with previous studies of SIDM halos hosting central galaxies \cite{cruz21, fz22, sq23, ymzhong23, zzc24, dnyang24, straight25}.  Notably, we find that within the physically relevant parameter space ($\sigma/m \lesssim10\ \rm cm^2/g$), the presence or absence of core formation is primarily determined by the stellar mass threshold and shows no strong dependence on the specific cross section value. This supports our hypothesis that the bypassed core formation is driven by the initial configuration of the halo's DM velocity dispersion profile, specifically, the initial negative temperature gradient induced by the deep baryon potential.

One potential alternative explanation to this no-core behavior is that the core still forms, just at a smaller radius than the 2.5 kpc aperture. In such a scenario, the monotonic increase in $M(<2.5\rm kpc)$  would not reflect the evolution of the halo center itself, but rather the accumulation of mass at intermediate radii resulting from the outward DM flow during core formation (see Sec. \ref{sec:iso-b} and \cite{gutcke25} for detailed discussion). To investigate this possibility, we examine the enclosed mass evolution at radii smaller and larger than 2.5 kpc in Appendix \ref{appx:aperture} (Figs. \ref{fig:mass-varyaperture-dmo} through \ref{fig:mass-varyaperture-iso-2thirdstars}). This analysis, extending down to 1 kpc, confirms that the fiducial $M_\star$ case exhibits no evidence of core formation across all tested cross sections. The $2/3M_\star$ case displays only a mild mass reduction ($\lesssim10\%$, compared to $\gtrsim$40\% in DMO) within 1 kpc, suggesting that this configuration lies near the transition boundary of the bypassed core-formation regime in the parameter space of Fig. \ref{fig:SHMR}.

In this subsection, we have explored the phenomenon of bypassed core formation in MW-like systems by fixing the halo properties while varying the stellar mass (and thus the central baryon potential). While it is qualitatively expected that this feature becomes more prominent in more massive galaxies, our limited simulation suite prevents us from definitively disentangling whether the stellar mass fraction or the total stellar mass is the dominant governing factor. For instance, it remains to be seen whether cluster-sized systems, which possess larger total stellar masses but lower stellar mass fractions, exhibit similar behavior. Furthermore, both the DM halo concentration and the spatial concentration of the stellar mass distribution are likely critical parameters, with the latter already demonstrated by the compact versus diffuse stellar comparison. For the fiducial DM concentrations adopted in pair14 ($c_{\rm MW}\approx 12.1$, $c_{\rm M31}\approx 7.4$ from TNG; see Table \ref{table:pair14}), we perform a simple sensitivity test in Appendix \ref{appdx:conc_test} in which both halos are instead initialized with $c\approx 9$ from the mass-concentration relation in \cite{diemer19}. We find that bypassed core formation and the compact vs. diffuse dichotomy in the LG runs remain qualitatively unchanged; only the core-collapse timescale shifts modestly when the MW concentration is lowered. A comprehensive study spanning this parameter space is required to fully characterize the evolutionary tracks of SIDM galaxies beyond the bright-dwarf regime explored in \cite{zzc24}. We leave such an investigation to future work.

\subsection{Evolution of DM within 20 kpc}\label{sec:iso-b}

To establish a proper baseline for the diffuse-stellar configuration prior to the Local Group encounter, Fig. \ref{fig:iso-tevo-b} and \ref{fig:iso-tevo-e} show the evolution of the enclosed DM mass within 20 kpc. This radius corresponds to the initial $r_{\rm half}$ of the diffuse stellar component and lies well outside the central DM density core (see Fig. \ref{fig:final-pro-a}). At this intermediate radius,  the dynamical impact of SIDM differs from the behavior governing the central core and is notably less drastic. In the DMO scenarios, core formation in the center pushes DM outward. Conversely, at larger radii where the temperature gradient is negative (i.e., outside the velocity dispersion peak in Fig. \ref{fig:final-pro-f}), the negative heat capacity of the halo drives an inward gravothermal mass flow. The aggregate effect of these two radial bulk motions results in a local density enhancement (a `bump') in the DM density profile, as evidenced by the DMO cases in Fig. \ref{fig:final-pro-d} (see also Fig. 2 of \cite{gutcke25}). This feature is also reflected in the normalized enclosed mass profile (Fig. \ref{fig:final-pro-e}), marking the region subject to net mass accumulation due to gravothermal infall from the outer halo. This mass excess peaks near 20 kpc, coinciding with the chosen $r_{\rm half}$ of the diffuse stellar component. As the core expansion phase proceeds, this enhancement grows, driving a monotonic increase in $M_{\rm DM}(<20\rm kpc)$ across all SIDM DMO simulations (scattered points in Fig. \ref{fig:iso-tevo-b}). For the largest cross section 10 $\rm cm^2/g$, this mass increase amounts to 10$\%$, a noticeable but subdominant effect compared to the drastic evolution occurring within the central 2.5 kpc.

Comparatively, MW analogs with a diffuse stellar component share the same bump feature in density/mass enhancement at 20 kpc, as shown in Fig. \ref{fig:iso-tevo-e} and \ref{fig:final-pro-e}. In contrast, for cases with a compact stellar component, the evolution of $M_{\rm DM}(<20\rm kpc)$ diverges from the DMO baseline (Fig. \ref{fig:iso-tevo-b}). While the 1 $\rm cm^2/g$ and 3 $\rm cm^2/g$ cases show a monotonic increase in enclosed mass within the central 2.5 kpc, they exhibit negligible change in $M_{\rm DM}(<20\rm kpc)$. Conversely, the 10 $\rm cm^2/g$ case (cyan thin line), which reaches the core-collapse threshold at $t\sim12$ Gyr, displays a decrease in $M_{\rm DM}(<20\rm kpc)$ during the late stages of collapse, opposite to the enhancement seen in the DMO and diffuse-stellar cases. This trend is also corroborated by the density and mass profiles of the final snapshot in Figs. \ref{fig:final-pro-d} and \ref{fig:final-pro-e}. Thus, while core collapse is characterized by significant DM mass inflow in the halo center, it concurrently induces a net DM mass outflow at intermediate radii ($\sim 20$ kpc in this MW analog) on the order of a few percent.

\subsection{Evolution of the stellar half-mass radius}\label{sec:iso-c}

Unlike DM which can have self-interaction, the stars in these gas-free simulations behave as collisionless tracers, responding purely to changes in the total gravitational potential produced by the DM and the stellar mass itself. Therefore, the evolution of the stellar half-mass radius $r_{\rm half}$ serves as a dynamical proxy for the mass within $r_{\rm half}$. While Sec. \ref{sec:iso-a} and Sec. \ref{sec:iso-b} utilize fixed apertures (set to the initial $r_{\rm half}$ values of 2.5 kpc and 20 kpc) to quantify DM evolution, Fig. \ref{fig:iso-tevo-c} and Fig. \ref{fig:iso-tevo-f} illustrate the evolution of the stellar $r_{\rm half}$ itself. As expected, the stellar structural evolution correlates closely with the trends observed in the enclosed DM mass.

In the cases with a compact stellar component ($r_{\rm half}=2.5$ kpc; Fig. \ref{fig:iso-tevo-c}), the CDM galaxy remains largely stable, exhibiting only minor expansion ($\lesssim5\%$ by the end of the simulation). This slight diffusion is attributable to numerical relaxation effects associated with the force resolution limit (see \cite{zzc24}, Sec. IIB for a detailed discussion). The SIDM galaxies, instead of puffing up as seen in the SIDM dwarfs in \cite{zzc24}, contract over time, with the magnitude of this contraction scaling with the SIDM cross section. This behavior mirrors the DM evolution within 2.5 kpc (Fig. \ref{fig:iso-tevo-a}, solid lines), where the enclosed mass increases monotonically across all cross sections, reflecting the bypassed core-formation phase.

However, for the cases with a diffuse stellar component ($r_{\rm half}=20$ kpc; Fig. \ref{fig:iso-tevo-f}), the $r_{\rm half}$ evolution is no longer governed by the DM in the halo's central region, but by the enclosed mass at larger radii $M(\lesssim20\rm kpc)$, which corresponds to the density/mass `bump' region identified in Sec. \ref{sec:iso-b} (see Fig. \ref{fig:final-pro-d} and \ref{fig:final-pro-e}). As the SIDM core forms and expands in this diffuse-stellar case, the DM mass enclosed within this intermediate region slightly increases, as shown in Fig. \ref{fig:iso-tevo-e}. The magnitude of this enhancement scales with the cross section, reaching $\lesssim10\%$ for 10 $\rm cm^2/g$. Consequently, the stellar $r_{\rm half}$ exhibits a minor contraction on the order of a few percent, as shown in Fig. \ref{fig:iso-tevo-f}.

In summary, both the compact and diffuse MW analogs exhibit a decrease in the stellar half-mass radius $r_{\rm half}$ as they co-evolve with the SIDM halo. While the stellar structural evolution tracks the deepening of the DM potential in both scenarios, the underlying drivers of the DM mass increase differ. In the compact case, the contraction is driven by the density increase in the halo center, resulting from the bypassed core-formation phase for this MW analog. In the diffuse-stellar case, the contraction of $r_{\rm half}$ is driven by the density enhancement (or `bump') accumulating outside the SIDM core. Unlike the compact case, this mechanism induces only a marginal reduction in $r_{\rm half}$. These isolated evolutionary baselines set the stage for the following section, where we analyze the evolution of these MW analogs within the dynamic environment of the Local Group.

\section{The past-pericenter scenario of the Local Group}\label{sec:lgsim}

\begin{figure*}
    \centering
    \includegraphics[width=\textwidth, clip,trim=0.2cm 0cm 0.2cm 0cm]{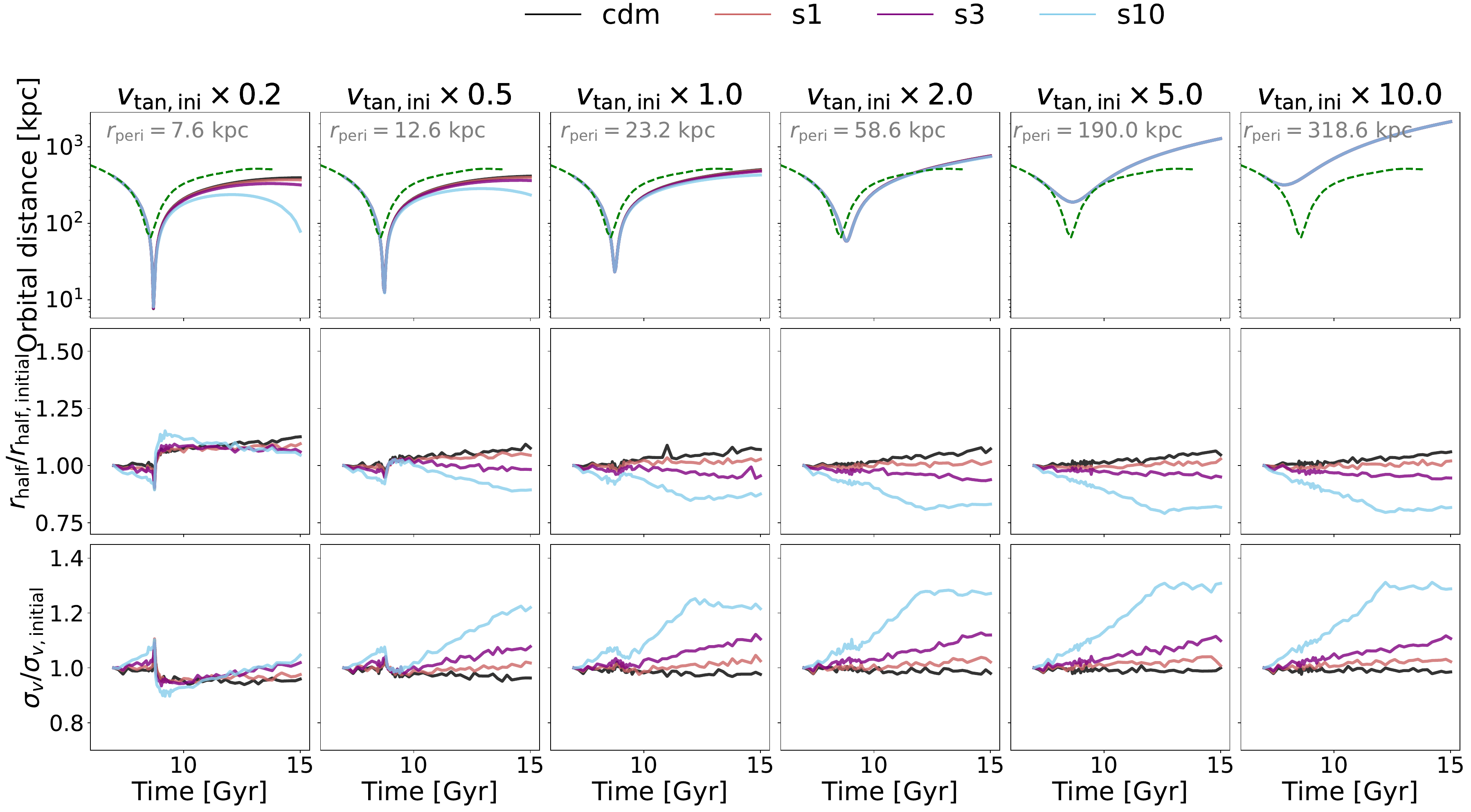}
    \caption{For the compact-stellar case of the MW analog (initial stellar $r_{\rm half}=2.5$ kpc): the time evolution of the LG analog's orbital distance (top row), the MW analog's stellar $r_{\rm half}$ (middle row) and radial velocity dispersion within $r_{\rm half}$ (bottom row). Each column represents a variant of the LG analog, where we vary the initial tangential velocity $v_{\rm tan, ini}$ to manually control the pericenter distance. Both the multiplication factor on $v_{\rm tan, ini}$ and the corresponding $r_{\rm peri}$ measured in simulation are labeled in plot. The green dashed line shows the original TNG orbit of this LG analog, and colored lines represent the simulated orbits under different DM models, with color scheme the same as in Fig. \ref{fig:iso-timeevo}.}
    \label{fig:bigplot-compact}
\end{figure*}

\begin{figure*}
    \centering
    \includegraphics[width=\textwidth, clip,trim=0.2cm 0cm 0.2cm 0cm]{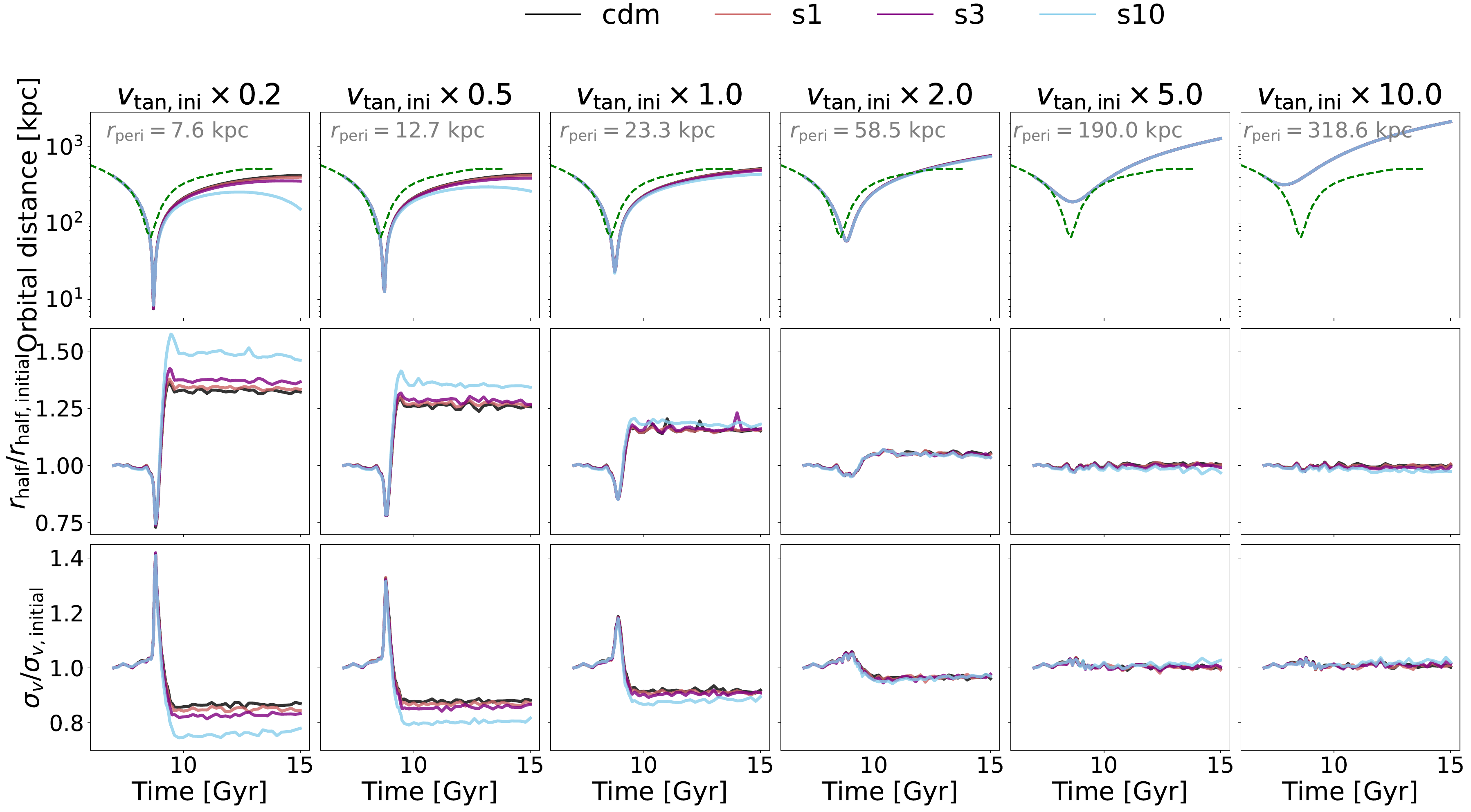}
    \caption{Similar to Fig. \ref{fig:bigplot-compact}, but for the diffuse-stellar case of the MW analog with initial stellar $r_{\rm half}=20$ kpc. }
    \label{fig:bigplot-diffuse}
\end{figure*}

The evolution of the MW in SIDM models is further complicated when the Local Group environment is taken into consideration, particularly in the context of a possible past pericenter encounter, as proposed by \cite{hartl24}. This section utilizes idealized simulations to investigate this past-pericenter scenario for LG analogs and evaluate its compatibility with SIDM constraints. Specifically, we aim to test two key aspects of this encounter: the macroscopic orbital evolution of the system under different SIDM cross-sections, and the dynamical robustness of the central galaxies during a strong pericentric passage. For our simulation setup, we adopt the properties of the `pair14' LG analog from TNG, of which the MW constituent was simulated in isolation in the preceding section. `pair14' was selected because, among all LG analogs with a past pericenter in TNG, its orbital history is most accurately reproduced by our idealized simulations (see Sec. \ref{sec:orbitsim}). We initialize the displacements and relative velocities for the MW and M31 analogs using information from TNG snapshots at early times instead of rewinding orbits from the present day, because the latter generally fails to capture the early pericenter (see Fig. \ref{fig:orbit-pairs-peris}, especially `pair0' and `pair14'). We select the TNG snapshot corresponding to cosmic time $t=7.0$ Gyr (redshift $z\sim0.75$), extract the initial displacement $r_{\rm ini}$, tangential velocity $v_{\rm tan, ini}$ and radial velocity $v_{\rm rad, ini}$, and incorporate them into the IC for our idealized simulations (for more details on selecting the TNG snapshot to extract IC information, see Appendix \ref{appdx:whichic}). By extracting the initial conditions directly from TNG, Sec. \ref{sec:orbitsim} establishes `pair14' as a cosmologically motivated baseline, demonstrating that such a past-pericenter scenario occurs naturally. Building upon this realistic anchor, we now proceed to systematically explore the parameter space around this baseline to investigate the dynamical robustness of the system. We emphasize that these full orbital simulations begin from the original, unevolved initial conditions at $t=7.0$ Gyr. Because the halos are actively evolving under SIDM thermodynamics, the pericenter is reached before the intrinsic SIDM core-collapse or core-expansion processes are fully complete. This allows us to capture the complex, simultaneous co-evolution of SIDM thermodynamics and external tidal disruption. We note that the exact structural outcome inherently depends on the chosen initial time $t_{\rm ini}$, which dictates the `pre-evolution time' the system undergoes prior to the encounter. A systematic investigation of how this pre-evolution time influences subhalo-host interactions is detailed in Appendix E of \cite{zzc22}. In this work, by comparing these full orbital runs against the strictly isolated baselines established in Sec.~\ref{sec:iso}, we can effectively disentangle the tidal encounter effects from the intrinsic SIDM evolution.

A concern regarding this pericenter scenario is the potential tidal disruption of the central galaxies. SIDM cores, due to the shallower central potentials, could further exacerbate this susceptibility.  To systematically investigate these effects, we vary the initial tangential velocity $v_{\rm tan, ini}$ by scaling the fiducial TNG value by factors of $[0.2, 0.5, 1.0, 2.0, 5.0, 10.0]$, thereby controlling the pericenter distance. The primary objective of this section is to determine the minimum pericenter distance compatible with the survival of the MW analog's central galaxy (see also \cite{sawala25} for the future-pericenter scenario). Following the methodology of Sec. \ref{sec:iso}, for each velocity configuration, we construct two distinct initial stellar distributions, either with an initial $r_{\rm half}$ of 2.5 kpc or 20 kpc. These configurations are designed to represent the compact disk/bulge and the diffuse stellar halo, respectively, as well as bracketing the range of central potential depths contributed by stellar components. To obtain conservative estimates of the disruption effects, we treat the less massive MW analog as our primary probe of tidal destruction, while fixing the more massive M31 analog's stellar distribution to the compact case ($r_{\rm half}=2.7$ kpc). This maximizes M31's tidal gravitational field while making it highly resilient to disruption itself. Because M31's own evolution simply mirrors the behavior of the compact-stellar MW cases, we hereafter focus exclusively on the potentially observable structural changes of the MW analog, whose diverse components provide tighter falsifiable constraints for this past-pericenter scenario with different SIDM models. 
We upload visualizations of these past-pericenter simulations for the readers' reference\footnote{Supplementary video URL: \url{https://www.youtube.com/playlist?list=PLqe-K0erSag_7dwUKX1om8I1BZUGCEkbJ}}.

Fig. \ref{fig:bigplot-compact} and Fig. \ref{fig:bigplot-diffuse} present the time evolution of the MW analog's properties for the compact and diffuse stellar components, respectively. In both figures, the top row displays the MW--M31 separation, the middle row shows the MW analog's stellar half-mass radius $r_{\rm half}$ (normalized to its initial value), and the bottom row shows the MW analog's stellar radial velocity dispersion $\sigma_v$ within $r_{\rm half}$ (also normalized). Comparing the compact and diffuse stellar configurations at a fixed $v_{\rm tan, ini}$, the LG orbital history remains virtually identical, indicating that the macroscopic orbital trajectories are largely insensitive to the internal stellar compactness. In fact, these trajectories are also broadly robust against variations in the dark matter physics: altering the SIDM cross-section produces noticeable deviations only during the closest encounters ($v_{\rm tan, ini}\times0.2$ and $v_{\rm tan, ini}\times0.5$). In these deep pericentric passages, larger cross-sections result in stronger orbital decay. This accelerated decay arises because SIDM particle collisions between the MW and M31 analogs introduce an effective collisional drag that decelerates the halos \cite{kahlhoefer14}. Additionally, SIDM-induced evaporation \cite{mv12, zzc22} leaves stripped material along the orbital trajectory, which further enhances self-friction \cite{miller20}.

To measure $r_{\rm half}$ and $\sigma_v$ at each snapshot, we utilize all stellar particles originally associated with the MW analog in the IC, regardless of their boundedness at later times. This is for two reasons: a) determining particle membership via halo finders is intrinsically difficult during close pericenter passages such as those we explore, especially when the two systems have similar masses; and b) this provides a conservative analysis, since including the unbound star particles will only exaggerate the change in the galaxy's properties.  To resolve the pericenter distances as accurately as possible, we deliberately increase the snapshot output frequency near the expected pericenter time, and mark the corresponding pericenter distance $r_{\rm peri}$ in the plots. Lower $v_{\rm tan, ini}$ scaling factors correspond to lower LG angular momentum and thus closer pericenter distances. Conversely, in the limit of large $v_{\rm tan, ini}$, we expect the MW analog's evolution to converge to the isolated case presented in Sec. \ref{sec:iso}.

For the compact-stellar cases (Fig. \ref{fig:bigplot-compact}), within each individual dark matter model, the structural evolution is highly robust against the encounter kinematics: across the $v_{\rm tan, ini}$ scaling factors of $[1.0, 2.0, 5.0, 10.0]$, the final $r_{\rm half}$ and $\sigma_v$ values remain nearly identical for a given cross-section.
Even in the $v_{\rm tan, ini}\times1.0$ and $v_{\rm tan, ini}\times2.0$ runs with the largest SIDM cross section of 10 $\rm cm^2/g$, we observe only transient perturbations in both $r_{\rm half}$ and $\sigma_v$ during the pericenter passage at $t\sim8$ Gyr; however, the system subsequently re-relaxes. This suggests that despite a close encounter with the M31 analog ($r_{\rm peri}\sim20$ kpc), the structural evolution of the MW analog remains primarily governed by its internal dynamics. Even in the extreme case of the closest pericenter (7.6 kpc), we observe only percent-level deviations in the evolution of the MW analog for CDM and 1 $\rm cm^2/g$ cases. In contrast, the 3 $\rm cm^2/g$ and 10 $\rm cm^2/g$ cases are more strongly affected, as the tidal interaction and evaporation (see more detailed discussion in \cite{zzc22}) disrupt their ongoing core-collapse process. We caution, however, that these results are derived from our simulation setup with only DM and stars. The inclusion of gas hydrodynamics might fundamentally alter these dynamics.

In contrast, for the diffuse-stellar cases (Fig. \ref{fig:bigplot-diffuse}), the MW analog's stellar component is significantly more susceptible to the pericenter encounter, exhibiting deviations in the final $r_{\rm half}$ and $\sigma_v$ when $r_{\rm peri}\lesssim 60$ kpc (corresponding to $v_{\rm tan, ini}\times2.0$). This differs from the compact-stellar MW analogs for two reasons: a) the diffuse distribution is designed to represent the extended stellar halo with $r_{\rm half}\sim20$ kpc, which is inherently more susceptible to tidal forces during the encounter; and b) the SIDM halo remains in the cored regime (unlike the core-collapsing compact-stellar cases; see Sec. \ref{sec:iso-a}), resulting in a shallower central potential that offers less gravitational binding. Consequently, for cases between $v_{\rm tan, ini}\times0.2$ to $v_{\rm tan, ini}\times1.0$, $r_{\rm half}$ increases and $\sigma_v$ decreases following the encounter, driven by the weakening of the central potential.

\begin{figure}
    \centering
    \begin{subfigure}[t]{\columnwidth}
        \centering
        \includegraphics[width=\textwidth, clip,trim=0.2cm 0cm 0.2cm 0cm]{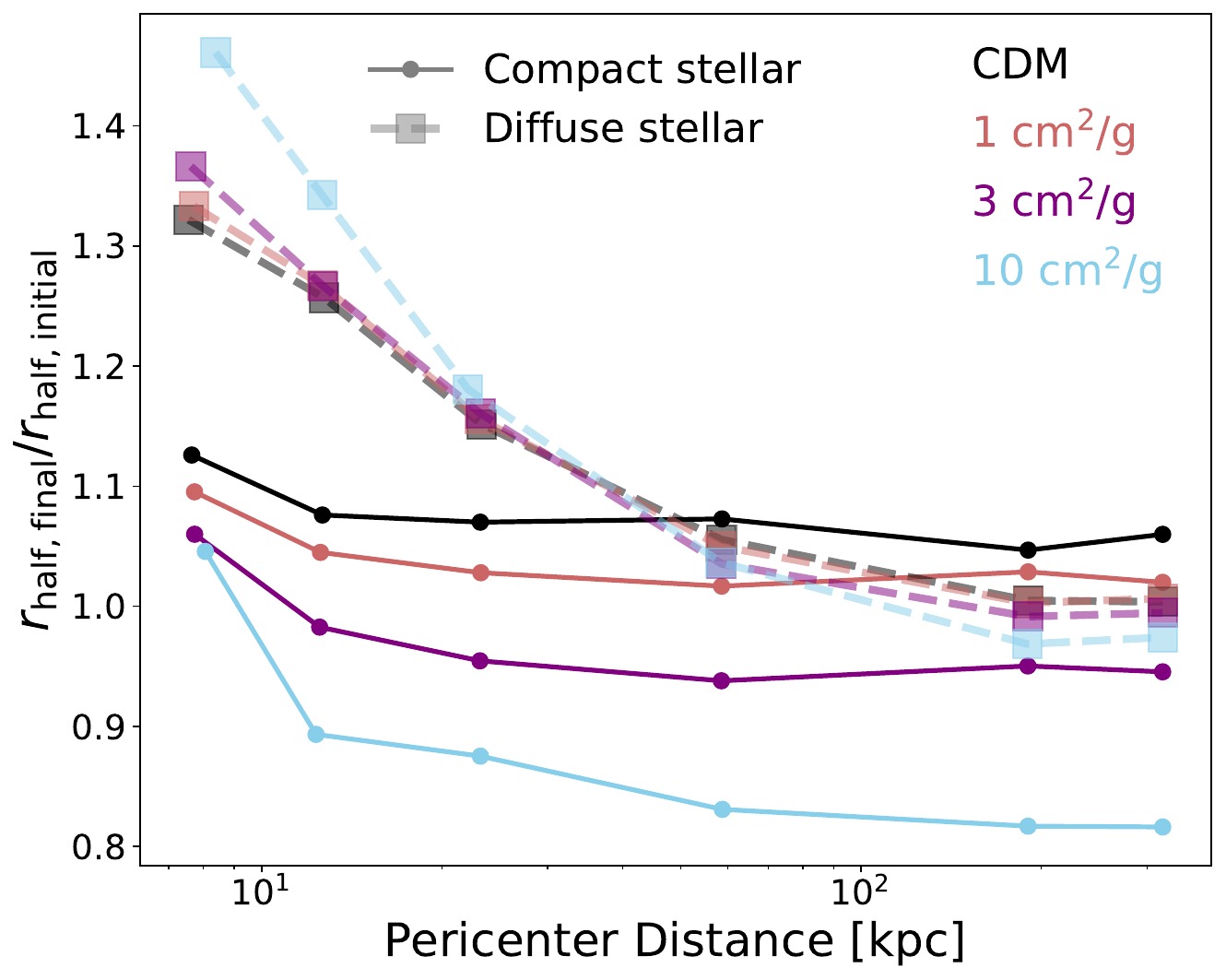}
        \caption{}
        \label{fig:effect-rhalf}
    \end{subfigure}
    ~
    \begin{subfigure}[t]{\columnwidth}
        \centering
        \includegraphics[width=\textwidth, clip,trim=0.2cm 0cm 0.2cm 0cm]{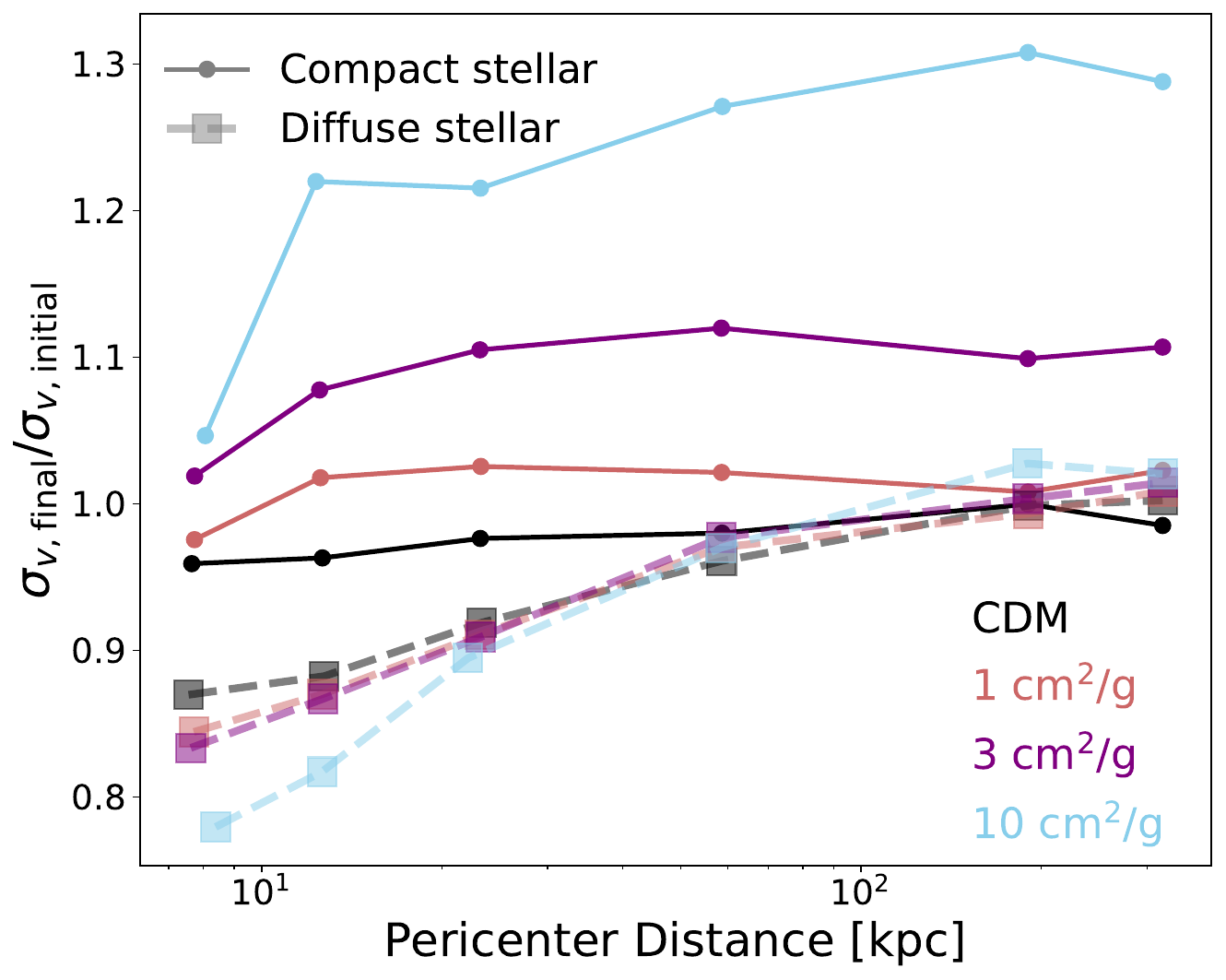}
        \caption{}
        \label{fig:effect-vdisp}
    \end{subfigure}
    \caption{The pericenter effects on the MW analog's stellar $r_{\rm half}$ (top) and velocity dispersion $\sigma_v(<r_{\rm half})$. The solid lines show the compact-stellar MW analog, and the dashed lines show the diffuse-stellar MW analog. Different colors represent different DM models, with the color scheme same as in Fig. \ref{fig:iso-timeevo} and also labeled in plot.}
    \label{fig:effects}
\end{figure}

Figure \ref{fig:effects} summarizes these post-pericenter effects as a function of $r_{\rm peri}$. For the compact-stellar cases, the structural evolution is primarily governed by the intrinsic SIDM evolution (dependent on cross section), with the past pericenter playing a negligible role unless $r_{\rm peri}\lesssim20$ kpc.
Conversely, for the diffuse-stellar cases, the intrinsic SIDM evolution has minimal impact, with variations in cross section producing only percent-level differences in the large  $r_{\rm peri}$ limit. This is because the initial $r_{\rm half}$ of 20 kpc lies well outside the SIDM core, and is therefore insensitive to the evolution of the halo center (see more detailed discussion in Sec. \ref{sec:iso-b}). However, pericenter effects become increasingly prominent as $r_{\rm peri}$ decreases below $\sim 100$ kpc. 
Therefore, in a realistic LG past-pericenter scenario, we expect the disk/bulge component to remain intact unless $r_{\rm peri}\lesssim20$ kpc, whereas the stellar halo would be susceptible to disruption for $r_{\rm peri}\lesssim100$ kpc. While larger SIDM cross sections may enhance the disruption of the stellar halo, they simultaneously drive stronger core collapse in the MW center.

Thus far, our analysis has focused on the post-pericenter properties of the MW analog, specifically after the system has re-relaxed following the encounter. However, the non-equilibrium dynamics during this violent encounter, while complex and secondary to our primary focus, warrant a brief discussion. We utilize the diffuse-stellar case for this illustration, as it provides a cleaner physical picture free from the complicating effects of ongoing core collapse. As shown in Fig. \ref{fig:bigplot-diffuse}, for cases ranging from $v_{\rm tan, ini}\times0.2$ to $v_{\rm tan, ini}\times1.0$, the stellar $r_{\rm half}$ exhibits an initial contraction followed by a subsequent expansion. Both phases occur in an abrupt, violent manner. This behavior contrasts sharply with the smooth evolution typical of tidal stripping scenarios, such as those observed in dwarf galaxies \cite{zzc24}. In this major-merger, close-pericenter scenario, the deepening of the joint gravitational potential as the halos overlap leads to a strong compression of the MW analog's central galaxy \cite{renaud09}. This tidal compression may further facilitate gas condensation and enhance the star formation rate \cite{renaud14}. Following compression, the impulsive heating from the rapidly changing tidal field during this short encounter, also known as tidal shock \cite{ostriker72, abenson22, xdu24}, injects kinetic energy into the stellar particles, making them less bound. The structural response to this shock manifests over a dynamical timescale, $t_{\rm dyn}\sim r_{\rm half}/\sigma_v \sim 0.3$ Gyr. Consequently, both Fig. \ref{fig:bigplot-diffuse} and the visualization 
movie\footnote{Supplementary video URL: \url{https://www.youtube.com/playlist?list=PLqe-K0erSag_7dwUKX1om8I1BZUGCEkbJ}}
display a delayed `explosion' or rapid expansion of the galaxies following the pericenter passage. Moreover, since the tidal heating rate scales as $r^2$ \cite{pullen14, sq20}, the outer DM halo is more severely affected by tidal shocks than the central galaxy. We therefore anticipate that the analysis of satellite galaxies in MW analogs could yield tighter constraints on the LG past-pericenter scenario. We reserve this investigation for future work (Hartl et al. in prep.).

\section{Summary and discussion}\label{sec:summary}

In this work, we investigate Local Group (LG) analogs selected from the Illustris TNG300-1 simulation, with a specific focus on systems exhibiting a past pericenter and the compatibility of this scenario with SIDM. We performed idealized N-body re-simulations of each LG analog (Sec. \ref{sec:orbitsim}), initializing the MW and M31 analogs as spherically symmetric halos isolated from large-scale structure. Our analysis of the orbital histories reveals that only 2 of the 15 LG analogs identified with a past pericenter \cite{hartl24} successfully reproduce this feature in idealized simulations. This implies that the orbital evolution of the majority of past-pericenter analogs is strongly influenced by their cosmological environment.

We subsequently focused on `pair14', the candidate that best reproduces the TNG orbit in our idealized setup. Before introducing the complex dynamics of the MW--M31 encounter, we first decoupled the internal SIDM-baryon interplay from the external LG tidal forces by analyzing the MW analog's evolution in isolation (Sec.~\ref{sec:iso}). The MW analog is populated with stellar particles to form a two-species (DM+stars) N-body simulation \cite{zzc24}, with two types of initial stellar distribution considered: a `compact-stellar' case (initial $r_{\rm half}=2.5$ kpc) designed to represent the disk and bulge, and a `diffuse-stellar' case (initial $r_{\rm half}=20$ kpc) representing the stellar halo. This allowed us to establish a baseline for the MW analog's evolution under different DM models prior to the onset of LG pericenter effects. 

For the compact-stellar case (representing the stellar disk/bulge), we find that the MW analog's stellar $r_{\rm half}$ decreases monotonically over time across all SIDM models, even for cross sections as low as 1 $\rm cm^2/g$, which is a trend absent in the CDM runs. This contrasts with our previous study of SIDM dwarf galaxies \cite{zzc24}, which predicted a universal core-formation phase characterized by initial expansion prior to any subsequent contraction. We attribute this to the dense central stellar distribution, which imposes an initial negative temperature gradient on the dark matter halo, effectively initializing the system in a state of ongoing core collapse. Our controlled tests with varying stellar masses confirm that this purely dynamical phenomenon is highly sensitive to the stellar mass fraction. Because this fraction naturally peaks at the MW mass scale, this bypassed core-formation regime may be unique to MW-like systems (see also \cite{sameie21, rose22, despali25} for complementary hydrodynamical studies). 

Conversely, in the diffuse-stellar case (representing the stellar halo), the system still undergoes minor contraction, but through a different mechanism. This occurs because, while an SIDM core forms due to the lower stellar concentration, the relevant $r_{\rm half}$ lies outside this core region. At this radius, the SIDM enclosed mass profile exhibits a localized enhancement (or `bump') resulting from the accumulation of dark matter displaced from the expanding core meeting material infalling from larger radii due to heat outflow. This leads to a deepening potential and, consequently, a slight contraction of the stellar halo.

Building upon the characterization of the MW analog's evolution in isolation, we proceed to embed the system within the Local Group environment using idealized simulations.  Our objective is to constrain the minimum allowed past pericenter distance compatible with the survival of the MW across different dark matter models. We modulate the LG pericenter by systematically varying the initial tangential velocity between the MW and M31 analogs, while holding all other LG properties fixed. In the compact-stellar case, we find that the MW's structural evolution is governed primarily by its internal SIDM thermodynamics, with pericenter effects becoming discernible only for $r_{\rm peri}\lesssim20$ kpc. Conversely, for the diffuse-stellar case, the MW proves significantly more susceptible to the past pericenter, with encounters of $r_{\rm peri}\lesssim100$ kpc leaving distinct imprints on the stellar distribution. Notably, this tidal response remains largely degenerate across the tested SIDM models $\sigma/m\lesssim10\ \rm cm^2/g$. Therefore, in the absence of a full treatment of hydrodynamics, we expect the MW's disk/bulge to be sensitive to the specific SIDM physics but robust against pericenter passages $>20$ kpc, whereas the stellar halo appears relatively insensitive to the SIDM model but is highly vulnerable to disruption by pericenter passages $<100$ kpc.

Beyond the scope of the current study, we identify several avenues for future research and model refinement:

\begin{itemize}
    \item \textit{Baryonic physics:} In this work, we performed N-body simulations containing only DM and star particles, lacking a hydrodynamical treatment of gas physics. Previous studies have investigated the evolution of MW-mass systems in SIDM using hydrodynamic simulations, including \cite{rose22} and \cite{despali25} with IllustrisTNG (\texttt{Arepo} code; subgrid modeling of stellar feedback), and \cite{sameie21} with FIRE-2 (\texttt{GIZMO} code; explicit, burstier stellar feedback). The evolution of the central stellar distribution is contributed by two competing baryonic mechanisms: adiabatic contraction, which increases the central density driven by gas cooling, star formation, and the subsequent steepening of the potential; and stellar feedback, which drives mass outflows and facilitates core formation. Both \cite{rose22} and \cite{despali25} report a phenomenon similar to that observed here: SIDM cores fail to form in MW-mass systems in hydrodynamic simulations, a result attributed to the interplay between adiabatic contraction and SIDM thermodynamics (see also \cite{ragagnin24} for similar results on cluster scales). In \cite{sameie21}, which employs the burstier feedback of FIRE-2, MW-mass systems display a slightly more diverse range of central stellar distributions. Notably, our work demonstrates that even in the complete absence of gas-driven adiabatic contraction, SIDM dynamics alone can generate systematically cuspier density profiles in MW-mass systems compared to CDM. This suggests that the stellar mass fraction and concentration may play a more dominant role than active hydrodynamical processes in driving this `bypassed core formation' phenomenon, though full hydrodynamical simulations remain essential for physical consistency.

    \item \textit{Stellar geometry}: In this work, we model the compact stellar component using a spherically symmetric Hernquist profile, rather than a realistic oblate disk geometry. We argue that this approximation is unlikely to qualitatively alter our main conclusions regarding the bypassed core-formation phenomenon, for the following reason. The gravothermal fluid model that underpins SIDM halo evolution \cite{balberg02, essig19} is formulated under spherical symmetry, where the evolution is driven by the radial temperature gradient set by the enclosed mass profile $M(<r)$. Under the assumption that this radial driver remains dominant in mildly non-spherical geometries, we note that a thin exponential disk (scale length $R_d \sim 2.5$ kpc, scale height $z_0 \sim 0.3$ kpc) produces a spherically-averaged enclosed mass at small radii comparable to, or exceeding, that of our Hernquist profile, owing to the vertical compression of mass into the disk midplane. This suggests that our setup may provide a conservative estimate of the bypassed core-formation effect. However, we caution that the impact of non-spherical potentials --- including both the disk geometry and the intrinsic triaxiality of dark matter halos --- on SIDM gravothermal evolution has not been rigorously quantified in the literature. Furthermore, generating dynamically stable ICs for N-body simulations including a disc component remains technically challenging \cite{dice}. A systematic investigation of these geometric effects therefore constitutes an important avenue for future work.

    \item \textit{Satellite galaxies:} We have demonstrated that the central galaxy of the MW analog remains largely insensitive to a potential past pericenter of the LG, provided $r_{\rm peri}>20$ kpc for the disk/bulge component or $r_{\rm peri}>100$ kpc for the stellar halo (see Fig. \ref{fig:effects}). However, as visualized in the supplementary video, the outer envelopes of the MW's DM halo undergo significant disruption during the encounter. This implies that the satellite populations and distributions of the MW and M31 may place tighter constraints on this past-pericenter scenario, or conversely, explain observational anomalies such as the anisotropic M31 satellite distribution described in \cite{kanehisa25}. A deeper understanding of these effects requires dedicated simulations of dwarf satellites tracking the complete merger history of these LG analogs, which we reserve for future work (Hartl et al. in prep.).

    \item \textit{Mass accretion:} Due to the idealized nature of our experimental setup, we do not model the continuous cosmological mass accretion of the two Local Group constituents. However, given the relatively late start time of the simulations ($t_{\rm ini}=7$ Gyr, $z\sim0.75$), a majority ($\gtrsim70\%$) final halo mass is typically already assembled by $t_{\rm ini}$ \cite{zhao09, fakhouri10, correa14}. Consequently, the omission of smooth mass accretion is expected to have a subdominant impact on the evolution of the LG analogs. Nevertheless, this simplification may contribute to the residual discrepancies observed between the re-simulated idealized orbits and the original TNG trajectories (e.g., `pair14' in Fig. \ref{fig:orbit-pairs-peris}). On the other hand, initializing the halos with their present-day masses at $t_{\rm ini}$ effectively overestimates the depth of the potential wells during the past encounter. This approach allows us to evaluate the pericenter effects in a conservative manner, as it maximizes the tidal forces and the resulting disruption experienced by the central galaxy.

    \item \textit{Gas Dynamics and Tidal Dwarfs:} The N-body framework employed in this study inevitably precludes the capture of complex gas-dynamical phenomena relevant to the past-pericenter scenario. During the close approach of the MW and M31 halos, their extended gas disks (spanning $\sim30$ kpc \cite{levine06, nakanishi16}), which are significantly more spatially extended than their stellar counterparts, become subject to strong tidal elongation. This interaction can generate `bridge'-like tidal structures, serving as potential formation sites for tidal dwarf galaxies characterized by low dark matter fractions and high metallicities \cite{toomre72, bournaud06}. In scenarios involving the closest pericenter passages, where gas disks may undergo high-velocity $\gtrsim500$ km/s, this mechanism could populate the MW-M31 orbital track with a trail of dark-matter-deficient dwarfs. This process mirrors the `bullet-dwarf' scenario \cite{shin20, vdk22, lee23}, which currently provides the leading theoretical explanation for the formation of the dark-matter-deficient galaxies DF2 and DF4 \cite{df2, df4, df4shen}. We reiterate that a comprehensive understanding of these issues requires full hydrodynamical simulations. Such simulations are intrinsically complex and computationally demanding, placing them beyond the scope of this work.

\end{itemize}

In conclusion, evaluating the viability of a past-pericenter encounter in the Local Group with SIDM requires understanding the complex interplay between SIDM thermodynamics, deep baryonic potentials, and environmental tidal disruption. Although our idealized simulations do not include a full treatment of gas physics, our findings from a pure dynamics perspective imply that the dense baryonic center of the Milky Way can drive the SIDM halo into a bypassed core-formation regime, a phenomenon highly sensitive to the stellar mass fraction and likely unique to MW-mass systems.	
This suggests that the compact central galaxy would remain highly robust against tidal disruption even during close pericentric passages.	In contrast, the diffuse stellar halo and the extended dark matter structures are more susceptible to the encounter. Therefore, future efforts to distinguish between Cold and Self-Interacting Dark Matter in the Local Group must account for this complex interplay between thermodynamics, baryonic contraction, and environmental disruption.

\begin{acknowledgments}

LES is supported by the U.S. DOE
Grant DE-SC0010813. The simulations in this work were conducted on the Grace Cluster of High Performance Research Computing at the Texas A\&M University. 

\end{acknowledgments}

\appendix

\section{Evolution of enclosed DM mass at different radii}\label{appx:aperture}

\begin{figure*}
    \centering
    \includegraphics[width=\textwidth, clip,trim=0.2cm 0cm 0.2cm 0cm]{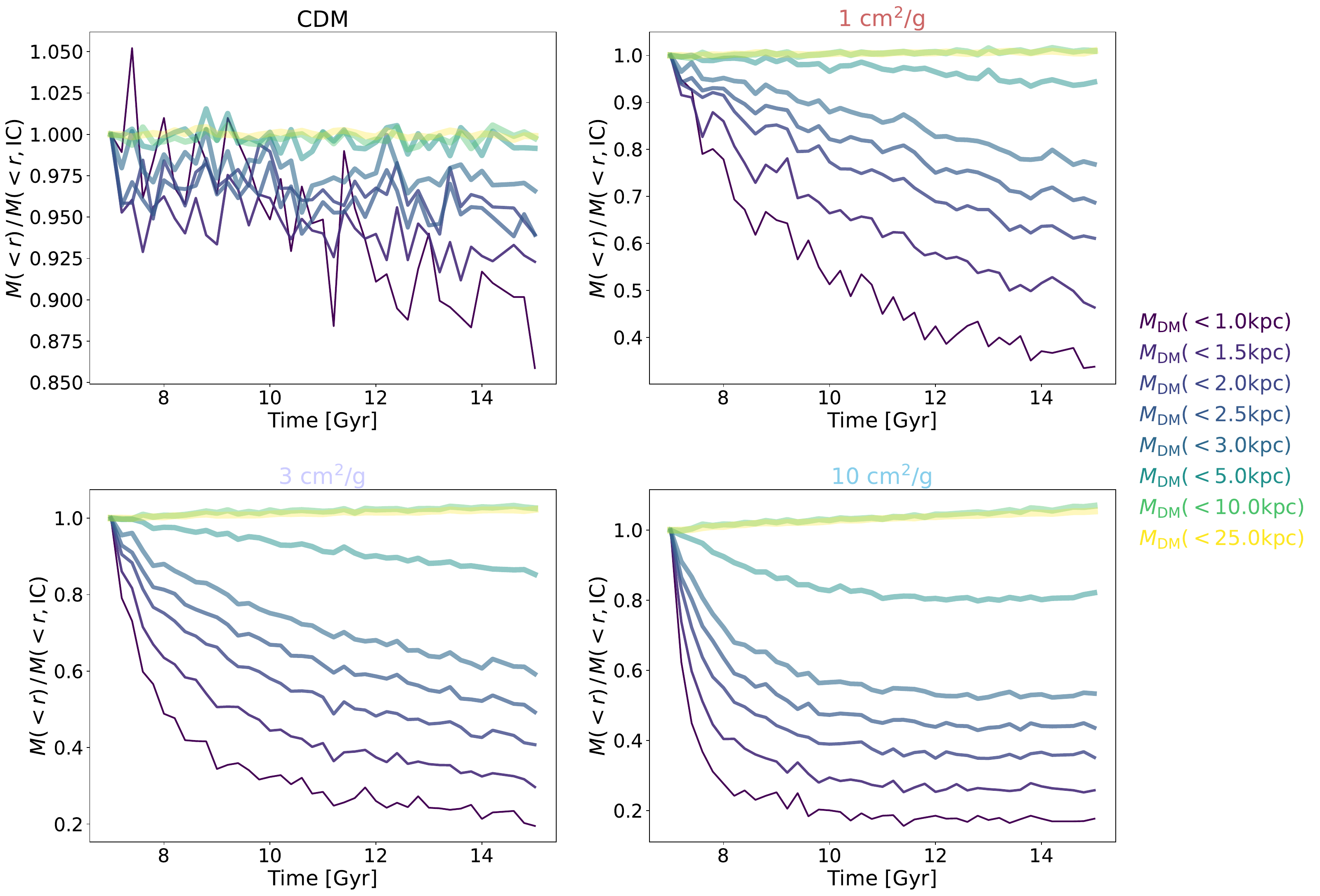}
    \caption{Evolution of the enclosed DM mass at different radii (color scheme as labeled on the right side of the plot), for the DMO case of the MW analog in isolation. Each panel show a different DM model, as shown in its title.}
    \label{fig:mass-varyaperture-dmo}
\end{figure*}

\begin{figure*}
    \centering
    \includegraphics[width=\textwidth, clip,trim=0.2cm 0cm 0.2cm 0cm]{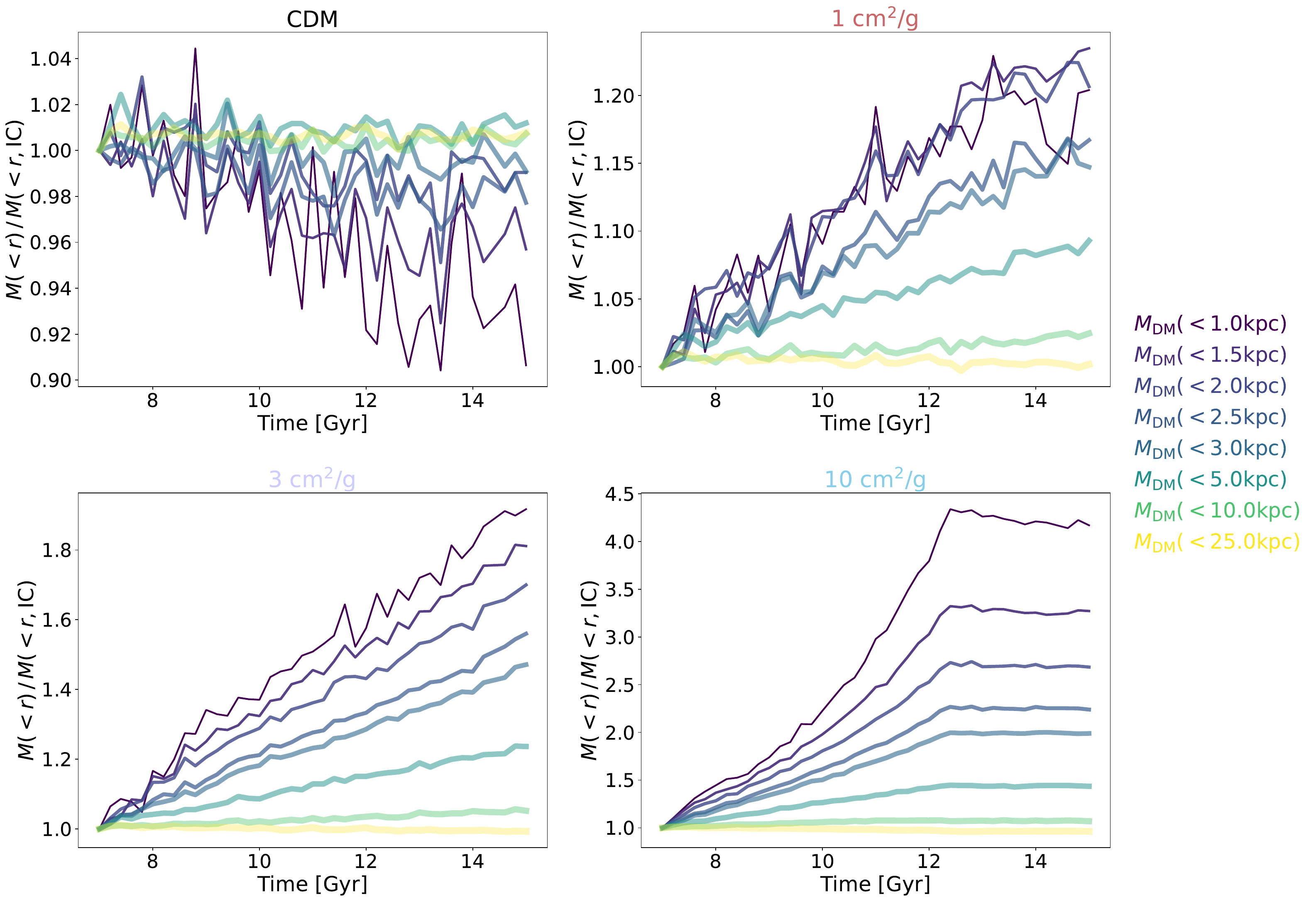}
    \caption{Same as Fig. \ref{fig:mass-varyaperture-dmo}, but for the MW analog with the default stellar mass and compact distribution (initial $r_{\rm half}=2.5$ kpc). }
    \label{fig:mass-varyaperture-iso}
\end{figure*}

\begin{figure*}
    \centering
    \includegraphics[width=\textwidth, clip,trim=0.2cm 0cm 0.2cm 0cm]{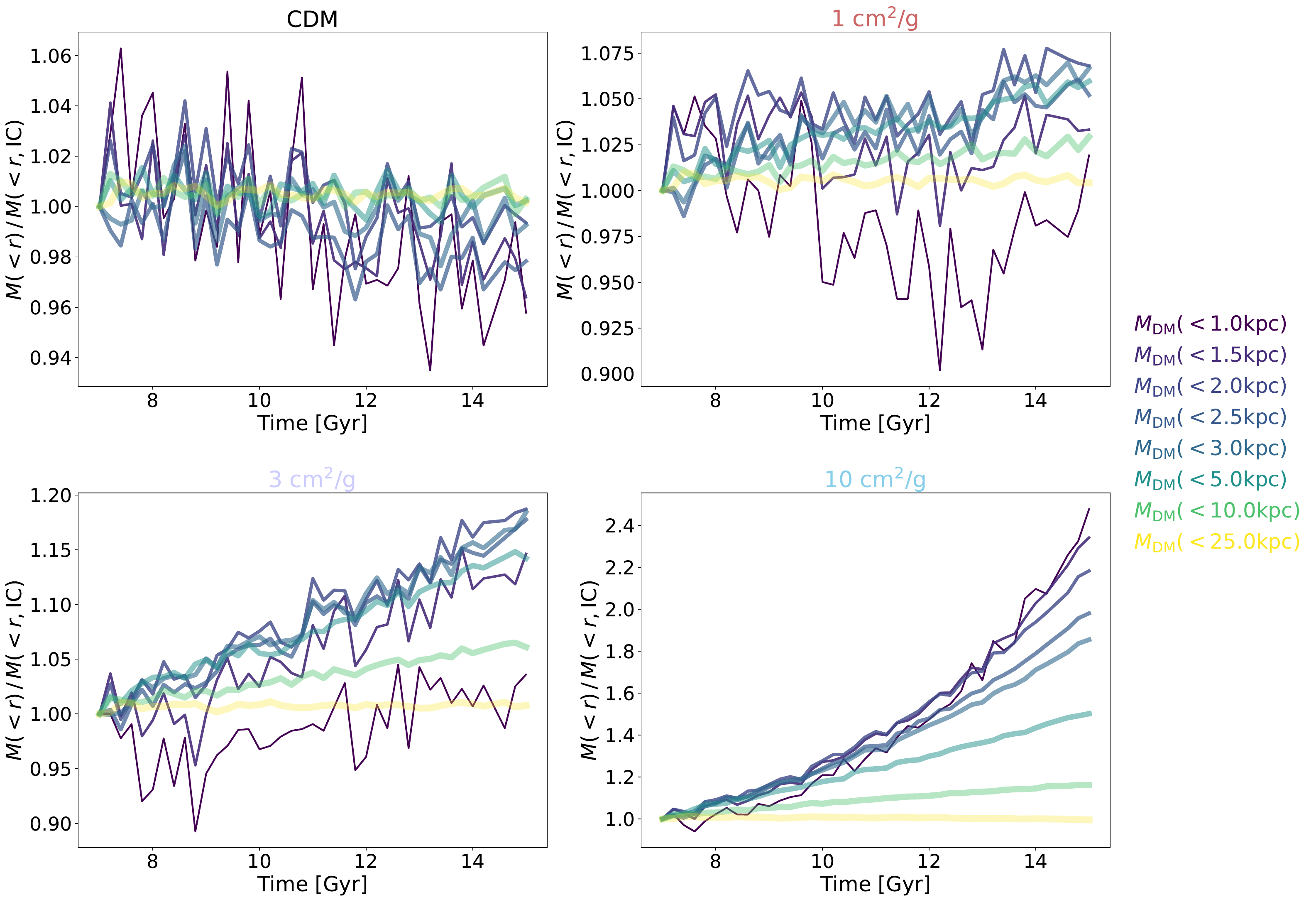}
    \caption{Same as Fig. \ref{fig:mass-varyaperture-dmo}, but for the MW analog with 2/3 default stellar mass and compact distribution (initial $r_{\rm half}=2.5$ kpc).}
    \label{fig:mass-varyaperture-iso-2thirdstars}
\end{figure*}

In this appendix, we present the time evolution of the enclosed dark matter mass at various finely spaced radii, ranging from 1.0 kpc to 25.0 kpc. Figures~\ref{fig:mass-varyaperture-dmo}, \ref{fig:mass-varyaperture-iso}, and \ref{fig:mass-varyaperture-iso-2thirdstars} illustrate these radial mass profiles for the dark-matter-only (DMO) baseline, the fiducial compact-stellar case, and the $2/3 M_\star$ compact-stellar variant, respectively. 

As discussed in Sec.~\ref{sec:iso}, this detailed radial breakdown extends our analysis inward to 1.0 kpc, allowing us to rule out the possibility that a smaller core forms unresolved within the 2.5 kpc aperture. These results confirm that the fiducial compact-stellar MW analog genuinely bypasses the core-formation phase across all central radii. In contrast, the $2/3 M_\star$ case exhibits only a mild mass reduction ($\lesssim 10\%$) at the innermost 1.0 kpc, confirming that it lies near the transition boundary of the bypassed core-formation regime.

\section{Selection of the TNG snapshot as IC}\label{appdx:whichic}

\begin{figure*}
    \centering
    \begin{subfigure}[t]{0.3\textwidth}
        \centering
        \includegraphics[width=\textwidth, clip,trim=0.2cm 0cm 0.2cm 0cm]{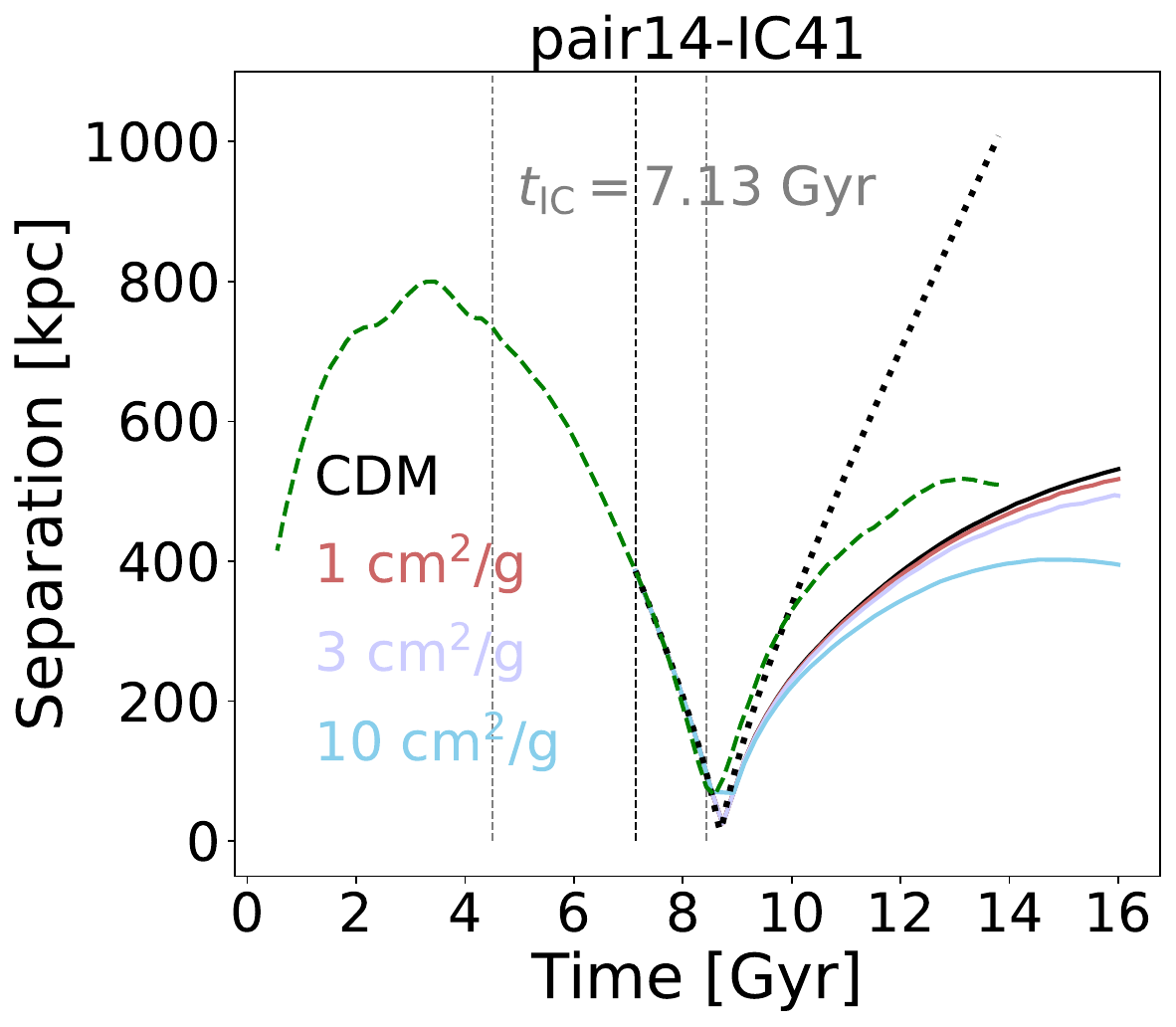}
        \caption{}
        \label{fig:tngic-a}
    \end{subfigure}
    ~
    \begin{subfigure}[t]{0.3\textwidth}
        \centering
        \includegraphics[width=\textwidth, clip,trim=0.2cm 0cm 0.2cm 0cm]{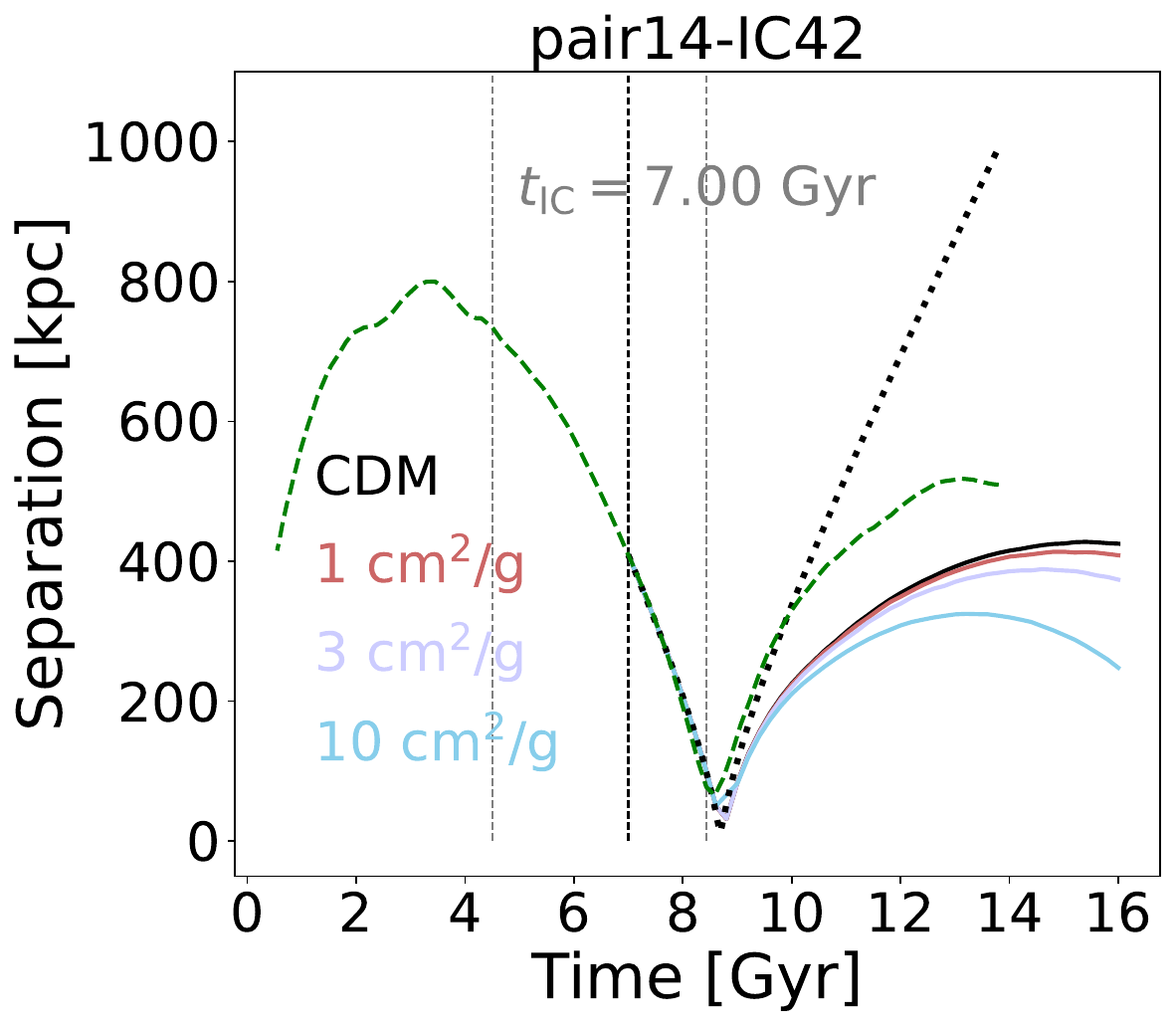}
        \caption{}
        \label{fig:tngic-b}
    \end{subfigure}
    ~
    \begin{subfigure}[t]{0.3\textwidth}
        \centering
        \includegraphics[width=\textwidth, clip,trim=0.2cm 0cm 0.2cm 0cm]{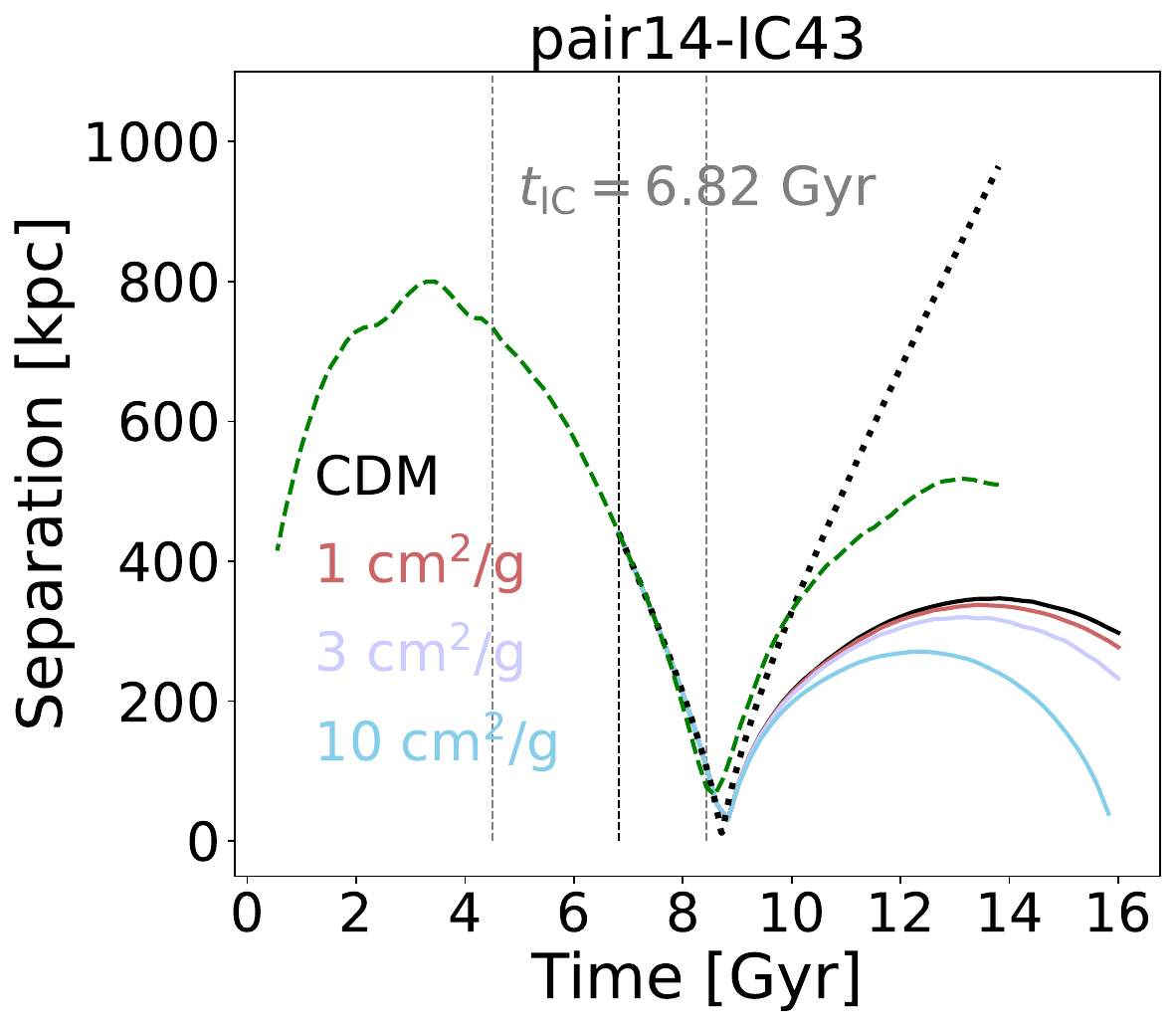}
        \caption{}
        \label{fig:tngic-c}
    \end{subfigure}
    \caption{Orbital separation histories of the `pair14' LG analog using initial conditions extracted from different TNG snapshots: $t_{\rm IC} = 7.13$ Gyr (a), 7.00 Gyr (b), and 6.82 Gyr (c). This comparison demonstrates how the precise choice of $t_{\rm IC}$ influences the resulting orbital evolution.}
    \label{fig:tngic}
\end{figure*}

The initial simulation time $t_{\rm ini}$ is selected following the general procedure outlined in Sec. \ref{sec:method}, with an additional refinement of comparing simulated LG orbits initialized from several adjacent TNG snapshots. We empirically identify the optimal $t_{\rm ini}$ that balances the fidelity of the re-simulated orbital history with the requirement of a negative (approaching) present-day radial velocity. As demonstrated in Fig. \ref{fig:tngic}, the precise choice of the IC time (and thus the corresponding initial state vectors $r_{\rm ini}$, $v_{\rm tan, ini}$ and $v_{\rm rad, ini}$) can noticeably influence the resulting orbital evolution. Consequently, we adopt the snapshot corresponding to $t\approx7.0$ Gyr, which yields the most consistent orbital reconstruction.

\section{Summary of Simulation Suites}\label{appdx:sim_summary}

To provide a self-contained overview of our numerical methodology, we summarize the three distinct simulation suites in Table \ref{tab:sim_summary}. Because the physical motivations, resolutions, and initial conditions for orbit reconstruction (Sec. \ref{sec:orbitsim}), isolated co-evolution (Sec. \ref{sec:iso}), and full tidal encounters (Sec. \ref{sec:lgsim}) differ by design, their detailed setups are introduced sequentially in the main text. This appendix serves as a centralized reference, outlining the kinematics, composition, target systems, particle resolutions, and the primary scientific aim for each simulation phase. For explicit values of individual particle masses and force softening lengths, we refer the reader to Table \ref{table:allpairs} and \ref{table:pair14} in the main text.

Concentrations are assigned according to the mass-concentration relation in \cite{diemer19} for all 15 analogs in Sec. \ref{sec:orbitsim}, but from TNG FoF group catalogs for pair14 in Sec. \ref{sec:iso} and \ref{sec:lgsim} (see Sec. \ref{sec:orbitsim} for rationale). Appendix \ref{appdx:conc_test} tests that the main conclusions are insensitive to this pair14 choice.

\begin{table*}[th!]
\centering
\caption{Overview of the numerical simulation suites. Particle masses and softening lengths are given in the main text (Secs.~\ref{sec:orbitsim} and~\ref{sec:iso}) and in Table~\ref{table:pair14}. A simple robustness test of the Pair~14 concentration choice is presented in Appendix~\ref{appdx:conc_test}.}
\label{tab:sim_summary}
\renewcommand{\arraystretch}{1.5}
\begin{tabular}{l c c c c c p{4.5cm}}
\hline \hline
\textbf{Section} & \textbf{Kinematics} & \textbf{Composition} & \textbf{Target System(s)} & \textbf{Concentration} & \textbf{Total Particles} & \textbf{Main Aim} \\
\hline
\textbf{II} & Orbital & DM-only & 15 LG analogs & Mass-$c$ relation (\cite{diemer19}) & $5 \times 10^5$ & Orbit reconstruction \& candidate selection \\
\textbf{III} & Isolated & DM + Stars & MW analog (pair14) & TNG catalog & $\approx 8.9 \times 10^5$ & Isolated baselines \& SIDM-stellar co-evolution \\
\textbf{IV} & Orbital & DM + Stars & pair14 (MW \& M31) & same as Sec.~III & $\approx 2.0 \times 10^6$ & Orbital encounter \& MW structural response \\
\hline \hline
\end{tabular}
\end{table*}

\section{Robustness to halo concentration assignment}\label{appdx:conc_test}

Our fiducial pair14 simulations in Secs. \ref{sec:iso} and \ref{sec:lgsim} adopt halo concentrations taken from the TNG FoF group catalogs at $z=0$ ($c_{\rm M31}\approx 7.4$, $c_{\rm MW}\approx 12.1$; Table \ref{table:pair14}). As discussed in Sec. \ref{sec:orbitsim}, the orbital-reconstruction survey instead uses concentrations drawn from the mass-concentration relation in \cite{diemer19}. Here we perform a simple sensitivity test to verify that our main physical conclusions do not depend on this specific concentration choice.

We re-run the isolated evolution (Sec. \ref{sec:iso}) and the full pair14 Local Group simulations (Sec. \ref{sec:lgsim}) with an alternative setup in which \emph{both} halos are initialized using concentrations drawn from \cite{diemer19} at their respective masses ($c_{\rm M31}\approx 9.1$, $c_{\rm MW}\approx 9.3$). All other IC properties, particle numbers, and softening lengths are held fixed. We compare this setup against the default TNG-concentration runs for the four DM models considered in the main text (CDM and SIDM with $\sigma/m = 1$, 3, and $10\,{\rm cm^2/g}$).

\begin{figure*}
    \centering
    \includegraphics[width=0.48\textwidth]{iso-encmass-t-2.5kpc-nonexten.pdf}
    \includegraphics[width=0.48\textwidth]{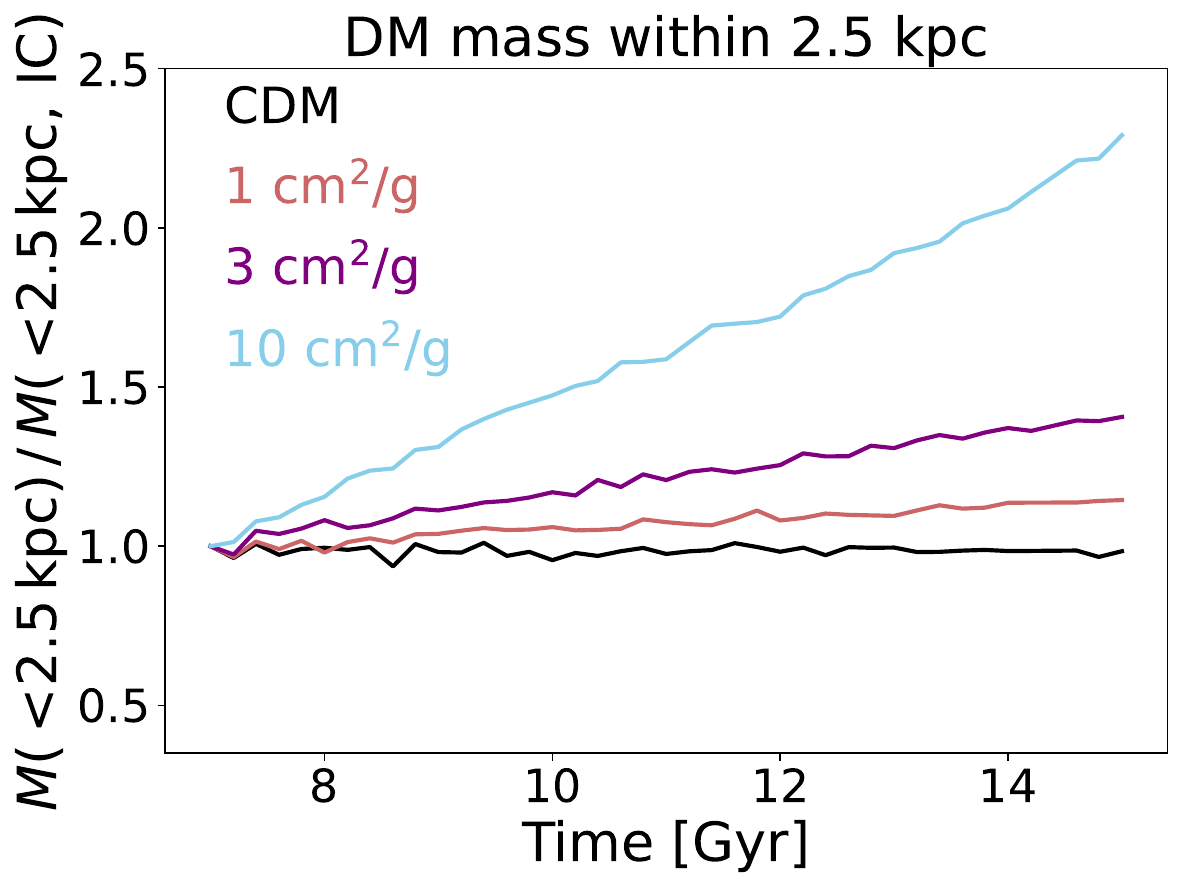}
    \caption{Isolated evolution of the compact MW analog ($r_{\rm half}=2.5$ kpc): enclosed DM mass within 2.5 kpc, normalized to the initial value. Left: fiducial TNG group concentrations ($c_{\rm MW}\approx 12.1$, $c_{\rm M31}\approx 7.4$). Right: both halos drawn from \cite{diemer19} ($c\approx 9$ at each halo mass). Color scheme as in Fig. \ref{fig:iso-timeevo}.}
    \label{fig:conc-iso}
\end{figure*}

\begin{figure*}
    \centering
    \includegraphics[width=0.85\textwidth]{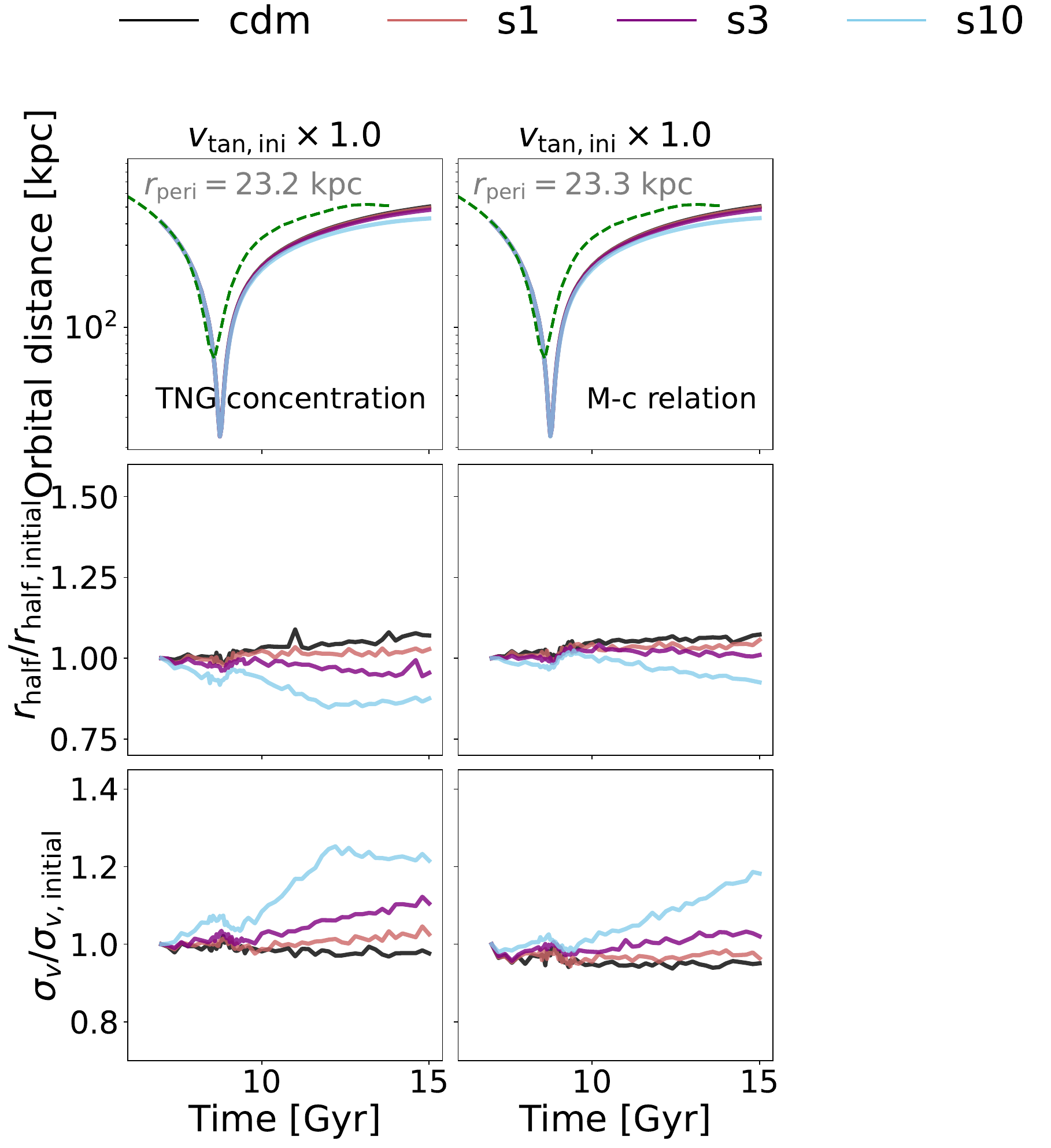}
    \caption{Full pair14 LG simulation, compact stellar component ($r_{\rm half}=2.5$ kpc). Left: fiducial TNG concentrations; Right: concentrations drawn from the mass-concentration relation in \cite{diemer19}. Top: orbital separation; middle and bottom: normalized stellar $r_{\rm half}$ and $\sigma_v$ within $r_{\rm half}$ for $v_{\tan,{\rm ini}}\times 1.0$.}
    \label{fig:conc-lg-compact}
\end{figure*}

\begin{figure*}
    \centering
    \includegraphics[width=0.85\textwidth]{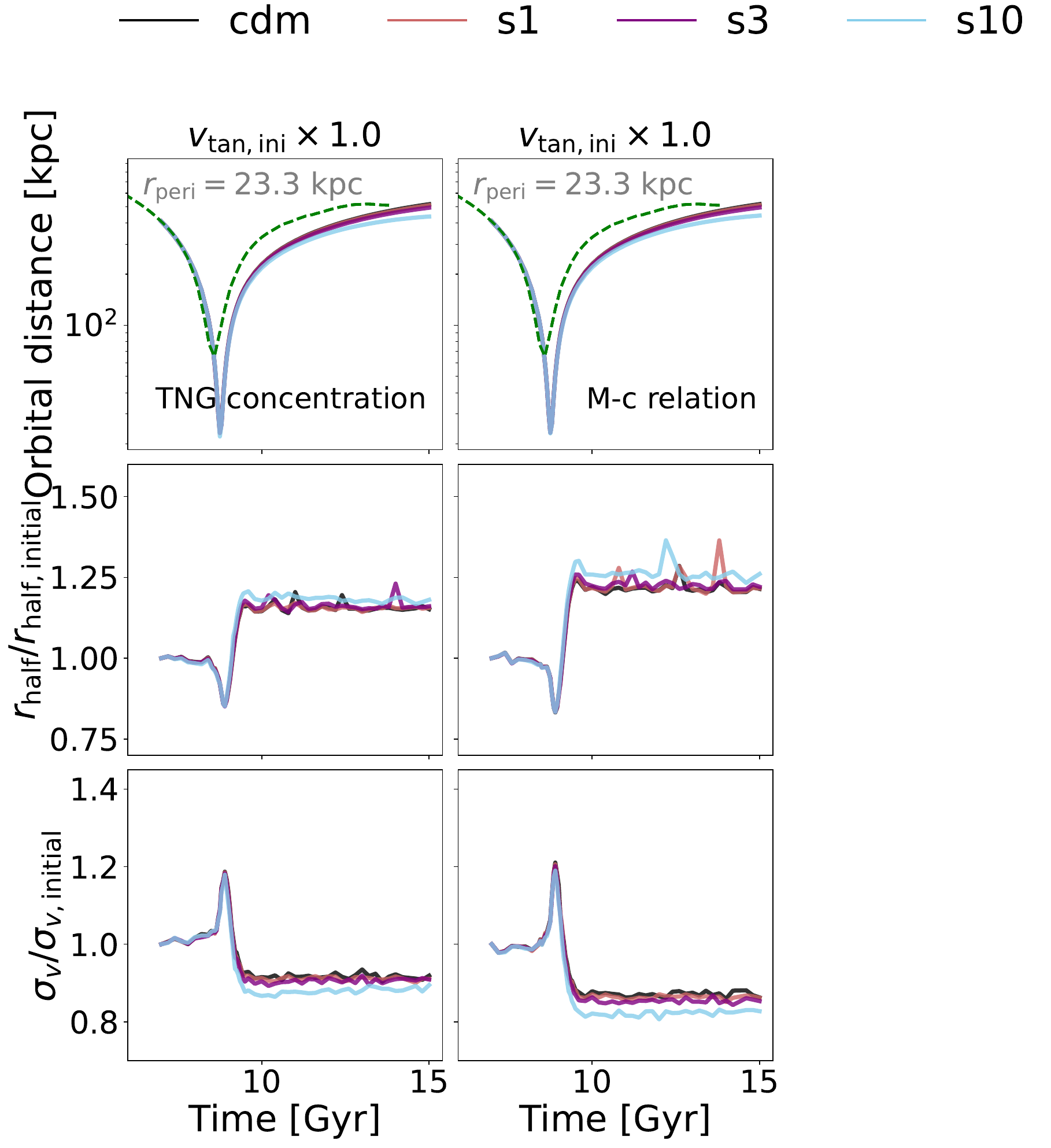}
    \caption{Same as Fig. \ref{fig:conc-lg-compact}, but for the diffuse stellar component ($r_{\rm half}=20$ kpc).}
    \label{fig:conc-lg-diffuse}
\end{figure*}

For the \textit{isolated} evolution (Fig. \ref{fig:conc-iso}), the qualitative behavior is unchanged: the compact-stellar MW analog still exhibits bypassed core formation and proceeds toward core collapse across all SIDM models examined. The only notable quantitative difference is a modest shift in the core-collapse timescale when the MW concentration is lowered from $c\approx 12.1$ to $c\approx 9.3$, consistent with the known sensitivity of gravothermal collapse to halo concentration \cite{essig19, zzc22}.
For the \textit{full LG} simulations (Figs. \ref{fig:conc-lg-compact} and \ref{fig:conc-lg-diffuse}), the orbital trajectories are nearly identical between the two concentration setups ($r_{\rm peri}\approx 23.2$--$23.3$ kpc for the $v_{\tan,{\rm ini}}\times 1.0$ case shown). The two structural conclusions of the main text remain robust: (i) for the compact stellar component, the final $r_{\rm half}$ and $\sigma_v$ are set primarily by intrinsic SIDM thermodynamics rather than by the concentration choice; and (ii) for the diffuse stellar component, the structural properties remain comparatively more sensitive to the pericenter encounter while being largely insensitive to the specific SIDM cross section. We therefore regard the fiducial TNG-based concentration assignment as adequate for the conclusions presented in the main text, while noting that a full exploration of the $(c_{\rm MW}, c_{\rm M31})$ parameter space is left to future work.

\section{Supplementary figures}\label{appx:morefigs}

We include additional supplementary figures in this appendix.

\begin{figure}
    \centering
        \includegraphics[width=0.48\textwidth, clip,trim=0.2cm 0cm 0.2cm 0cm]{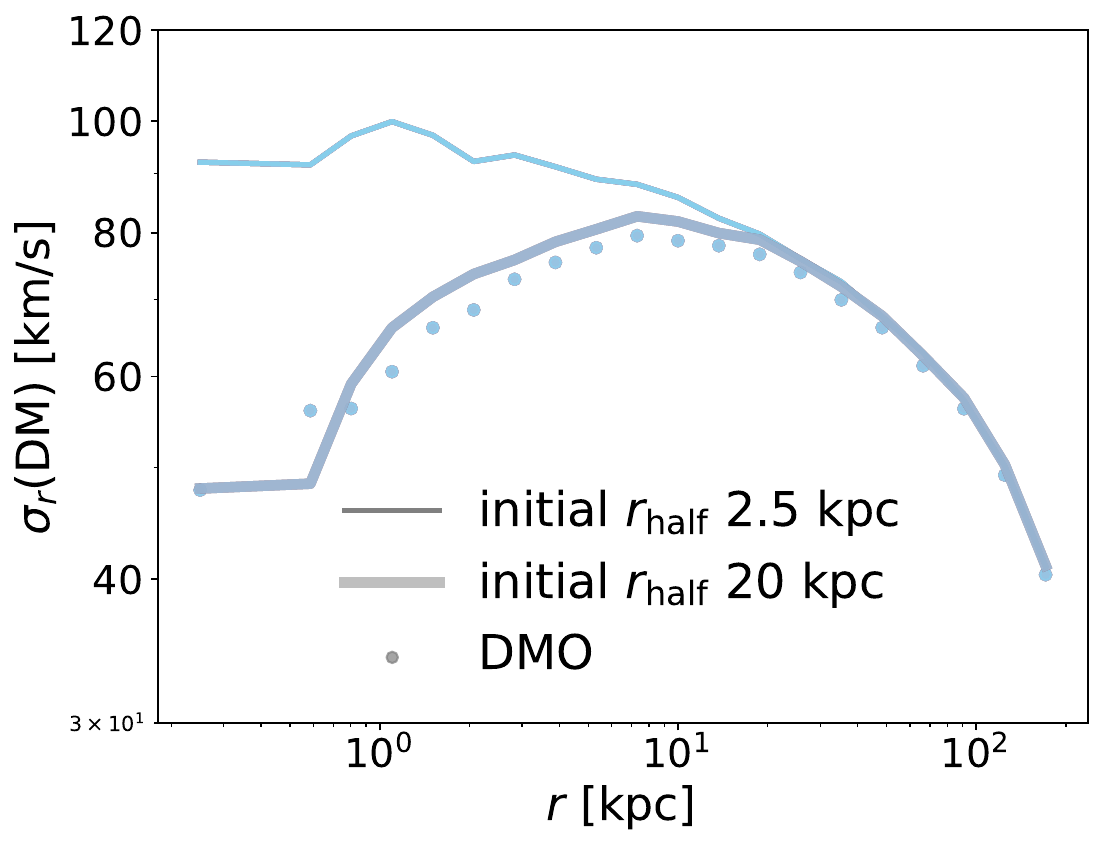}
    \caption{The velocity dispersion profiles of the isolated MW analog, for the initial conditions at $t=7$ Gyr. This figure represents the un-evolved, initial configuration of Fig. \ref{fig:final-pro-f}.}
    \label{fig:ic-sigmav}
\end{figure}


\bibliography{apssamp}

\end{document}